\documentclass[]{aastex631}
\usepackage{amsmath}
\usepackage{tabularx}
\usepackage{tablefootnote}

\newcommand{\mytilde}{\raise.19ex\hbox{$\scriptstyle\sim$}}

\shorttitle{Dark Matter in Radio Relic Merging Clusters}
\shortauthors{Finner et al.}
\graphicspath{{./}{figures/}}
\begin{document}

\title{Weak-Lensing Characterization of the Dark Matter in 29 Merging Clusters that Exhibit Radio Relics}

\correspondingauthor{Kyle Finner}
\email{kfinner@caltech.edu}

\author[0000-0002-4462-0709]{Kyle Finner}
\affiliation{IPAC, California Institute of Technology, 1200 E California Blvd., Pasadena, CA 91125, USA}
\affiliation{Department of Astronomy, Yonsei University, 50 Yonsei-ro, Seodaemun-gu, Seoul 03722, Republic of Korea}
\author[0000-0002-5751-3697]{M. James Jee}
\affiliation{Department of Astronomy, Yonsei University, 50 Yonsei-ro, Seodaemun-gu, Seoul 03722, Republic of Korea}
\affiliation{Department of Physics and Astronomy, University of California, Davis, One Shields Avenue, Davis, CA 95616, USA}
\author[0000-0001-5966-5072]{Hyejeon Cho}
\affiliation{Department of Astronomy, Yonsei University, 50 Yonsei-ro, Seodaemun-gu, Seoul 03722, Republic of Korea}
\affiliation{Center for Galaxy Evolution Research, Yonsei Univesity, 50 Yonsei-ro, Seodaemun-gu, Seoul 03722, Republic of Korea}
\author[0000-0002-2550-5545]{Kim HyeongHan}
\affiliation{Department of Astronomy, Yonsei University, 50 Yonsei-ro, Seodaemun-gu, Seoul 03722, Republic of Korea}
\author[0000-0002-1566-5094]{Wonki Lee}
%\altaffiliation{Sejong Fellow}
\affiliation{Department of Astronomy, Yonsei University, 50 Yonsei-ro, Seodaemun-gu, Seoul 03722, Republic of Korea}
\author[0000-0002-0587-1660]{Reinout J. van Weeren}
\affiliation{Leiden Observatory, Leiden University, P.O. Box 9513, 2300 RA Leiden, The Netherlands}
\author[0000-0002-0813-5888]{David Wittman}
\affiliation{Department of Physics and Astronomy, University of California, Davis, One Shields Avenue, Davis, CA 95616, USA}
\author[0000-0002-3683-9559]{Mijin Yoon}
\affiliation{Leiden Observatory, Leiden University, P.O. Box 9513, 2300 RA Leiden, The Netherlands}



\begin{abstract}
We present a multiwavelength analysis of 29 merging galaxy clusters that exhibit radio relics. For each merging system, we perform a weak-lensing analysis on Subaru optical imaging. We generate high-resolution mass maps of the dark matter distributions, which are critical for discerning the merging constituents. Combining the weak-lensing detections with X-ray emission, radio emission, and galaxy redshifts, we discuss the formation of radio relics from the past collision. For each subcluster, we obtain mass estimates by fitting a multi-component NFW model with and without a concentration-mass relation. Comparing the two mass estimate techniques, we find that the concentration-mass relation underestimates (overestimates) the mass relative to fitting both parameters for high- (low-) mass subclusters. We compare the mass estimates of each subcluster to their velocity dispersion measurements and find that they preferentially lie below the expected velocity dispersion scaling relation, especially at the low-mass end (\mytilde$10^{14}\ M_\odot$). We show that the majority of the clusters that exhibit radio relics are in major mergers with a mass ratio below 1:4. We investigate the position of the mass peak relative to the galaxy luminosity peak, number density peak, and BCG locations and find that the BCG tends to better trace the mass peak position. Finally, we update a golden sample of 8 galaxy clusters that have the simplest geometries and can provide the cleanest picture of the past merger, which we recommend for further investigation to constrain the nature of dark matter and the acceleration process that leads to radio relics.
\end{abstract}


%% Sections are demarcated by \subsection and \subsection, respectively.

\section{Introduction} \label{sec:intro}
On large scales of the universe, dark matter, gas, and stars are organized into galaxy clusters that are connected by filaments: the cosmic web \citep{1980peebles, 1996bond}. Galaxy clusters grow through the hierarchical buildup of mass where low-mass structures form and merge to produce more massive structures. Major growth events occur when galaxy clusters merge.

Merging galaxy clusters provide an extreme environment to study the universe in regions where the most energetic events since the Big Bang occur \citep{2002sarazin}. The merging of galaxy clusters is driven by the gravitational force. It begins with clusters detaching from the cosmic expansion and starting a Gyrs plunge toward each other. As the outskirts of the galaxy clusters come in contact (a few virial radii separated), the intracluster medium (ICM), able to undergo electromagnetic interactions, feels a ram pressure as gas particles collide \citep[see for example][for gas density profiles of clusters]{2012eckert}. The relative velocity of the galaxy clusters increases as they approach pericenter and particle interactions become more frequent. The clusters begin moving at supersonic velocities with Mach numbers in the range of 1 to 5 \citep{2003gabici}. At these velocities, ram pressure becomes an important perturber of the ICM causing a significant momentum exchange between the gas of the two clusters and consequently a deceleration. Unlike the ICM, galaxies, being sparse in the cluster environment, are effectively collisionless and follow a path that gravity defines. This leads to the open and alluring question as to the nature of dark matter, the most massive component of galaxy clusters. The Bullet cluster (1E 0657-56) is a prime example of a cluster that is valuable to understanding the nature of dark matter. The cluster demonstrates a clear spatial separation of the ICM from the dark matter and galaxies \citep{2004markevitch, 2006clowe}. This configuration provides a laboratory to study the nature of dark matter \citep[e.g.,][]{2008randall}. %If dark matter interacts only through the gravitational force, then it is expected to be spatially coherent to the galaxies \citep{1988frenk}. However, if it can interact (for example through self interactions), then it may also slow in its orbit relative to the galaxies \citep{2017kim}.

A result of clusters colliding at supersonic velocities is the formation of shocks. These shocks can be classified as equatorial shocks that propagate perpendicular to the merger axis and bow (or axial) shocks that propagate along the merger axis \citep{2018ha}. Shocks are visible in X-ray observations as an abrupt surface brightness drop but one must take care to not confuse them with contact discontinuities \citep[e.g.,][]{2007markevitch}. Shocks are also sometimes visible in radio observations as radio relics (also known as cluster radio shocks), which in some of the best cases appear as giant (up to a few Mpc) arc-shaped radio emission \citep[see][for a review]{2019vanweeren}.

Radio relics are extended synchrotron emission from charged particles that are accelerated by the magnetic fields of merger-induced shocks. It is postulated that diffusive shock acceleration (DSA) is the primary acceleration mechanism. However, the inefficiency of DSA to accelerate electrons from a thermal distribution to energies that are sufficient to see the observed radio relic brightness has been an issue given the low Mach number of these shocks \citep{2007brunetti, 2012kang, 2020botteon}. The re-acceleration of previously energized electrons is the favored scenario with evidence compounding \citep{2014bonafede, 2017vanweeren, 2022lee_zwcl1447}. The rarity of radio relics, with not all merging clusters hosting them, remains another unresolved issue. Simulations have shown that projection may be somewhat to blame for the rarity of radio relics and the interpretation of radio emission \citep{2012vazza, 2013skillman, 2024lee}. Furthermore, if re-acceleration is required, then a pre-existing population of suprathermal electrons should be present for the passing shock to accelerate.

Merging clusters are an ideal laboratory to study the nature of dark matter, high-energy astrophysics, and the evolution of the large-scale structure of the universe. However, observations provide only a single snapshot into the Gyrs-long process. Therefore, to gain an understanding of the progression of the merger, multiwavelength observations and a large sample of cluster mergers are needed. X-ray and radio observations provide insight into the ICM. Optical and IR observations track the galaxies. Dark matter, which by definition does not emit light, can be detected through its gravitational potential via the gravitational lensing effect.

Weak gravitational lensing (WL hereafter) is a statistical analysis of galaxy shapes. WL analysis permits the detection of the total mass distribution of galaxy clusters \citep[for a review see][]{2020umetsu}. It provides valuable information on the centroid and morphology of the dark matter. It can discern substructures and detect the connection to the large-scale structure \citep{2015eckert, 2024hyeonghan}, which is critical to understanding the formation and evolution of clusters. Additionally, it can be used to estimate the masses of the substructures. By combining the WL results with multiwavelength observations of the ICM and stars, the past collision of merging clusters can be reconstructed.

This work entails a WL analysis of galaxy clusters that exhibit radio relics. The primary goal of the work is to identify and characterize the substructures that are merging using the WL effect. In Section \ref{Section:data}, observations, data reduction, and photometry are described. Section \ref{Section:theory} presents an overview of the WL pipeline, which covers WL formalism, point-spread function (PSF) modeling, source selection, lensing efficiency, convergence mapping, substructure identification, and mass estimation. Results for each cluster are described in Section \ref{Section:results} and the clusters are put into the context of the multi-wavelength literature. Section \ref{Section:discussion} discusses the sample of clusters as a whole and presents a statistical analysis of the cluster masses. We summarize our conclusions in Section \ref{Section:conclusions}.

All magnitudes are presented in the AB magnitude system. In all calculations and presentations, a flat $\Lambda$CDM cosmology is assumed with $H_0$=70 km s$^{-1}$ Mpc$^{-1}$, $\Omega_m=0.3$ and $\Omega_\Lambda$=0.7. Masses are defined as the mass within a sphere of radius $R_\Delta$, where $R_\Delta$ is the radius at which the average density within is $\Delta$ times the critical density of the universe at the redshift of the cluster.

\subsection{Radio Relic Sample}\label{intro:mc2}
The sample of clusters that are analyzed in this work exhibits megaparsec-scale radio emission. The identification of the radio relics was done in past radio studies such as \cite{2011vanweeren} and \cite{2012feretti}. Some of the clusters have bow-shaped radio emission that clearly resembles a spherical shock. However, some have patchy radio emission that may have arisen from other particle acceleration phenomena. The study of radio relics is a very active field of astronomy and new insights into the nature of the radio relics are occurring daily. We include the most up-to-date information of the radio relics in our interpretations of the mergers.

By design, the galaxy clusters studied in this work are merging systems. This selection should be considered when making conclusions. The detection of these clusters is dependent on the brightness of the radio emission, and therefore the sample is probing bright radio sources. Furthermore, the existence of the radio relics is an indication that the clusters are likely post-pericenter systems \citep{2012vazza, 2013skillman, 2024lee}. Conclusions in this work are made with the understanding that the clusters belong to a small subset of galaxy clusters and may not be representative of all merging clusters.

Table \ref{table:MC2} lists the radio relic merging clusters that are studied here. The sample of radio relic clusters range in redshift from 0.07 to 0.54. This range of redshifts is ideal for ground-based WL due to the dependence of the lensing signal on the distances of the lens and background galaxies. \citet[][hereafter G19]{2019bgolovich} performed a thorough dynamical analysis of the clusters utilizing a vast catalog of spectroscopic redshifts. Their study identified subclusters using Gaussian mixture modeling (GMM) and derived velocity dispersions. Combining X-ray and radio observations with the GMM-defined subclusters, they developed merger scenarios for each of the systems. A key finding of their study is that the majority of these clusters are merging near the plane of the sky. \cite{2018wittman} provide quantitative constraints on the viewing angle of some of the clusters.

\begin{table*}[!ht]
\def\arraystretch{0.8}
\centering
\caption{Overview of 29 merging cluster sample}
\resizebox{\textwidth}{!}{%
\begin{tabular}{llllll}
\hline
\hline
Cluster             & Short name & R.A.    & Decl.     & Redshift & Discovery Band \\
\hline
1RXS J0603.3+4212   & 1RXSJ0603  & 06:03:13.4 & +42:12:31 & 0.226    & Radio          \\
Abell 115           & A115       & 00:55:59.5 & +26:19:14 & 0.193    & Optical        \\
Abell 521           & A521       & 04:54:08.6 & -10:14:39 & 0.247    & Optical        \\
Abell 523           & A523       & 04:59:01.0 & +08:46:30 & 0.104    & Optical        \\
Abell 746           & A746       & 09:09:37.0 & +51:32:48 & 0.214    & Optical        \\
Abell 781           & A781       & 09:20:23.2 & +30:26:15 & 0.297    & Optical        \\
Abell 1240          & A1240      & 11:23:31.9 & +43:06:29 & 0.195    & Optical        \\
Abell 1300          & A1300      & 11:32:00.7 & -19:53:34 & 0.306    & Optical        \\
Abell 1612          & A1612      & 12:47:43.2 & -02:47:32 & 0.182    & Optical        \\
Abell 2034          & A2034      & 15:10:10.8 & +33:30:22 & 0.114    & Optical        \\
Abell 2061          & A2061      & 15:21:20.6 & +30:40:15 & 0.078    & Optical        \\
Abell 2163          & A2163      & 16:15:34.1 & -06:07:26 & 0.201    & Optical        \\
Abell 2255          & A2255      & 17:12:50.0 & +64:03:11 & 0.080     & Optical        \\
Abell 2345          & A2345      & 21:27:09.8 & -12:09:59 & 0.179    & Optical        \\
Abell 2443          & A2443      & 22:26:02.6 & +17:22:41 & 0.110     & Optical        \\
Abell 2744          & A2744      & 00:14:18.9 & -30:23:22 & 0.306    & Optical        \\
Abell 3365          & A3365      & 05:48:12.0 & -21:56:06 & 0.093    & Optical        \\
Abell 3411          & A3411      & 08:41:54.7 & -17:29:05 & 0.163    & Optical        \\
CIZA J2242.8+5301   & CIZAJ2242  & 22:42:51.0 & +53:01:24 & 0.189    & X-ray          \\
MACS J1149.5+2223   & MACSJ1149  & 11:49:35.8 & +22:23:55 & 0.544    & X-ray          \\
MACS J1752.0+4440   & MACSJ1752  & 17:52:01.6 & +44:40:46 & 0.365    & X-ray          \\
PLCK G287.0+32.9    & PLCKG287   & 11:50:49.2 & -28:04:37 & 0.383    & SZ             \\
PSZ1 G108.18-11.53  & PSZ1G108   & 23:22:29.7 & +48:46:30 & 0.335    & SZ             \\
RXC J1053.7+5452    & RXCJ1053   & 10:53:44.4 & +54:52:21 & 0.072    & X-ray          \\
RXC J1314.4-2515    & RXCJ1314   & 13:14:23.7 & -25:15:21 & 0.247    & X-ray          \\
ZwCl 0008.8+5215    & ZwCl0008   & 00:11:25.6 & +52:31:41 & 0.104    & Optical        \\
ZwCl 1447+2619      & ZwCl1447   & 14:49:28.2 & +26:07:57 & 0.376    & Optical        \\
ZwCl 1856.8+6616    & ZwCl1856   & 18:56:41.3 & +66:21:56 & 0.304    & Optical        \\
ZwCl 2341+0000      & ZwCl2341   & 23:43:39.7 & +00:16:39 & 0.270     & Optical \\
\hline
\end{tabular}}
\label{table:MC2}
\begin{minipage}{13cm}
   \small Source: \cite{2019agolovich}
\end{minipage}
\end{table*}

\section{OBSERVATIONS AND DATA REDUCTION} \label{Section:data}
Our analysis combines WL with X-ray, radio, and optical observations to provide a multi-wavelength insight into past cluster-cluster merger events that resulted in radio relics. In this section, we describe the multi-wavelength observations that we utilize in our analysis.

\subsection{Subaru Observations}\label{section:subaru_observations}
The observations used in this study to make WL measurements were obtained with the Subaru telescope. The Subaru telescope has a primary mirror diameter of 8.2~m and is an optical/infrared telescope located at the top of Mauna Kea on Hawaii. The optical observations used in this study come from two instruments: Suprime-Cam \citep{2002miyazaki} and Hyper Suprime-Cam \citep[HSC;][]{2018miyazaki, 2018komiyama, 2018kawanomoto, 2018furusawa}. The Suprime-Cam is a $5 \times 2$ array of $4\mathrm{k} \times 2\mathrm{k}$ Hamamatsu CCDs with a $34\arcmin \times 27\arcmin$ field of view. In 2017, the Suprime-Cam was decommissioned and the HSC became the primary imager. The HSC is a 116 CCD instrument (104 science detectors) with a 1.5 degree diameter field of view.

The imaging presented in this work are summarized in Table \ref{table:subaru_imaging}. For about 1/3 of the clusters, archival imaging is available. For those that lack Subaru imaging, additional observations were gathered by PI: D. Wittman and more recently by PI: H. Cho. These observations were performed with WL analysis in mind. A dithering and rotation technique was executed to minimize degradation of the images from the diffraction spikes caused by bright stars, bad columns, cosmic rays, and bleeding trails. This imaging technique has been shown to increase the number of usable galaxies available for WL studies \citep{2015jee, 2016jee}.

WL analysis requires a careful modeling of the PSF because it mimics the lensing effect. For ground-based observations, the PSF is predominately affected by the atmospheric seeing. Subaru observations were obtained in some of the best seeing conditions with typically sub-arcsecond seeing achieved (Table \ref{table:subaru_imaging}). For each cluster, we selected the filter for WL analysis by carefully balancing exposure time (maximize) and seeing (minimize) with a preference given to the redder filter when the choice was close. Redder filters are preferred in WL analysis because the redder light from galaxies tends to be less clumpy than the blue, star-forming light and is thus better fit with a smooth model \citep{2018lee}. The filter selected for WL is highlighted in bold font in Table \ref{table:subaru_imaging}.

\subsection{Subaru Data Reduction}\label{section:subaru_reduction}
Data reduction requires special care to produce a WL quality image. Subaru \texttt{ SDFRED} 1 (pre 2009) and 2 (post 2009) packages\footnote{http://subarutelescope.org/Observing/Instruments/SCam/sdfred} were used for the basic reduction steps of overscan subtraction, bias correction, flat-fielding, and distortion correction for the Suprime-Cam images. SExtractor \citep{1996bertin} was run on each frame to prepare a catalog of objects for SCAMP \citep{bertinscamp}. SCAMP was applied to correct for residual distortions and to refine the World Coordinate System (WCS) for each frame. We used the Pan-STARRS photometric catalog \citep{2016chambers} as a reference in SCAMP to compute the astrometry. For the HSC images, the LSST Science Pipelines \citep{2018bosch, 2019bosch, 2022jenness} were used for overscan/bias/dark subtraction, flat fielding, astrometric correction, etc.

The final step in creating a WL-quality image is to co-add the frames into a mosaic image. The best $S/N$ mosaic is created by mean averaging the frames. However, the frames are prone to cosmic rays, saturation trails, and other detector effects. For this reason, a two-step process was used to co-add the frames. First, SWarp \citep{bertinswarp} was applied to the frames to generate a median-stacked mosaic image. In preparation for co-addition, SWarp translates, rotates, and distortion corrects the frames. These individual \texttt{RESAMP} frames are critical to our PSF modeling (Section \ref{section:psf modeling}) and were stored for later use. After resampling, the frames were carefully aligned using the WCS solution from SCAMP and co-added by SWarp into a median-stacked mosaic image. The median-stacking process outputs a weight file for each component frame. We utilized the weight files to remove the aforementioned spurious signals. The median-stacked mosaic image was compared to the \texttt{RESAMP} frames and pixels from the \texttt{RESAMP} frames that deviated more than 3 times from the root mean square (rms) of the median-stacked image were set to zero in the corresponding weight file. A second pass of SWarp was done by weight-averaging the input frames, and a mean-stacked mosaic image was created. This process was performed on the filter with the longest exposure time first, and then the header information was applied to the remaining filters to ensure the same alignment and footprint.

\begin{figure*}[!ht]
    \centering
    \includegraphics[width=\textwidth]{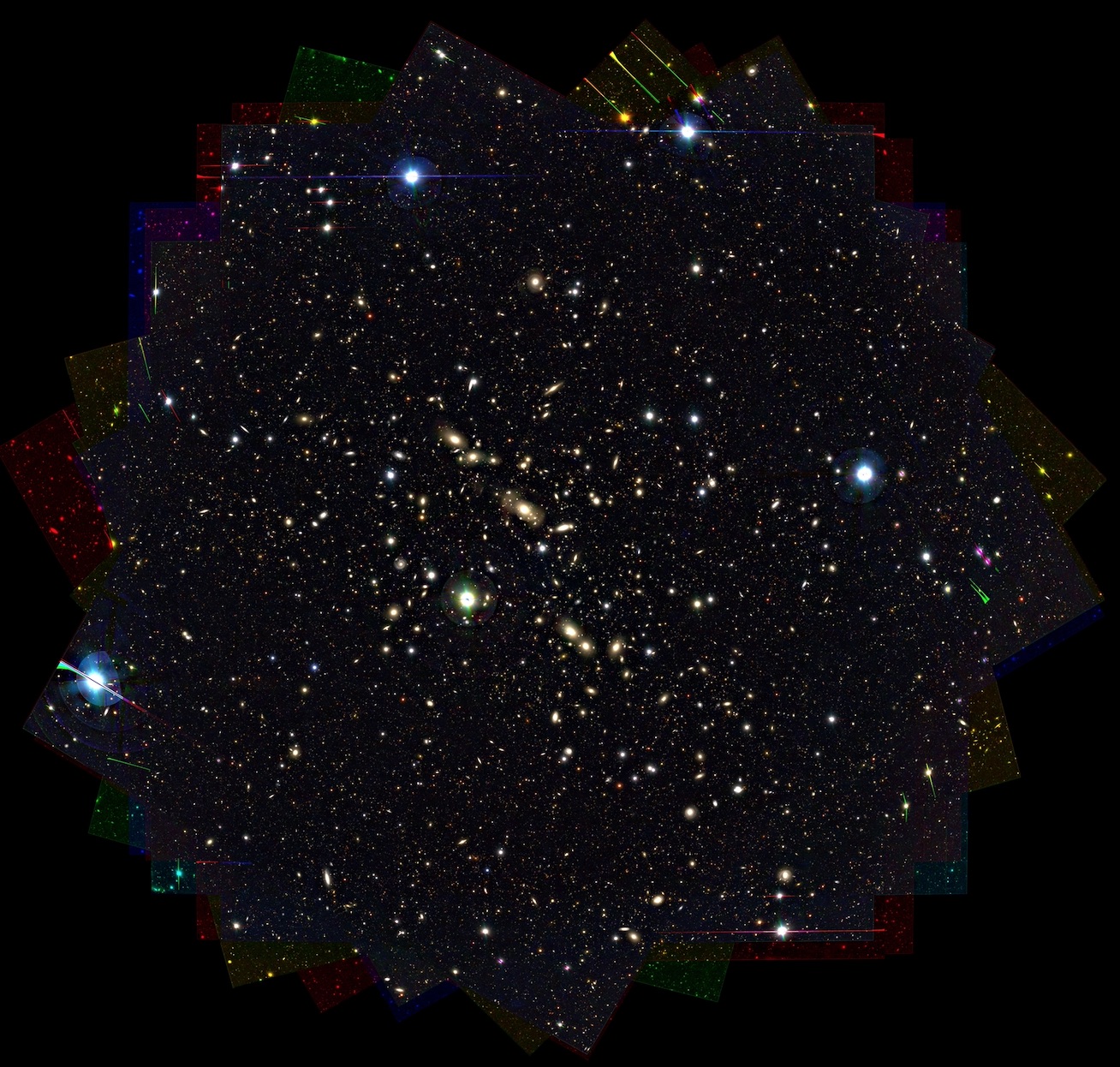}
    \caption{The Subaru color image of the galaxy cluster A2061 provides an example of the quality of the imaging (blue=g-band, green=r-band, red=i-band). The rotation and dithering technique of the observations is noticeable at the perimeter of the image and is beneficial to WL analysis because it minimizes the diffraction spikes caused by stars and allows a better sampling of the PSF. }
    \label{fig:abell2061_full}
\end{figure*}

Figure \ref{fig:abell2061_full} presents a color version of the co-added mosaic of A2061. The three channels of the RGB image are $i$, $r$, and $g$, respectively. The rotation and dithering technique is noticeable around the edges of the mosaic and the benefit of the technique is apparent by comparing the bright star in the center to the stars near the edge. The Suprime-Cam's large field of view provides ample coverage to perform a WL analysis of a galaxy cluster (\mytilde4 Mpc diameter field of view at z=0.1).

% \usepackage[normalem]{ulem}
\begin{table*}[!ht]
\def\arraystretch{0.8}
\centering
\caption{Subaru imaging}
\resizebox{\textwidth}{!}{%
\begin{tabular}{lllllll}
\hline
\hline
Cluster               & Filters & Dates                                & Seeing & Exposure & Source Density & $\left<\beta\right>$ \\
& & & [arcsec] & [s] & [arcmin$^{-2}$] \\
\hline
1RXSJ0603 & g, \textbf{r}      & 2014/02/25, 2014/02/25 & 0.57, \textbf{0.57} & 720, \textbf{2800} & 31 & 0.66          \\
A115             & \textbf{V}, i      & 2003/09/25, 2005/10/03 & \textbf{0.58}, 0.65 & \textbf{1530}, 2100 & 24 & 0.54        \\
A521             & V, R, \textbf{i}      & 2001/10/14, 2001/10/15 & 0.59, 0.65, \textbf{0.59} & 1800, 1620, \textbf{2040} & 44 & 0.63         \\
A523             & g, \textbf{r}      & 2014/02/26 & 1.00, \textbf{0.78} & 720, \textbf{2880} & 19 & 0.81       \\
A746             & g, \textbf{r2}    & 2023/01/16,20,22 & 1.26, \textbf{0.75} & 3480, \textbf{4320} & 24 & 0.63          \\
A781             & V, \textbf{i}      & 2010/03/14, 2010/03/15   & 0.90, \textbf{0.80}  & 3360, \textbf{2160} & 28 & 0.55         \\
A1240            & g, \textbf{r}      & 2014/02/25   & 0.67, \textbf{0.58}   & 720, \textbf{2880} & 43 & 0.73         \\
A1300            & g, \textbf{r}      & 2014/02/26   & 0.89, \textbf{0.88} & 720, \textbf{2913} & 23 & 0.55          \\
A1612            & g, \textbf{i}      & 2014/02/25, 2010/04/11    & 0.62,\textbf{ 0.65} & 2880, \textbf{1920} & 19 & 0.70         \\
A2034            & g, \textbf{R}      & 2005/04/11, 2007/06/19    & 0.82, \textbf{0.90} & 720, \textbf{12880} & 35 & 0.82      \\
A2061            & g, \textbf{r}, i    & 2013/07/13, 2014/02/26   & 0.68, \textbf{0.67}, 0.65  & 720, \textbf{2550}, 4120 & 44 & 0.87 \\
A2163            & V, \textbf{R}      & 2009/04/30, 2008/04/07    & 0.70, \textbf{0.75}        & 2100, \textbf{4500} & 27 & 0.69  \\
A2255            & B, R, \textbf{i}      & 2007/08/14, 2008/07/30 & 0.98, 1.00, \textbf{0.64}  & 1260, 2520, \textbf{1200} & 26 & 0.86    \\
A2345            & V, \textbf{i}      & 2010/06/10, 2010/11/10, 2005/10/03 & 0.70, \textbf{0.72} & 3600, \textbf{2100} & 17 & 0.68       \\
A2443            & $-$      & $-$ & $-$ & $-$ & $-$ & $-$       \\
A2744            & B, R, \textbf{z}      & 2013/07/15, 2013/07/16 & 1.00, 1.14, \textbf{0.79}  & 2100, 3120, \textbf{3600} & 25 & 0.52 \\
A3365            & g, r, \textbf{i}      & 2014/02/25             & 0.97, 0.71, \textbf{0.62}  & 720, 720, \textbf{2880} & 20 & 0.81       \\
A3411            & g, r, \textbf{i}      & 2014/02/25             & 0.8, 0.82, \textbf{0.77}   & 1000, 720, \textbf{2880} & 13 & 0.69     \\
CIZAJ2242     & g, \textbf{i}      & 2013/07/13                & 0.63, \textbf{0.55}        & 720, \textbf{2880} & 14 & 0.62         \\
MACSJ1149     & V, \textbf{R}      & 2003/04/5, 2005/03/05, 2010/03/18 & 0.90, \textbf{0.86} & 2520, \textbf{5490} & 26 & 0.40       \\
MACSJ1752 & g, r, \textbf{i}      & 2013/07/13               & 0.62, 0.64, \textbf{0.73} & 2520, 720, \textbf{4400} & 26 & 0.62    \\
PLCKG287   & g, \textbf{r}      & 2014/02/26                & 0.81, \textbf{0.97} & 720, \textbf{2880} & 27 & 0.55       \\
PSZ1G108    & g, \textbf{i2}      & 2017/08/20, 2017/08/19        & 0.65, \textbf{0.72}        & 1440, \textbf{2880} & 15 & 0.48    \\
RXCJ1053      & g, \textbf{r}      & 2014/02/26                & 0.83, \textbf{0.92}        & 720, \textbf{2910} & $-$ & $-$          \\
RXCJ1314      & g, \textbf{r}      & 2014/02/25                & 0.86, \textbf{0.71}        & 720, \textbf{2880} & 18 & 0.64      \\
ZwCl0008  & g, \textbf{r}      & 2013/07/13                & 0.52, \textbf{0.57}        & 720, \textbf{2880} & 24 & 0.81       \\
ZwCl1447        & g, \textbf{r}, i      & 2014/02/26             & 0.91,\textbf{ 0.76}, 0.55  & 720, \textbf{2880}, 720 & 44 & 0.52        \\
ZwCl1856    & g, \textbf{r}      & 2015/09/12                & 0.70, \textbf{0.65}        & 720, \textbf{2520} & 31 & 0.55      \\
ZwCl2341    & g, \textbf{r}      & 2013/07/13                & 0.49, \textbf{0.50}        & 720, \textbf{2880} & 20 & 0.69          \\
\hline
\end{tabular}}
\label{table:subaru_imaging}
\begin{minipage}{14cm}
   \small The filter used for WL is in bold font.
\end{minipage}
\end{table*}

\subsection{Keck/DEIMOS Observations} \label{section:spectroscopy}
Throughout this study, we utilize redshifts that were measured from Keck DEep Imaging Multi-Object Spectrograph (DEIMOS) observations. The survey and data reduction of the observations are thoroughly described in \cite[][]{2019agolovich}. In G19, they performed an analysis of the spectroscopic redshifts and identified subclusters utilizing a GMM method.

\subsection{MMT Hectospec Observations and Data Reduction}
\label{section:MMT}
Some of the clusters in the G19 work have insufficient numbers of spectroscopic redshifts to discern substructures. We collected MMT/Hectospec \citep{2005fabricant} fiber observations (PI: K.Finner) at the 6.5~m monolithic-mirrored MMT telescope on Mount Hopkins, Arizona. Cluster targets were selected based on the scarcity of their spectroscopic coverage and on the prospect of them having detectable large scale filaments. The MMT/Hectospec is an efficient instrument for simultaneously achieving these two scientific goals because it has a one-degree diameter field of view and 300 fibers. However, fibers collide at distances of about $20\arcsec$, which limits how densely the cores of clusters can be sampled, especially for those at higher redshifts.

We observed A521, A746, A1240, and A2443 with approximately 1.5 hours of exposure time each. Two configurations of the Hectospec instrument were designed for each galaxy cluster utilizing the 270 grating (spectral range 3650 - 9200 \AA). The goal of the observations was to securely measure redshifts for galaxies that reside in the clusters. We plotted color-magnitude diagrams (CMDs) from Subaru imaging for A521, A746, and A1240 and SDSS imaging for A2443. The spectroscopically confirmed cluster galaxies in the CMD form a red sequence. We fit a line to the red sequence to select cluster member candidates that fall within a color range of $g - r \pm 0.1$ with the $r$-band brightness limit set to $21.5$.

Raw spectra were processed with the \texttt{hs\_pipeline\_wrap} command in HSRED 2.1\footnote{https://github.com/MMTObservatory/hsred} to produce sky-subtracted and variance-weighted spectra. The IRAF add-on RVSAO \citep{1998kurtz} was used to cross-correlate a set of template spectra and estimate redshifts. To select robust redshift estimates, we only retained estimates with a cross-correlation $r$-value $ > 4$. The MMT/Hectospec redshift catalogs are published in \cite{2020yoon_a521} for A521, \cite{2022cho} for A1240, \cite{2023hyeonghan_a746} for A746, and Kim et al. in prep. for A2443.

\subsection{Archival Radio and X-ray Observations}
In this study, we rely on archival radio and X-ray observations to investigate the relation of dark matter and ICM. Table \ref{table:xray_radio} summarizes the sources of the radio and X-ray data. Radio images were kindly provided by the authors listed in the Radio References column. X-ray data were retrieved from the respective archive and processed.

\textit{Chandra} observations were downloaded from the Chaser\footnote{https://cda.harvard.edu/chaser/} archive. Utilizing the Python version of the CIAO\footnote{https://cxc.cfa.harvard.edu/ciao/} package, the multiple visits for each cluster were reprocessed and combined into a broad flux image containing emission from 0.5-7~keV. For this reduction, we followed the Diffuse Emission tutorial.

XMM-\textit{Newton} images were downloaded from the XMM-\textit{Newton} science archive\footnote{http://nxsa.esac.esa.int/nxsa-web/}. The images were reduced and combined with the SAS pipeline following the procedures for diffuse extended sources described in the XMM-ESAS Cookbook\footnote{http://heasarc.gsfc.nasa.gov/docs/xmm/esas/cookbook/}. The final combined images have an energy range of 0.5-7~keV.

Unless otherwise stated, the X-ray imaging processed through these methods are used purely for qualitative analysis.

\begin{table*}[!ht]
\def\arraystretch{0.8}

\centering
\caption{Archival radio and X-ray imaging utilized in this study.}

\resizebox{\textwidth}{!}{%
\begin{tabular}{lllll}
\hline
\hline
Cluster   & Radio Image   & Radio References          & X-ray Telescope & Exposure (ks) \\

\hline
1RXSJ0603 & GMRT 610 MHz  & \cite{2012vanweeren_toothbrush} & Chandra         & 250           \\
A115      & LOFAR 150 MHz   & \cite{2022botteon_lofar}     & Chandra         & 360           \\
A521      & MeerKAT 1.3 GHz & \cite{2022knowles}      & Chandra         & 170           \\
A523      & VLA 1.4 GHz   & \cite{2011vanweeren}      & Chandra         & 30            \\
A746      & LOFAR 150 MHz & \cite{2022botteon_lofar}  & XMM-\textit{Newton}         & 184            \\
A781      & LOFAR 150 MHz  & \cite{2022botteon_lofar}        & Chandra         & 48            \\
A1240     & LOFAR 150 MHz   & \cite{2022botteon_lofar} & Chandra         & 52            \\
A1300     & GMRT 325 MHz  & \cite{2013venturi}        & Chandra         & 100           \\
A1612     & GMRT 325 MHz  & \cite{2011vanweeren}      & Chandra         & 31            \\
A2034     & LOFAR 150 MHz & \cite{2022botteon_lofar}      & Chandra         & 261           \\
A2061     & LOFAR 150 MHz & \cite{2022botteon_lofar}      & Chandra         & 32            \\
A2163     & VLA 1.4 GHz   & \cite{2001feretti_a2163}     & Chandra         & 90            \\
A2255     & WSRT 350 MHz  & \cite{2009pizzo_a2255}  & XMM-\textit{Newton}      & 42            \\
A2345     & VLA 1.4 GHz   & \cite{2009bonafede_a2345}    & XMM-\textit{Newton}      & 93            \\
A2443     & VLA 325 MHz   & \cite{2011cohen}    & Chandra         & 116           \\
A2744     & MeerKAT 1.3 GHz  & \cite{2022knowles}     & Chandra         & 132           \\
A3365     & VLA 1.4 GHz   & \cite{2011vanweeren}      & XMM-\textit{Newton}      & 161           \\
A3411     & GMRT 325 MHz  & \cite{2017vanweeren}  & Chandra         & 215           \\
CIZAJ2242 & WSRT 1382 MHz & \cite{2010vanweeren_sausage}  & Chandra         & 206           \\
MACSJ1149 & LOFAR 150 MHz  & \cite{2021bruno_macsj1149}    & Chandra         & 372           \\
MACSJ1752 & LOFAR 150 GHz  & \cite{2022botteon_lofar} & XMM-\textit{Newton}      & 13            \\
PLCKG287  & GMRT 325 MHz  & \cite{2014bonafede_plckg287}    & Chandra         & 200           \\
PSZ1G108  & GMRT 323 MHz  & \cite{2015degasperin} & Chandra         & 27            \\
RXCJ1053  & WSRT 1382 MHz & \cite{2011vanweeren}      & Chandra         & 31            \\
RXCJ1314  & MeerKAT 1.3 GHz   & \cite{2022knowles}    & XMM-\textit{Newton}      & 110           \\
ZwCl0008  & WSRT 1382 MHz & \cite{2011vanweeren_zwcl0008} & Chandra         & 411           \\
ZwCl1447  & GMRT 700 MHz   & \cite{2022lee_zwcl1447}      & Chandra         & 30            \\
ZwCl1856  & LOFAR 150 MHz & \cite{2021jones_zwcl1856} & XMM-\textit{Newton}      & 12            \\
ZwCl2341  & GMRT 610 MHz  & \cite{2009vanweeren_zwcl2341} & Chandra         & 227          \\
\hline
\label{table:xray_radio}
\end{tabular}}
\end{table*}

\subsection{Object Detection and Photometry}\label{detection}
The detection of the WL signal requires a statistical analysis of galaxy shapes. A main source of noise that contributes to a WL analysis is shape noise, which is caused by the dispersion of the intrinsic ellipticity of galaxies. To reduce the contribution from shape noise, a large number of galaxies is needed. Therefore, we optimize our object detection method to robustly detect as many galaxies as possible.

We performed photometry with SExtractor \citep{1996bertin} in dual-image mode. As the name implies, dual-image mode uses two images. The detection image is kept constant while the measurement image is varied between runs to ensure that the resulting filter-specific catalogs have the same objects. For each cluster, we created a detection image by weight-averaging all available filters together. When running SExtractor, a weight image for the detection image was provided that was created by weight-averaging the SWarp weight images together. Measurements were taken on the measurement image (one for each filter) with an rms image provided for weighting. The rms image was created by multiplying the weight image of the filter by the background rms of its mosaic and masking spurious pixels. Our configuration of SExtractor set \texttt{DETECT\_MINAREA} to 5 pixels and \texttt{DETECT\_THRESH} to 2. The settings for deblending were set to \texttt{DEBLEND\_NTHRESH} of 32 and \texttt{DEBLEND\_MINCOUNT} of $10^{-4}$ to maximize detection of overlapping objects.

\begin{figure*}[!ht]
    \centering
    \includegraphics[width=\textwidth]{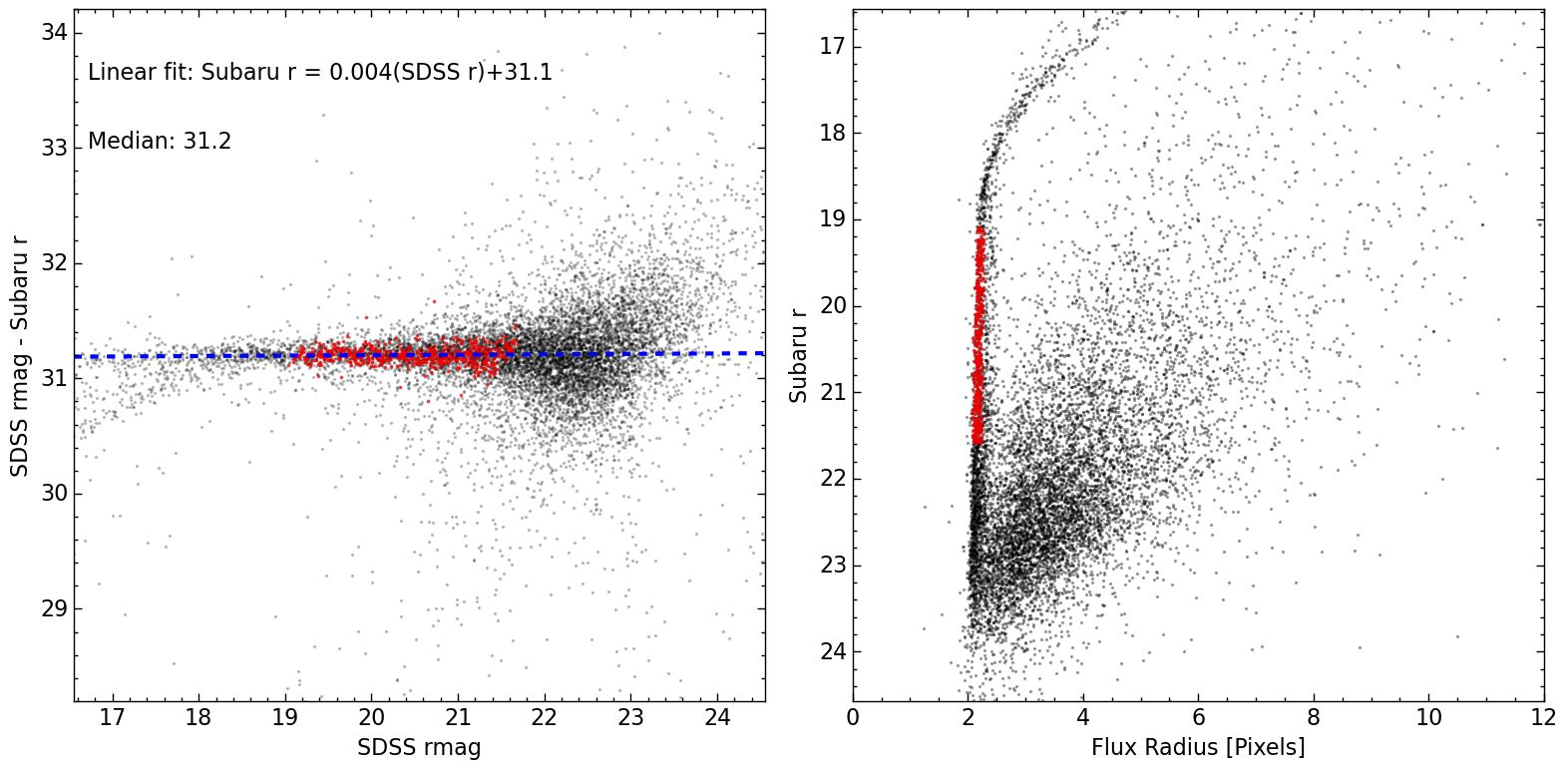}
    \caption{Photometric zeropoint calibration of A2061 (left panel). The SDSS catalog is used as a reference. Stars (red circles) are selected from the size-magnitude plot (right panel) and matched with stars in the reference catalog. The magnitude offset as a function of SDSS $r$-band magnitude is plotted and a linear fit to the stars is made to determine the zeropoint. The same stars are utilized in PSF modeling.}
    \label{fig:zp_calibration}
\end{figure*}

The Subaru photometric zero-point was calibrated by matching stars to an external star catalog. The first choice was the SDSS DR14 photometric catalog but when not available the Pan-STARRS DR1 catalog was used. In all cases, the difference in filter throughput was accounted for by performing synthetic photometry \citep{2005sirianni}. To derive the correct photometric zeropoint, the magnitude of stars from the reference catalog were compared to the SExtractor \texttt{MAG\_AUTO} measurements. The left panel of Figure \ref{fig:zp_calibration} displays the magnitude difference as a function of magnitude for the stars in the A2061 $r$-band image. Instead of relying on the full population of stars, we selected unsaturated stars from the stellar locus as shown in the right panel. The magnitude difference follows a linear relation over the region shown and in most cases has a slope close to 0. We repeated this method for each filter and applied the linear calibration to the Subaru magnitudes.

\section{WL THEORY AND METHODOLOGY}\label{Section:theory}
\subsection{Gravitational Lensing Formalism} \label{section:lensing theory}
\cite{2001bartelmann}, \cite{2013hoekstra}, and \cite{2022meneghetti} provide comprehensive reviews of gravitational lensing theory. For brevity, we state the critical ideas and nomenclature that are required to understand our analysis.

The gravitational lensing effect is caused by the deflection of light by a gravitational potential.  The deflection of light leads to a shift in the apparent position of the light source, a galaxy in our case. The relation between the true source position $\boldsymbol{\eta}$ and the observed position $\mathbf{x}$ is
\begin{equation}
\boldsymbol{\eta} = \mathbf{x} - \boldsymbol{\alpha}(\mathbf{x}),
\end{equation}
where
\begin{equation}
\boldsymbol{\alpha}(\mathbf{x}) = \frac{1}{\pi} \int \kappa(\mathbf{x'})\frac{\mathbf{x}-\mathbf{x'}}{|\mathbf{x}-\mathbf{x'}|^2}d^2\mathbf{x}
\end{equation}
is the scaled deflection angle. The scaling depends on the convergence

\begin{equation}
\kappa(\mathbf{x}) = \frac{\Sigma(\mathbf{x})}{\Sigma_c},
\label{eq:kappa}
\end{equation}
which is a dimensionless quantity of the projected mass density divided by the lensing critical density. The lensing critical density is
\begin{equation}
 \Sigma_c = \frac{c^2 D_s}{4\pi G D_l D_{ls} } ,
\label{eq:sigma_c}
\end{equation}
where $D_l$ is the angular diameter distance to the lens, $D_s$ is the angular diameter distance to the source, $D_{ls}$ is the angular diameter distance from lens to source, $c$ is the speed of light, and $G$ is the gravitational constant. The ratio $\beta = D_{ls}/D_s$ is referred to as the lensing efficiency.

The tidal gravitational field causes an anisotropic distortion, called the shear ($\gamma$), that stretches galaxy images tangential to the local gravitational potential gradient. When the gravitational lensing effect is in the weak regime ($\kappa \ll 1$, $\gamma \ll 1$), single galaxy images appear and the distortions are small. \cite{1993kaiser} show that the shear and convergence are related by the convolutions

\begin{equation}
\kappa(\mathbf{x}) = \frac{1}{\pi} \int D^*(\mathbf{x} - \mathbf{x'})\gamma(\mathbf{x'})d^2\mathbf{x'} ,
\label{eq:kappa_convolve}
\end{equation}

\begin{equation}
\gamma(\mathbf{x}) = \frac{1}{\pi} \int D(\mathbf{x} - \mathbf{x'})\kappa(\mathbf{x'})d^2\mathbf{x'} ,
\end{equation}
where $D(\mathbf{x}) = -1/(x_1 - ix_2)^2$ is the convolution kernel.

Observationally, we detect the reduced shear

\begin{equation}
    g=\frac{\gamma}{1-\kappa},
\end{equation}
which is the combination of the convergence and the shear. The distortions caused by WL may be expressed by the Jacobian matrix

\begin{equation}
A = (1-\kappa)\begin{bmatrix}
1-g_1 & -g_2 \\ -g_2 & 1 + g_1
\end{bmatrix}.
\label{eq:jacobian}
\end{equation}
The reduced shear is encoded as a complex term ($g = g_1 + ig_2$) where positive (negative) values of $g_1$ distort galaxy images along the $x$ ($y$) directions and positive (negative) values of $g_2$ distort images along the $x=y$ ($x=-y$) directions. In this work, we define the ellipticity (shape) of galaxies as $e = (a-b)/(a+b)$, where $a$ and $b$ are the semi-major and semi-minor axes, respectively. Each observed galaxy has a measured ellipticity that includes its intrinsic shape and the local shear
\begin{equation}
e \approx e_{\mathrm{intrinsic}} + g.
\end{equation}
Under the assumption that galaxy images have random orientations, the averaged ellipticity of galaxy images is the reduced shear
\begin{equation}
    \langle e\rangle\approx g .
    \label{eq:average_eq_redshear}
\end{equation}

\subsection{Point Spread Function Modeling} \label{section:psf modeling}
WL requires the careful measurement of galaxy shapes. However, the turbulent atmosphere and the diffraction of light through the telescope causes a significant anisotropic blurring of the galaxy images. It is critical for a WL analysis to properly model and remove the effect of the PSF. For our analysis, we utilize a principal component analysis (PCA) approach to deriving a spatial and temporal PSF model. \cite{2007jee} showed that the PCA approach accounts for small and large scale structures of the PSF. PCA is also beneficial because it derives the basis functions from the data set itself and requires few components. The PCA technique that is used in this work has been applied to a variety of space- and ground-based observations \citep{2017finner, 2020finner, 2020hyeonghan, 2023afinner, 2023cfinner, 2024hyeonghan} including the James Webb Space Telescope (JWST) in \cite{2023bfinner}.

The dynamic atmosphere causes the PSF of Subaru to vary temporally and spatially. Furthermore, both the Suprime-Cam and the HSC have a large field of view, which leads to a complex PSF across the full mosaic that is difficult to interpolate and suffers from discontinuities at CCD boundaries \citep[in similar fashion to that shown in the Deep Lens Survey;][]{2013jee}. To resolve these issues, we create PSF models on a frame-by-frame basis and then stack them into a final PSF model that is usable for measurement from the coadded mosaic.

For brevity, we will list the steps of the PSF pipeline. For an in-depth explanation of the pipeline, we refer the reader to \cite{2017finner}. First, a PCA is performed on stars that best represent the PSF (isolated, bright but not saturated, etc.).
\begin{itemize}
    \item Select stars for each frame based on size and magnitude. The right panel of Figure \ref{fig:zp_calibration} illustrates the selection process for one cluster. Create 21 pixel by 21 pixel postage stamps of each selected star and shift them to be centered on the cutout.
    \item Store the mean star and a residual array (N stars by 441 pixels) that is created by subtracting the mean star from each individual star.
    \item Perform a PCA of the residual and keep the 21 components with the highest variance. The choice for keeping 21 components was found empirically by evaluating the PSF residuals as presented later in this section.
\end{itemize}
The PCA and the mean star can be utilized to generate a PSF at any position in the frame, which can then be stacked into PSF models for the coadded mosaic as follows.
\begin{itemize}
    \item Fit the PCA result with a third order polynomial.
    \item Add the fitted PCA result onto the mean star.
    \item Repeat for all objects of interest in the frame.
    \item For each object in the mosaic image, determine which frames compose the mosaic and stack the PSF models for those frames to a coadded PSF model.
\end{itemize}

\begin{figure}
    \centering
    \includegraphics[width=0.48\textwidth]{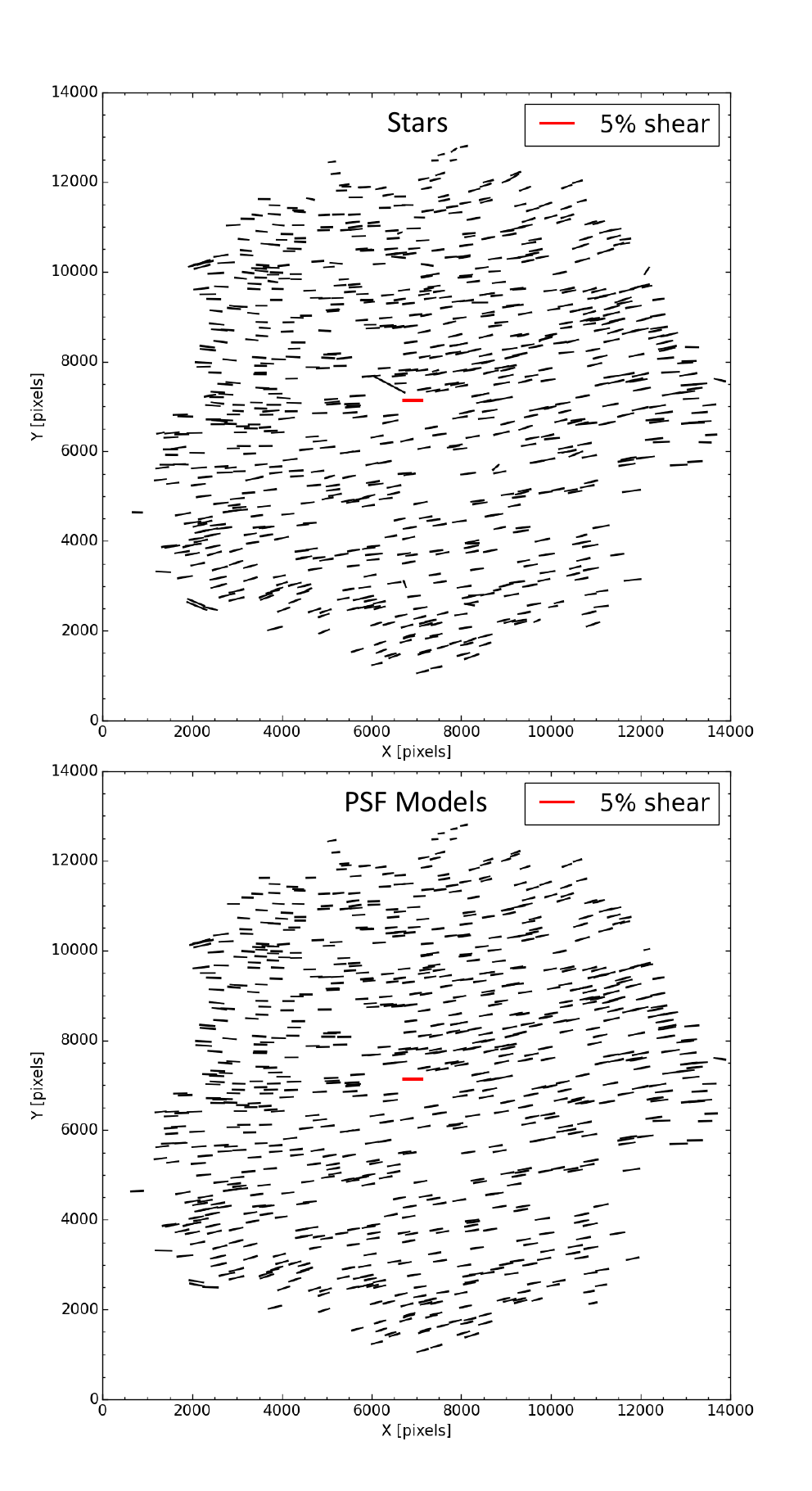}
    \caption{A2061 PSF distortion. Top: Measured ellipticity of a selection of stars from the mosaic coadded image. The lines represent the direction and magnitude of the ellipticity of stars. Outliers have not been removed and clearly do not follow the general trend of the PSF. Bottom: Ellipticity of the model PSFs at each star location. Models are designed in each component frame and stacked to a coadded PSF. Comparing the top and bottom panels shows that the measured stellar shape is well reproduced by the PSF model.}
    \label{fig:PSF_model}
\end{figure}

\begin{figure}
    \centering
    \includegraphics[width=0.48\textwidth]{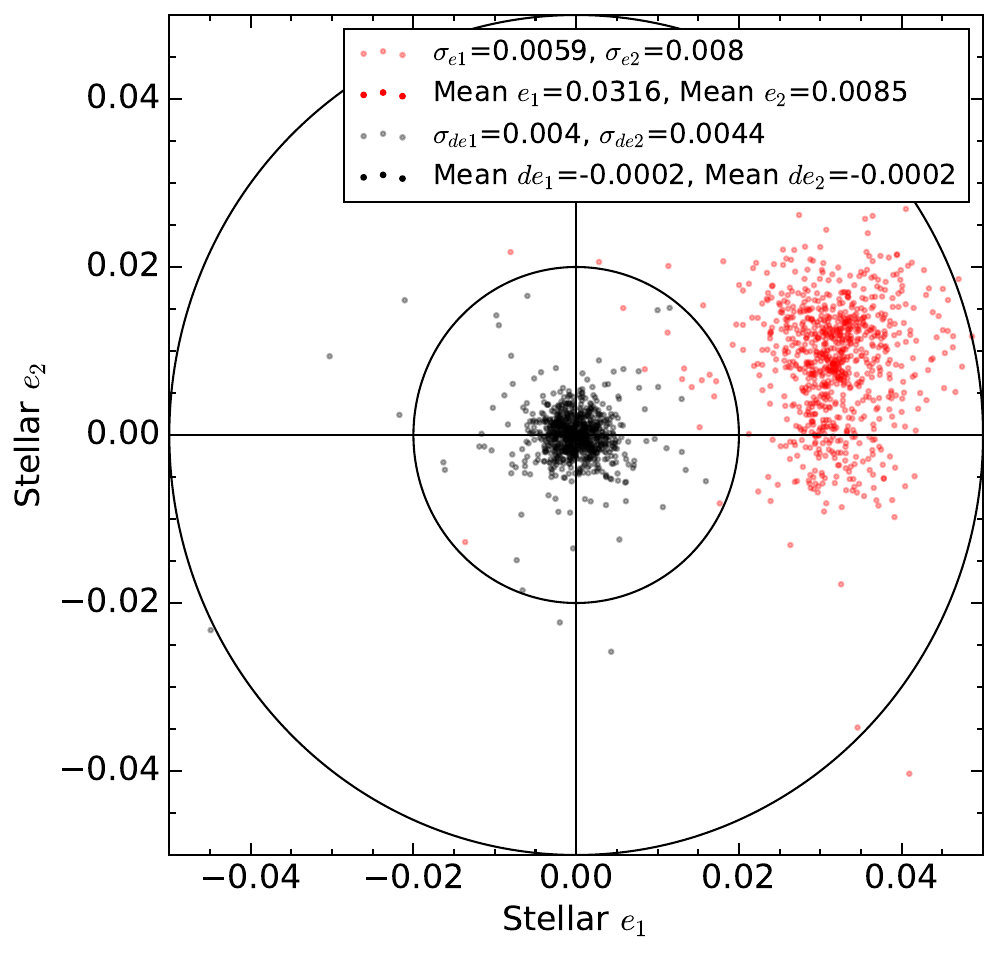}
    \caption{A2061 PSF model correction. Red circles denote the observed ellipticity of stars. Black circles are the corrected ellipticities where the PSF model ellipticities have been subtracted from the observed ellipticities of their respective stars. The figure demonstrates that the ellipticity is being corrected. The goal is to tighten the dispersion and shift the centroid to zero. As the legend shows, the dispersion is decreased by about two and the centroid is corrected to the $10^{-4}$ level. }
    \label{fig:psf_comparison}
\end{figure}

Figure \ref{fig:PSF_model} compares the ellipticity of stars (top) in A2061 to the corresponding PSF models (bottom). In most cases, the PSF models trace the magnitude and direction of the ellipticity of stars. Figure \ref{fig:psf_comparison} displays the statistics of the correction made by the PSF model. The red circles are the ellipticity of stars in the mosaic. By subtracting the ellipticity of the PSF model for each star, we get the residual (black circles). There are two desired effects that are important to see when comparing the measured ellipticity of stars and the residual. First, the distribution should shift toward 0 if the average ellipticity is being corrected. Second, the distribution should tighten if the spatial variation of the ellipticity is being corrected. We see that both effects are being corrected for. A2061 demonstrates the typical residual for our PSF models with a mean ellipticity residual of order $10^{-4}$ and standard deviation of $10^{-3}$. This level of accuracy is sufficient for WL
analysis of galaxy clusters.

\subsection{Galaxy Shape Measurement with PSF Correction} \label{section:shape measurement}
Detection of the WL effect requires a statistical analysis of the shapes of distorted galaxy images. We employed a model-fitting technique to measure the shapes of galaxies. For each cluster, we follow the same recipe for measuring galaxy shapes.

To measure the shape of a single galaxy, we cut out a postage stamp image of the galaxy from the mosaic image. A corresponding rms noise postage stamp $\sigma_{\mathrm{rms}}$ was also cut out from the rms mosaic image. It is important to consider the size of the postage stamp image. A large postage stamp image will contain the light from nearby objects, which may significantly alter the shape measurement. However, a small postage stamp may prematurely truncate the galaxy light profile and lead to truncation bias \citep{2018mandelbaum}. We chose to cut large postage stamp images that are eight times the size of the half-light radius as measured by SExtractor \citep{1996bertin}, with a 10 pixel floor for very small objects. Making use of the SExtractor segmentation map, we masked any nearby bright objects that could influence the shape measurement by setting the relevant pixels in the rms noise postage stamp to $10^6$.

We fit an elliptical Gaussian function, $G$, to the postage stamp image, $I$, while forward-modeling the corresponding PSF model, $P$, as follows:

\begin{equation}
    \chi^2 = \sum \left(\frac{I - G\circledast P}{\sigma_{\mathrm{rms}}}\right)^2,
    \label{eq:chi2_}
\end{equation}
where the summation is over the pixels of the postage stamp. The elliptical Gaussian function has seven free parameters: background, amplitude, position ($x$ and $y$), semi-major axis ($a$), semi-minor axis ($b$), and orientation angle ($\phi$). To fit the galaxies, we utilized a Python version of the Levenburg-Marquardt least-squares fitting code MPFIT. Uncertainties returned by MPFIT are determined from the Hessian matrix and assume a Gaussian distribution. We fixed the background, $x$, and $y$ to the SExtractor values of \texttt{BACKGROUND}, \texttt{XWIN\_IMAGE}, and \texttt{YWIN\_IMAGE} while fitting the remaining four parameters. Complex ellipticities were determined from the fitted values of $a$, $b$, and $\phi$ as

\begin{equation}
\begin{split}
e_1 &= \frac{a-b}{a+b}\cos 2\phi, \\
e_2 &= \frac{a-b}{a+b}\sin 2\phi.
\end{split}
\end{equation}

Galaxy shape measurement techniques suffer from biases that must be corrected for. The biases are typically encoded into a linear correction factor with a multiplicative and additive bias. We derived a calibration factor from simulations of our pipeline to correct for multiplicative bias. Following the technique called SFIT \citep{2013jee}, which was the best performing technique in the GREAT3 challenge \citep{2015mandelbaum}, we processed simulated images with our Subaru WL pipeline and compared the input shear to the measured shear. We found that a multiplicative calibration factor of 1.15 was necessary to correct the biases of our pipeline and that additive biases were low ($<10^{-4}$).

For each cluster, galaxy shape measurements were made in the WL image (see Table \ref{table:subaru_imaging}) and shape catalogs were compiled. The shape catalogs created through this procedure will be further culled in the following sections to a background source catalog that ideally contains only lensed galaxies.

\subsection{Source Selection} \label{section:source selection}
The ideal catalog for a WL analysis contains only galaxies that are at a greater distance than the lens (cluster) from the observer. Having distance measurements for each galaxy would make source selection trivial. However, gathering spectroscopic redshifts for all background galaxies is an immense undertaking. Photometric redshifts are a second option but a reliable redshift requires vast multiband imaging, which is not readily available for our sample of galaxy clusters. We instead rely on colors and magnitudes to separate foreground, cluster, and background galaxies.

\begin{figure}[!ht]
    \centering
    \includegraphics[width=0.5\textwidth]{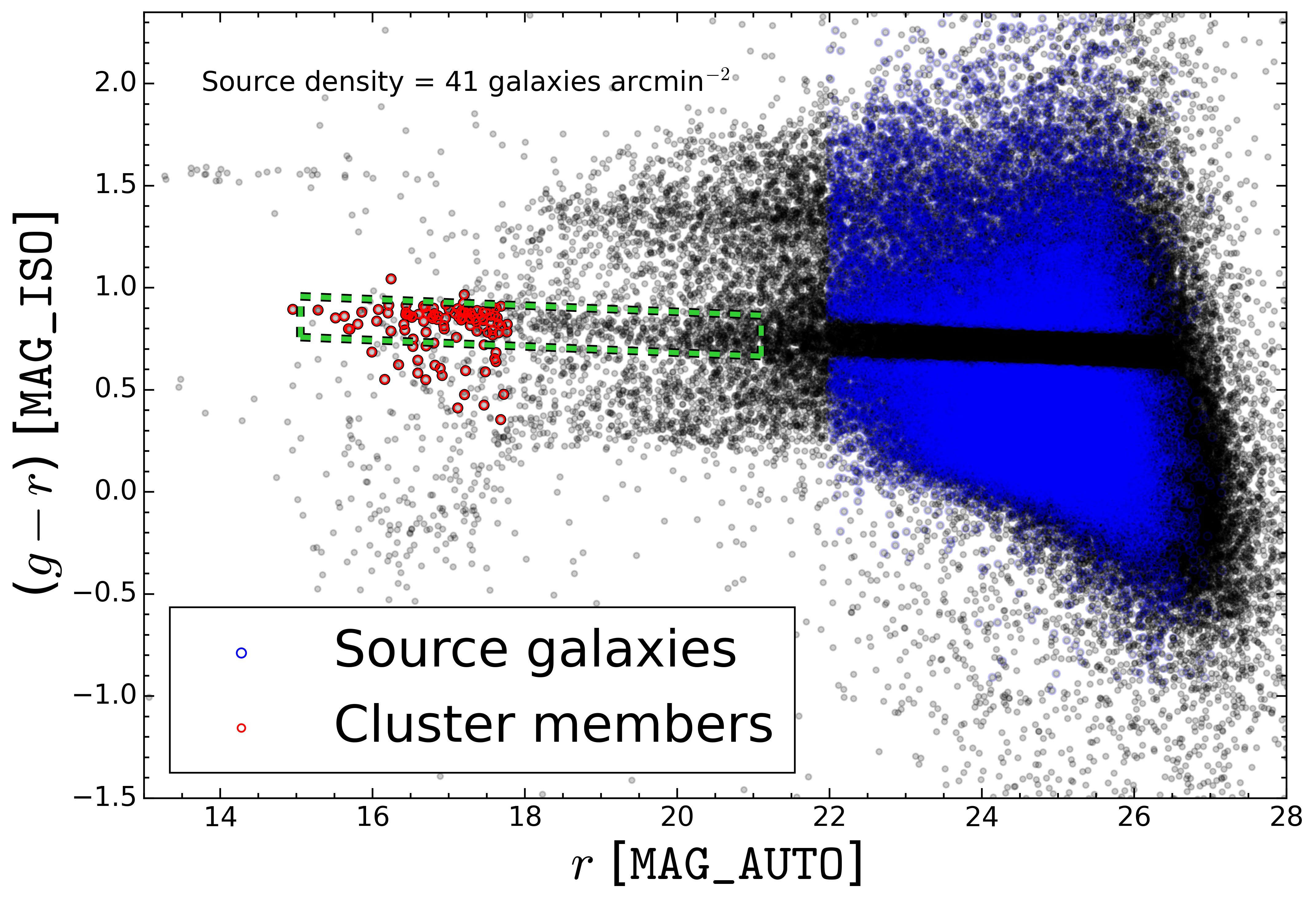}
    \caption{CMD for A2061 with $g-r$ [MAG\_ISO] color and $r$ [MAG\_AUTO] from SExtractor. Red circles represent the spectroscopically confirmed cluster member galaxies within $z\pm0.03$ of the cluster redshift. The green dashed box highlights the photometrically selected sample of cluster member candidates based on a linear fit to the red sequence. Background galaxies (blue circles) are selected following the criteria described in section \ref{section:source selection}. A red selection of source galaxies is included for A2061 because of its low redshift ($z=0.08$). }
    \label{fig:cmd_abell2061}
\end{figure}

There are multiple properties of galaxies that are useful for placing galaxies into these three categories. Galaxies residing in a cluster environment tend to have lower star formation rates, an overall older population of stars, and a large amount of dust compared to field galaxies \citep{1984dressler}. In addition, an accumulation of metals that are deposited by past star formation leads to a prominent feature in the spectral energy distribution (SED) of evolved cluster galaxies called the 4000\ \AA\ break. The 4000\ \AA\ break manifests a red-sequence relation in a CMD when observed with two filters that bracket the feature.  In the redshift range of our sample (0.07-0.54), the 4000\ \AA\ break is well bracketed by the $g$ and $r$ or $g$ and $i$ filters. Figure \ref{fig:cmd_abell2061} presents the CMD of A2061. The red sequence is clearly marked by the red circles of spectroscopically confirmed cluster galaxies. However, it can also be seen extending to fainter magnitudes within the green, dashed rectangle. We make a photometric selection of cluster galaxy candidates (dashed, green rectangle) by extending the red sequence to an $r$-band magnitude of 21. We opt to select galaxies that are within 0.1 in $g-r$ color to the linear fit, which encapsulates the majority of red-sequence galaxies. Following this recipe, we create a catalog of cluster galaxies that are used to plot galaxy luminosity and number density maps in Section \ref{Section:results}.

The color and magnitude properties of galaxies are also useful for selecting background galaxies. Galaxies behind the cluster are on average fainter than the cluster members as apparent brightness is proportional to inverse-squared distance. Identifying background galaxies based on color is more complicated. The cosmological expansion of space redshifts galaxy emission. However, there is the competing evolutionary effect where galaxies at greater distances appear on average to be younger than nearby galaxies. When viewed through the color of $g-r$ or $g-i$, the evolutionary effect tends to be stronger than the redshift effect beyond a redshift of 0.5, which leads to background galaxies being bluer than cluster and foreground galaxies \cite[see][for an example]{2018schrabback}. Therefore, we select galaxies that are bluer than the red sequence for clusters that are above redshift of 0.2. Below a redshift of 0.2, limiting the background galaxy catalog to only sources that are bluer than the red sequence leads to a low source density. Therefore, in these cases, we include both red and blue galaxies in our source catalog but reject the galaxies that exist in the color space that follows the red sequence. The CMD for A2061 in Figure \ref{fig:cmd_abell2061} shows this selection technique. In all cases, we set a brightness limit for the source catalog of 22nd magnitude because the likelihood of such bright galaxies being foreground or cluster galaxies is high.

In addition to these magnitude and color selection criteria, we constrain the background source catalog by our ability to measure their shapes. We ensure that spurious objects that are too small to be lensed galaxies are removed by constraining the semi-minor axis to be greater than 0.3 pixels. Highly elongated objects are rejected by forcing the measured ellipticity to be less than 0.9. We require the ellipticity error to be less than 0.3. Finally, only ``well fit'' objects from the MPFIT fitter are kept (MPFIT status of 1).

These constraints lead to a robust source catalog that carefully balances the purity and number density of the source (lensed) galaxy catalog. The source densities for each cluster are summarized in Table \ref{table:subaru_imaging}. The average source density is ~25 arcmin$^{-2}$.

\subsection{Redshift Estimation}\label{section:redshift}
In a WL analysis, each source galaxy provides a probe of the projected galaxy cluster potential. As is apparent in Equation \ref{eq:kappa}, the effectiveness of the gravitational lens varies on a source-by-source basis with the lensing efficiency ratio, $\beta$. Unfortunately, distances to each of the source galaxies in our sample are unavailable. To remedy this, we rely on the photometric redshift catalog \citep{2010dahlen} of the GOODS-S field as a reference for our source galaxy catalog. A version of the GOODS-S catalog that is modeled to represent the source catalog is used to quantify the effective redshift of the source catalog by the method described below. This technique is common in WL studies that do not have the luxury of redshifts for each source galaxy \citep[to name a few, ][]{2011jee, 2016okabe, 2018schrabback}.

\begin{figure}
    \centering
    \includegraphics[width=0.45\textwidth]{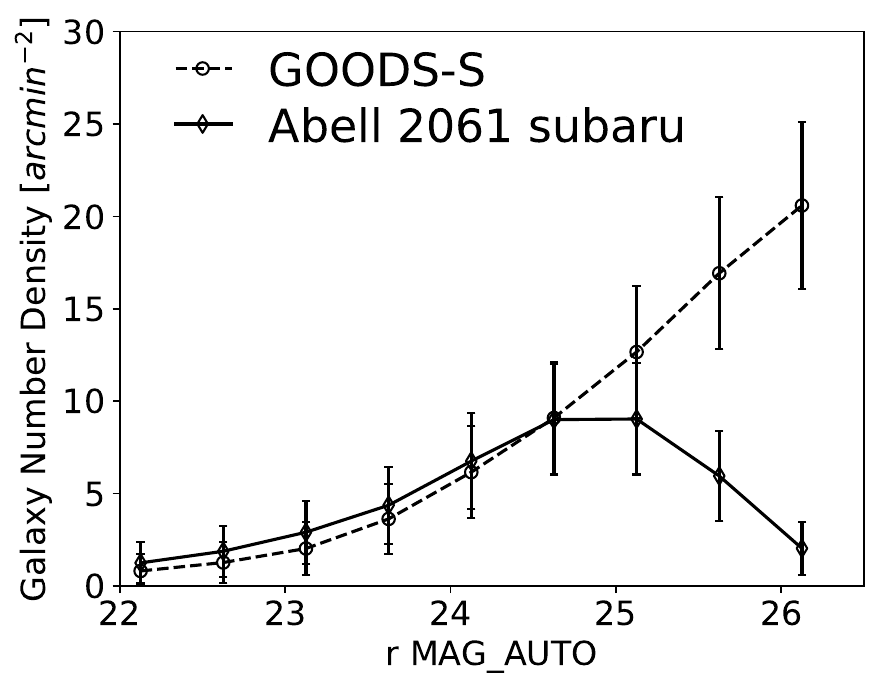}
    \caption{Galaxy number density by magnitude for the constrained (by color) GOODS-S reference catalog and A2061 source galaxy catalog. The GOODS-S observations are much deeper, which gives rise to the large difference in source density at faint magnitudes. To make up for the difference in depth, the GOODS-S catalog is weighted by the ratio of the bins when used as a reference catalog.}
    \label{fig:redshift_est}
\end{figure}

The GOODS-S reference catalog is constrained with the same color and magnitude criteria that are applied when selecting the source galaxies. Figure \ref{fig:redshift_est} compares the galaxy number density of the constrained GOODS-S catalog to the A2061 source catalog. The GOODS-S observations are much deeper than the Subaru observations and probe to much fainter magnitudes. To alleviate the difference in depth, the constrained GOODS-S catalog is weighted by the number density ratio for each bin. An effective $\beta$ for the source galaxies of the Subaru imaging is then inferred from the constrained and weighted GOODS-S catalog ensuring that any galaxy that is foreground is assigned a $\beta=0$ following

\begin{equation}
    \left<\beta\right> = \left<max\left(0,\frac{D_{ls}}{D_s}\right)\right>.
\end{equation}
The $\left<\beta\right>$ values for each cluster are tabulated in Table \ref{table:subaru_imaging}. Since the source galaxies are represented by a single $\left<\beta\right>$, a first-order correction \citep{1997seitz}

\begin{equation}
g' = \left[1 + \left(\frac{\left<\beta^2\right>}{\left<\beta\right>^2}-1\right) \kappa \right] g.
\end{equation}
is applied to the reduced shear to take the width of the distribution into consideration. Foregoing this correction can lead to an overestimation of cluster masses \citep[e.g.,][]{2000hoekstra}.

\subsection{Convergence Reconstruction} \label{section:mass density}
In Section \ref{section:lensing theory}, the basic inversion method for convergence reconstruction was introduced. The simplest method to recover the convergence is to average galaxy shapes in spatial bins across the field of view of the cluster. The spatial averaging of galaxy shapes produces a map of the reduced shear. Then, by the convolution of Equation \ref{eq:kappa_convolve}, the convergence distribution can be recovered. This convolution method is prone to edge effects that can artificially increase the lensing signal near the edge of the image. In this work, we utilize a code called FIATMAP \citep[][]{2006wittman} that performs the convolution in real-space rather than in the Fourier domain.

\subsection{Substructure Identification} \label{section:convergence peaks}
One of the goals of this study is to identify the merging subclusters that may be responsible for the formation of radio relics. For all of the clusters in this study, multiple peaks in the WL maps are expected. However, not all of these peaks should be taken as merging subclusters (substructures).

In order to identify the real WL peaks from the false detections, we utilize the multiwavelength data. Significant subclusters are expected to be massive enough to emit brightly in X-rays. Furthermore, they are expected to have bright cluster galaxies in the vicinity of their WL peaks. Therefore, we will identify subclusters as those with significant WL peaks ($S/N>3$) and nearby galaxy overdensities or X-ray brightness peaks.

\subsection{Mass Estimation} \label{section:mass estimation}
There are multiple methods to estimate the mass of a galaxy cluster. One method is to use a proxy for the mass and calibrate it with a robust mass estimate. These methods are called mass-scaling relations and rely on the emission from the gas and stars as a tracer of the cluster potential. One of the common scaling relations is the mass-richness relation, which positively correlates the number of cluster galaxies to the total mass of the system \citep[e.g.][]{2019murata}. Other scaling relations utilize the X-ray emission or the Sunyaev Zel'dovich \citep[SZ;][]{1972sunyaev} effect from the gas to estimate the mass \citep[e.g.][]{2019ge}. The galaxies and gas can also provide a mass estimate directly by assuming that they are in hydrostatic equilibrium (HSE). With that assumption, the gravitational potential can be equated to the outward pressure force of the gas. In this work, we derive mass from the galaxy velocity dispersion measurements presented in G19 by applying the scaling relation from \cite{2008evrard}:

\begin{equation}
    M_{200} = \left(\frac{\sigma_{DM}}{\sigma_{DM,15}}\right)^{1/\alpha}10^{15}\ \mathrm{M}_\odot,
\end{equation}
where $\sigma_{DM,15} = 1082\pm4$ km s$^{-1}$ and $\alpha=0.3361\pm0.0026$ are derived from cosmological simulations.

The validity of applying HSE when estimating the mass of a galaxy cluster has been questioned. \cite{2013suto}, and more recently \cite{2016biffi}, found that the HSE assumption for clusters in cosmological simulations had mass estimates that departed from the true cluster mass by as much as $30\%$. As expected, these studies find that the disturbed systems show the largest deviations from HSE.

Mass estimates based on the WL signal from galaxy clusters do not require an HSE assumption. Furthermore, the WL signal is caused by the complete mass (gravitational potential) of the cluster, which is predominantly dark matter. Thus, WL should be a more accurate probe of the mass of merging systems. However, there are recent investigations into the bias of fitting Navarro-Frenk-White \citep[NFW;][]{1997navarro} models to the WL signal of merging galaxy clusters that suggests that mass estimates may be biased at certain stages of the cluster merger process \citep{2023lee}.

\subsubsection{Multiple Halo Mass Estimation}
The clusters that are being analyzed in this paper contain multiple substructures. Hence, it would be improper to model them as a single object. Instead, we fit a multi-halo NFW model to the observed data to estimate the masses of each subcluster simultaneously. The fit is accomplished by predicting the reduced shear at every source galaxy position and then calculating the $\chi^2$ between the observed galaxy ellipticity and modeled shear

\begin{equation}
    \chi^2 = \sum_{i,j} \left(\frac{e_{i,j} - g_{i,j}}{\sigma_{i,j}}\right)^2
\label{eq:chi2_mass}
\end{equation}
where $i$ is summed over each galaxy and $j$ is summed over the two components of the shear (ellipticity). In this equation, $\sigma_{i,j}=(\delta e_{i,j}^2 + \sigma_{SN}^2)^{-0.5}$ is the ellipticity measurement uncertainty $\delta e_{i,j}$ and the shape noise $\sigma_{SN}^2$ added in quadrature. We fix the shape noise to $\sigma_{SN}=0.25$. The reduced shear is predicted following the NFW equations from \cite{2000wright}.

In this analysis, we use two different methods to constrain the mass with the NFW profile.  The first method utilizes a concentration-mass ($c-M$) relation. The $\chi^2$ in Equation \ref{eq:chi2_mass} is passed into the optimization package of MPFIT, which typically converges on the order of 10 iterations. Relations for $c-M$ are developed from cosmological simulations that cover a wide range of redshifts and cluster masses. One of the more common $c-M$ relations that is applied in WL studies is that of \cite{2008duffy}. Their $c-M$ relation is derived from \texttt{Gadget2} simulations with a box size of 400 Mpc and is as follows:

\begin{equation}\label{eq:duffy}
    M_{200} = 2\left(1+z\right)^{-B/A}\left(\frac{c}{5.71\pm0.12}\right)^{1/A} 10^{14}\ \mathrm{M}_\odot,
\end{equation}
where $A=-0.084\pm0.006$ and $B=-0.47\pm0.04$ for clusters in a redshift range of $0<z<2$. However, $c-M$ relations that are derived from simulations are subject to the limitations of the simulations. For instance, the limited box size and cosmology imprinted on the initial conditions affect the $c-M$ relation. This is manifested in the variety of $c-M$ relations that have been derived from various simulations \citep[e.g.][]{2008duffy, 2014dutton, 2019diemer}. We elect to use the \cite{2008duffy} relation in this work because it has been commonly applied in past WL studies and will ease comparison.

An alternative to using a $c-M$ relation is to fit both concentration and mass. However, there is a degeneracy between the concentration and mass that prevents this technique from converging for clusters that have low WL signal \citep{2017finner}. Therefore, we sample the $c-M$ parameter space with Markov Chain Monte Carlo (MCMC). We confine the MCMC to uniform priors that sufficiently cover the typical range of the mass ($10^{13}$ M$_\odot < M_{200} < 10^{16}\ M_\odot$) and concentration ($1<c<9$) for galaxy clusters. Masses are determined from the highest likelihood returned from Equation \ref{eq:chi2_mass} and uncertainties on the mass are calculated by marginalizing over the concentration. We will refer to this method as 2PNFW (for 2 parameter) from here on.

In both of these methods, multiple halos are simultaneously fit to the WL signal. We choose to fix each halo to its mass peak's corresponding BCG. The BCG is not necessarily the center of the cluster, but on average it is a good tracer of the cluster potential centroid \citep{2012zitrin} and is well defined. Another good choice is the WL derived mass peak, but there is a large positional uncertainty associated with it for low $S/N$ WL results. \cite{2022sommer} investigated the impact of miscentering on WL mass estimates and showed that it tends to lead to underestimates of the mass. In many of these merging cluster cases, the X-ray peak would be a poor choice for the center because it departs significantly from the dark matter density peak.

\begin{table}[!ht]
\caption {WL Subcluster Mass Estimates} \label{table:subaru_mass}
\centering
\footnotesize
\def\arraystretch{0.5}
\begin{tabularx}{\linewidth}{ l c c c c}
\hline
\hline
\\
Cluster & Ind. &  Duffy $M_{200}$ & 2PNFW $M_{200}$ & $\sigma_v$       \\
&& $10^{14}\ $M$_\odot$ & $10^{14}\ $M$_\odot$ & km s$^{-1}$ \\ \\
\hline \\
1RXSJ0603 N & A & $6.2\pm1.2$           & $12.9^{+3.3}_{-3.1}$   & $925\pm51$            \\
1RXSJ0603 S & A & $2.8\pm0.8$           & $1.9^{+1.0}_{-0.6}$   & $763\pm100$           \\
A115 N & B  & $1.4\pm0.5$           & $1.0^{+0.5}_{-0.3}$   & $1056\pm60$            \\
A115 S & B  & $3.0\pm0.8$           & $3.3^{+1.4}_{-1.0}$   & $1108\pm70$            \\
A521 C & C  & $3.5\pm0.9$           & $5.9^{+2.4}_{-1.9}$   & $990\pm85$            \\
A521 NW & C  & $1.1\pm0.5$           & $0.5^{+0.6}_{-0.2}$   &  $-$              \\
A521 SE & C  & $2.5\pm0.7$           & $1.9^{+1.2}_{-0.8}$   & $735\pm67$            \\
A523 N & D  & $2.7\pm0.9$           & $2.2^{+1.7}_{-1.0}$   & $814\pm60$            \\
A523 S & D  & $1.6\pm0.7$           & $1.8^{+1.8}_{-0.9}$   & $673\pm53$            \\
A746 S & E   & $6.3\pm1.5$           & $6.6^{+2.7}_{-1.9}$   & $1094\pm95$            \\
A781$^a$ East & F  & $3.3\pm0.8$           & $2.8^{+1.6}_{-1.0}$   &   $-$               \\
A781 Middle & F & $4.2\pm0.9$           & $5.0^{+2.9}_{-1.8}$   & $821\pm65$            \\
A781 Main & F & $3.5\pm0.8$           & $4.0^{+2.0}_{-1.4}$   & $830\pm54$            \\
A781$^a$ North & F & $1.7\pm0.6$           & $1.5^{+1.2}_{-0.6}$   &   $-$             \\
A1240 N & G & $2.6\pm0.6$           & $3.3^{+0.9}_{-1.1}$   & $706\pm52$            \\
A1240 S & G & $1.1\pm0.4$           & $0.8^{+0.5}_{-0.5}$   & $727\pm68$            \\
A1300 S & H & $11.0\pm 1.8$           & $12.8^{+2.7}_{-1.3}$   & $1205\pm51$            \\
A1612 E & I & $4.0\pm1.0$           & $6.6^{+2.9}_{-2.0}$   & $826\pm83$            \\
A1612 W & I & $1.8\pm0.7$           & $1.0^{+0.8}_{-0.5}$   & $662\pm118$           \\
A2034 N & J & $0.9\pm0.4$           & $0.6^{+0.4}_{-0.2}$   & $815\pm94$            \\
A2034 S & J & $3.6\pm0.7$           & $4.7^{+1.8}_{-1.3}$   & $726\pm57$            \\
A2061 N & K & $2.0\pm0.6$           & $3.1^{+1.8}_{-1.3}$   & $841\pm54$            \\
A2061 S & K & $1.6\pm0.5$           & $1.1^{+0.6}_{-0.4}$   & $-$            \\
A2163$^a$ N & L & $1.1\pm0.7$           & $0.8^{+0.6}_{-0.5}$   & $952\pm92$            \\
A2163 E & L & $9.2\pm1.8$           & $9.0^{+2.3}_{-2.8}$   & $1389\pm75$            \\
A2163 W & L & $0.8\pm0.6$           & $0.5^{+0.5}_{-0.3}$   & $741\pm75$            \\
A2255 E & M & $1.3\pm0.6$           & $1.1^{+0.8}_{-0.5}$   &   $-$             \\
A2255 W & M & $4.7\pm1.1$           & $5.1^{+3.2}_{-1.9}$   & $988\pm36$            \\
A2345$^a$ E & N & $4.7\pm1.2$           & $5.0^{+1.3}_{-1.0}$   & $1065\pm67$            \\
A2345 C & N & $3.2\pm1.0$           & $3.3^{+2.3}_{-1.5}$   & $1065\pm67$            \\
A2345 W & N & $2.1\pm0.8$           & $1.6^{+1.6}_{-0.8}$   &   $-$              \\
A2744 E & O & $7.2\pm1.8$          & $8.5^{+2.5}_{-2.5}$   &  $772\pm52$              \\
A2744 W & O & $6.6\pm2.3$           & $3.2^{+2.2}_{-1.5}$   &  $678\pm82$               \\
A2744 N & O & $1.0\pm0.8$          & $1.9^{+1.9}_{-1.0}$   &  $1154\pm70$              \\
A2744 S & O & $0.6\pm0.6$           & $0.8^{+0.6}_{-0.5}$   &   $666\pm52$              \\
A3411 W & P & $2.3\pm1.0$           & $3.9^{+3.9}_{-1.9}$   & $1204\pm84$            \\
A3411 E & P & $1.8\pm1.0$           & $1.0^{+1.3}_{-0.6}$   &   $-$             \\
A3412$^a$ & P & $1.1\pm0.8$           & $0.8^{+1.2}_{-0.5}$   & $1199\pm82$            \\
\hline

\end{tabularx}
\begin{minipage}{10.5cm}%\\
\small $68\%$ uncertainties reported\\
\small $^a$ denotes subclusters that are not involved in the merger
\end{minipage}
%$68\%$ uncertainties reported \\
%$^a$ clusters that are not involved in the merger

\end{table}

\begin{table}[!ht]
\renewcommand\thetable{4}
\caption{WL Subcluster Mass Estimates continued}
\centering

\def\arraystretch{0.5}
\begin{tabularx}{\linewidth}{ l c c c c}
\hline
\hline
\\
Cluster & Ind. & Duffy $M_{200}$ & 2PNFW $M_{200}$ & $\sigma_v$       \\
&& $10^{14}\ $M$_\odot$ & $10^{14}\ $M$_\odot$ & km s$^{-1}$ \\ \\
\hline \\
CIZAJ2242 N & Q  & $8.2\pm1.8$           & $12.3^{+3.9}_{-3.0}$   & $1102\pm69$            \\
CIZAJ2242 S & Q  & $9.3\pm1.8$           & $15.8^{+4.6}_{-3.5}$   & $1146\pm88$            \\
MACSJ1149 C & R   & $16.4\pm3.1$           & $13.5^{+3.5}_{-2.8}$   & $1295\pm62$            \\
MACSJ1752 NE & S  & $5.6\pm1.8$           & $5.5^{+1.9}_{-1.4}$   & $1006\pm68$            \\
MACSJ1752 SW & S  & $5.6\pm1.7$           & $5.1^{+2.0}_{-1.5}$   & $1038\pm77$            \\
PLCKG287 SE & T   & $1.7\pm0.7$           & $1.0^{+0.7}_{-0.6}$   & $680\pm87$            \\
PLCKG287 C  & T  & $20.4\pm1.9$          & $20.0^{+2.4}_{-2.2}$  & $1373\pm72$            \\
PLCKG287 NW & T   & $1.4\pm0.7$          & $1.0^{+0.7}_{-0.5}$  & $-$            \\
PSZ1G108 M  & U    & $6.9\pm2.1$           & $7.6^{+2.5}_{-2.5}$   &   $873\pm90$             \\
PSZ1G108 NE  & U   & $1.5\pm1.2$           & $1.4^{+1.2}_{-0.8}$   &   $-$             \\
RXCJ1314 E & V   & $2.3\pm1.0$           & $3.9^{+2.1}_{-1.9}$   & $799\pm73$            \\
RXCJ1314 W  & V  & $4.2\pm1.3$           & $2.5^{+1.2}_{-0.8}$   & $845\pm99$            \\
ZwCl0008 E & W  & $2.9\pm0.8$           & $2.8^{+1.5}_{-1.0}$   & $757\pm62$            \\
ZwCl0008 W & W  & $2.7\pm0.8$           & $2.8^{+1.5}_{-1.0}$   & $681\pm66$            \\
ZwCl1447 N & X & $2.7\pm0.8$           & $2.0^{+1.8}_{-0.9}$   & $1147\pm63$            \\
ZwCl1447 S & X & $1.0\pm0.5$           & $1.2^{+2.2}_{-0.8}$   &   $-$           \\
ZwCl1447$^a$ X & 24  & $2.2\pm0.7$           & $1.7^{+1.0}_{-0.6}$   &   $-$               \\
ZwCl1856 N & Y  & $1.6\pm0.8$           & $1.2^{+0.5}_{-0.5}$   & $934\pm117$           \\
ZwCl1856 S & Y  & $1.5\pm0.7$           & $1.0^{+0.4}_{-0.7}$   & $862\pm133$           \\
ZwCl2341 NW & Z  & $0.8\pm0.6$           & $0.6^{+0.7}_{-0.3}$   &   $359\pm39$            \\
ZwCl2341 C & Z & $3.1\pm1.2$           & $3.0^{+2.3}_{-1.3}$   &   $-$             \\
ZwCl2341 S & Z  & $1.3\pm0.7$           & $1.1^{+1.4}_{-0.6}$   & $801\pm46$    \\
\hline

\end{tabularx}
\begin{minipage}{10.5cm}%\\
\small $68\%$ uncertainties reported\\
\small $^a$ denotes subclusters that are not involved in the merger
\end{minipage}

\end{table}

\section{WEAK LENSING MASS DISTRIBUTIONS}\label{Section:results}
In this section, the WL analysis of 29 radio relic merging galaxy clusters are presented and discussed. A summary of the literature on these merging clusters is presented in G19. For this reason, we will try not to repeat the work of G19 but will describe the relevant merging features, summarize previous WL results, and present/compare our new WL results. Our goal is to provide new insight into the merging systems with the mapping of the dark matter.

For each cluster, we present a four-panel figure. The top left panel is the WL mass map $S/N$ contours plotted over the Subaru color image. Contours start at $2\sigma$ and increase in intervals of $1\sigma$ unless otherwise noted in the figure caption. These WL mass maps are available upon request to the authors.  We indicate each of the significant subclusters ($S/N > 3$) with a blue, dashed circle that is centered on the BCG that we have assigned to the subcluster. The radius of the circles are chosen to be $R_{3000}$ because it fits elegantly into the field of view, where $\Delta=3000$ is the density contrast. Each $R_{3000}$ is derived from the mass estimates following the masses of the $c-M$ relation fits in Table \ref{table:subaru_mass}. The top right panel shows the WL contours over the X-ray imaging with radio contours (green). The bottom-left and bottom-right panels have the WL contours plotted over the galaxy luminosity and number density maps, respectively, where these galaxies were selected following the method in Section \ref{section:source selection}.

Table \ref{table:subaru_mass} contains the mass estimates for all the subclusters that we identified as securely detected. As mentioned, WL mass estimates are done via the 2PNFW method with $c$ and $M$ as free parameters and via the $c-M$ relation of \cite{2008duffy}. In addition, the velocity dispersion measurements from G19 are included in the table. The velocity dispersion measurements are matched to the WL measurements by their projected separation from the location of the WL mass peak.

% THIS IS THE START OF THE FIGURES FOR EACH CLUSTER
\begin{figure*}[!ht]
    \centering
    \includegraphics[width=0.8\textwidth]{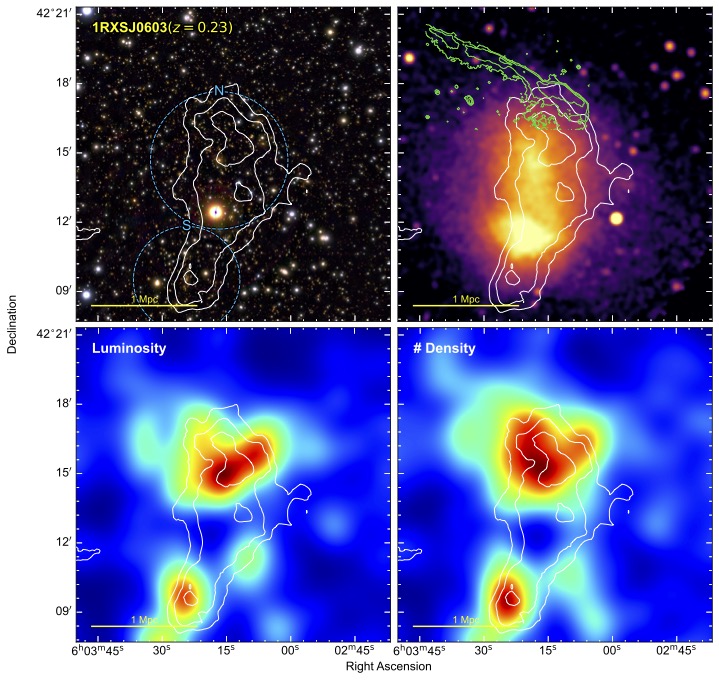}
    \caption{1RXSJ0603. A dissociative merger with an extremely large radio relic. \textit{ Top-left}: WL mass map (white contours) of the Toothbrush cluster over Subaru color image. \textit{Top-right}: WL mass over X-ray emission with radio relic plotted as green contours. \textit{Bottom-left}: WL over galaxy luminosity density. \textit{Bottom-right}: WL over galaxy number density.}
    \label{fig:toothbrush}
\end{figure*}

\subsection{1RXS J0603.3+4212 \texorpdfstring{($z=0.226$)}{TEXT}}
1RXSJ0603, also known as the Toothbrush cluster, is a well-studied merging cluster because of its 2 Mpc radio relic that resembles a toothbrush \citep{2012vanweeren_toothbrush, 2016stroe_toothbrush, 2016vanweeren_toothbrush, 2018rajpurohit_toothbrush, 2020rajpurohit_toothbrush, 2020degasperin_toothbrush}. The morphology of the relic was reproduced in hydrodynamic simulations \citep{2012bruggen_toothbrush}. \cite{2016vanweeren_toothbrush} suggested that re-acceleration is occurring at the Toothbrush relic and simulations agreed that re-acceleration is likely \citep{2016kang_toothbrush, 2017kang_toothbrush}. Radio measured Mach numbers for the Toothbrush relic range from 2.8 to 4.6 \citep{2012vanweeren_toothbrush, 2016vanweeren_toothbrush, 2018rajpurohit_toothbrush}.

The X-ray emission has an elongated morphology that stretches approximately 1.5 Mpc in a north-south direction with the radio relic lying to its north. The southern region of the X-ray emission has sharp edges that resemble ram-pressure stripping from a bullet-like core. \cite{2013ogrean_tooth} identified two distinct subclusters in the X-ray emission and provided evidence for three shocks with Mach numbers less than 2. \cite{2015itahana_toothbrush} measured the Mach number in the north shock to be \mytilde1.5.

A WL analysis of Subaru and \textit{HST} imaging is presented in \cite{2016jee}. Their study detected four substructures of which two were deemed significant and referred to as subclusters. These two subclusters are located at the BCGs in the north and south and are likely the subclusters that collided to create the radio relic. G19 identified four subclusters from galaxy redshifts with the two with the largest velocity dispersion corresponding to the largest WL signal detections. \cite{2016jee} assumed the \cite{2008duffy} $c-M$ relation and estimated the mass of the north (south) subclusters as $M_{200}=6.3^{+2.2}_{-1.6}\times10^{14}\ M_\odot$ ($M_{200}=2.0^{+1.2}_{-0.7}\times10^{14}\ M_\odot$).

\textbf{WL result:}
Our WL analysis is done on solely the Subaru imaging (Figure \ref{fig:toothbrush}). We detect the two primary subclusters that are presented in \cite{2016jee}. In addition, two other substructures from \cite{2016jee} are detected as an elongation from the northern mass peak and a separate peak directly to its south. We estimate the mass of the north (south) subclusters to be $M_{200}=6.2\pm1.2\times10^{14}\ M_\odot$ ($M_{200}=2.8\pm0.8\times10^{14}\ M_\odot$), which are consistent with the \cite{2016jee} result.

\begin{figure*}[!ht]
    \centering
    \includegraphics[width=0.8\textwidth]{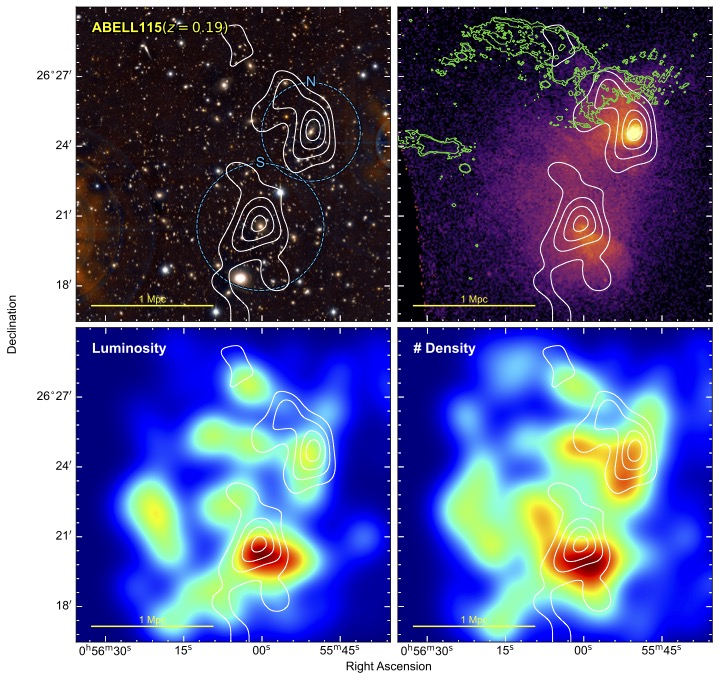}
    \caption{A115. A merger with peculiar ram-pressure stripped gas that suggests a large impact parameter. WL contours increase in steps of $0.5\sigma$.}
    \label{fig:a115}
\end{figure*}

\textbf{Merger insight:}
As is apparent from the offset between the mass peak (or BCG) and the X-ray emission in the south of the cluster, the Toothbrush cluster is a dissociative merger. The alignment of the radio relic and the elongated X-ray and WL distributions suggest that the N and S subclusters collided to form the shock. The ram-pressure stripped morphology of the X-ray emission is expected when the impact parameter is small. The Toothbrush cluster is a strong candidate to constrain the properties of dark matter because of the large separation of the mass peak from the X-ray peak but the cluster may be too complex.

\subsection{A115 \texorpdfstring{($z=0.193$)}{TEXT}}

A115 is a single radio relic cluster with double X-ray peaks \citep{1983beers_a115, 1984feretti_a115}. It stands out from the rest of the merging clusters because both subclusters appear to have ram-pressure stripped gas trailing behind cool cores, but the stripped gas does not align with the axis that the radio relic (shock) is moving. Since the cluster has a unique X-ray distribution, it has been the subject of quite a few studies. \cite{1999shibata_a115} presented evidence that A115 is a merging cluster with temperature variations measured by the Advanced Satellite for Cosmology and Astrophysics (ASCA). The dynamical analysis of \cite{2007barrena_a115} identified two structures in the galaxy distribution. \cite{2016botteon_a115} detected a shock in Chandra observations that is co-spatial with the radio relic and determined the Mach number to be $1.7\pm0.1$. A hot region was found between the subclusters in the X-ray temperature analysis of \cite{2018hallman_a115}. From VLA and GMRT observations, they calculated the radio Mach number to be 2.1.

\cite{2010okabe} included A115 in their WL analysis of 30 galaxy clusters, LoCuSS. Their WL mass map revealed two peaks. However, a \mytilde 300 kpc offset of the mass peak from the BCG was found in the northern cluster, which is an expected signature of an exotic dark matter model (ie. SIDM).

\begin{figure*}[]
    \centering
    \includegraphics[width=0.8\textwidth]{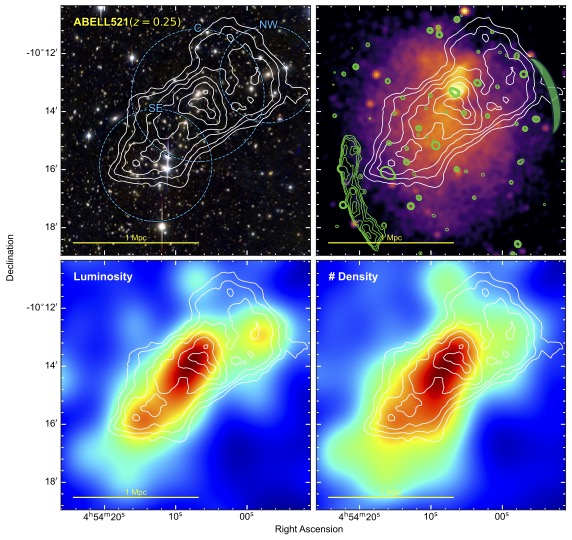}
    \caption{A521. A three subcluster system with a cool core, a radio halo, and two radio relics. The NW radio relic is faint and marked with a green crescent. WL contours increase in steps of $0.5\sigma$.}
    \label{fig:a521}
\end{figure*}

\textbf{WL result:}
Our WL analysis (Figure \ref{fig:a115}), published in \cite{2019kim_a115}, characterized the cluster as a bimodal merger with the northern cluster of mass $M_{200}=1.5\pm0.5 \times10^{14}\ M_\odot$ and southern cluster mass $M_{200}=3.0\pm0.8\times10^{14}\ M_\odot$. As Figure \ref{fig:a115} shows, both WL mass peaks are consistent with their X-ray peak and BCG counterparts, which is in contrast to the WL peak positions in \cite{2010okabe}. %Our investigation found that the offset was likely caused be a poor PSF model.

\textbf{Merger insight:} The BCG, X-ray brightness peaks, and mass peaks are co-spatial for each of the subclusters. The biggest mystery in A115 is reconciling the ram-pressure stripped tails seen in X-ray with the position of the radio relic. If one assumes that the tails are signifying the past direction of motion for the subclusters, it does not agree with an axial shock origin for the radio relic. Utilizing the WL measurements of \cite{2019kim_a115}, idealized \texttt{RAMSES} simulations of a two-cluster collision with a relatively large impact parameter by \cite{2020wonki} concluded that the ram-pressure stripped tails could be slingshot tails that have rotated relative to the original collision axis. Their simulations also reproduced the radio relic in the observed location. The large impact parameter of A115 sets it as unique merger that exhibits a radio relic.

\subsection{A521 \texorpdfstring{($z=0.247$)}{TEXT}}
\cite{2000arnaud_a521} showed that the X-ray emission of A521 has an irregular morphology with two peaks that are separated by about 500 kpc. \cite{2000maurogordato_a521} measured the radial velocities of 41 cluster galaxies and calculated a velocity dispersion of $1386^{+206}_{-139}$ km s$^{-1}$. \cite{2003ferrari_a521} provided evidence for a merging system by showing that the line-of-sight (LOS) velocity of the galaxies departs from a single Gaussian distribution. The X-ray emission detected by Chandra is elongated in a NW to SE direction with two major components \citep{2006ferrari_a521}. The radio relic is situated to the east of the cluster and elongates north to south \citep{2006giacintucci_a521, 2009dallacasa_a521}. The cluster has a distinct cool core that is co-spatial with the BCG and is compressed on its southern side \citep{2013bourdin_a521}. \cite{2013bourdin_a521} highlighted a bullet-like shape extending from the cool core, which may be further evidence of the ongoing merger. They detected a shock at the location of the radio relic in the XMM-\textit{Newton} X-ray observation and found the Mach number to be 2.4. MeerKAT observations from the cluster legacy survey \citep{2022knowles} are presented in Figure \ref{fig:a521}. A second radio relic is found in the NW of the cluster that may be a counter relic to the bright relic in the east (artificially placed arc in Figure \ref{fig:a521}). The nature of the diffuse radio emission is investigated further in \cite{2024santra_a521}.

G19 mention that the galaxies of the cluster can be divided into three subclusters with the primary in the center, one to the northwest, and one to the southeast. However, their GMM does not separate the northwest from the central subcluster and results in two components.

\textbf{WL result:}
Our work on A521 was published in \cite{2020yoon_a521}. The WL signal has three peaks (Figure \ref{fig:a521}). The most significant peak is consistent with the BCG and X-ray peak. In addition, WL peaks are found at the NW BCG (2nd) and the SE overdense galaxy region. Our mass estimates for the subclusters are $3.5\pm0.9$, $1.1\pm0.5$, and $2.5\pm0.7 \times 10^{14}\ M_\odot$ for the C, NW, and SE subclusters, respectively.

\textbf{Merger insight:}
The elongation of the mass distribution is slightly rotated with respect to the X-ray emission, which gives it a better agreement with the morphology of the radio relics than the X-ray emission has. The separation of the two subclusters from the central subcluster does not provide clear evidence to which subclusters collided to form the radio relics. Utilizing the wealth of information gained from the multiwavelength data, \cite{2020yoon_a521} tested merger scenarios with idealized simulations. Each tested merger scenario had its agreeing and disagreeing features with the observed features. The simulation showed that the subcluster that is now in the SE could have approached from the north and collided with the central cluster with a large impact parameter to form the shock that is observed at the SE radio relic position. That collision may have also caused the formation of the NW radio relic. Alternatively, the radio relics could originate from two different collisions.

\begin{figure*}[!ht]
    \centering
    \includegraphics[width=0.8\textwidth]{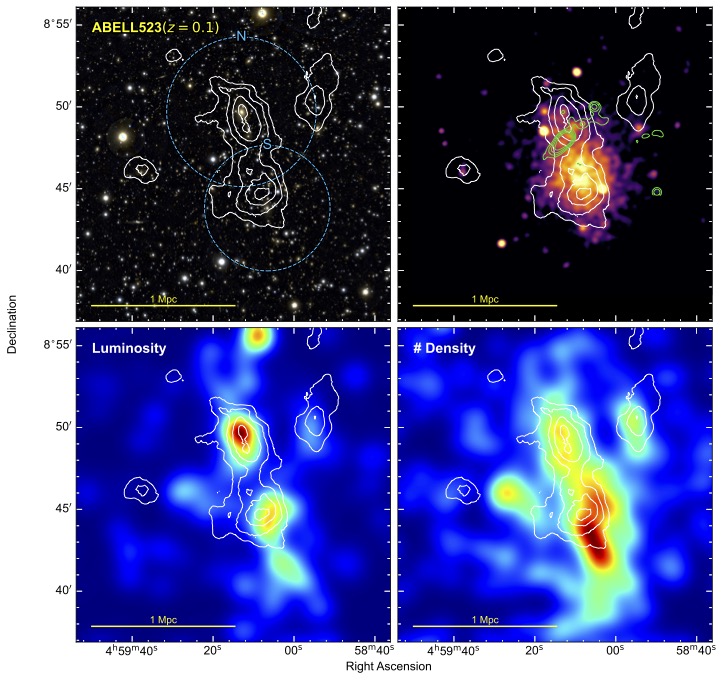}
    \caption{A523. A collision between two subclusters with radio emission in the middle. WL contours increase in steps of $0.5\sigma$.}
    \label{fig:a523}
\end{figure*}

\subsection{A523 \texorpdfstring{($z=0.104$)}{TEXT}}
A523 is a dissociative merger with the north ICM separated from the mass peak. \cite{2016girardi} specified that there are two BCGs with the brightest in the north and second brightest in the south. They found a galaxy overdensity directly west of the northern BCG and showed that two background structures straddle the cluster on the east and west. \cite{2019cova} investigated the X-ray emission of the cluster with XMM-\textit{Newton} and NuSTAR and detected the X-ray emission from the west component. The radio emission in A523 is not clearly a merger-induced bow shock and it is positioned between the BCGs (mass peaks) of the subclusters.  \cite{2022vacca_a523} analyzed LOFAR and VLA observations of A523 and showed that the radio features are quite complex.

\textbf{WL result:}
This is the first WL analysis of this cluster (Figure \ref{fig:a523}). Three mass clumps are detected from the Subaru imaging. The most significant detection is the north subcluster. The peak of the north subcluster is elongated north to south and encapsulates two bright cluster galaxies, of which the northern is brighter and elliptical and the southern is bluer and disky. The southern mass peak is situated to the south of the X-ray brightness peak and slightly north of the southern BCG. Our WL analysis also detects the northwest subcluster that was suggested in literature \citep{2016girardi, 2019cova}. The NW subcluster is likely not involved in the collision because the BCG associated with it is at a higher redshift of 0.13 (G19) and thus the mass peak has been omitted from the analysis. A two-halo NFW fit with peaks in the north and south give masses of $2.7\pm0.9$ and $1.6\pm0.7 \times10^{14}\ M_\odot$, respectively.

\textbf{Merger insight:}
Since the X-ray emission is elongated in the N-S direction and the radio emission is perpendicular to that, we predict that a collision occurred between the north and south subclusters. The mass estimate shows that the clusters that collided have a mass ratio of 3:1. It is interesting that the more massive cluster is in the north with the BCG but the X-ray emission peak is closer to the southern subcluster. The luminosity and number density maps also show an inversion with the northern cluster being brighter but the southern having more galaxies.

\begin{figure*}[!ht]
    \centering
    \includegraphics[width=0.8\textwidth]{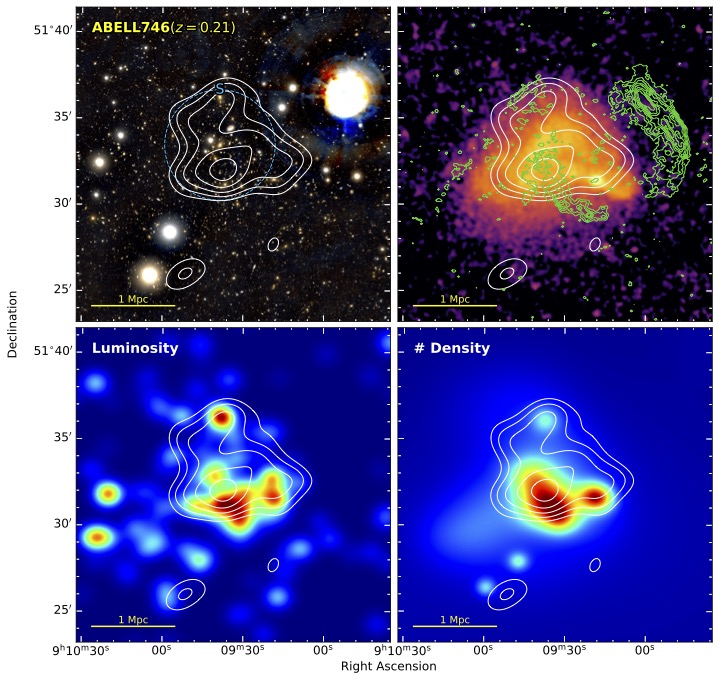}
    \caption{A746. A complex cluster merger with 4 diffuse radio sources. }
    \label{fig:a746}
\end{figure*}

\subsection{A746 \texorpdfstring{($z=0.214$)}{TEXT}}
A746 is a complex system with double relics, two isolated relics, a candidate radio halo, and many X-ray features \citep{2024rajpurohit_a746}. The X-ray and radio analysis of \cite{2024rajpurohit_a746} detected three merger-driven shock fronts. They estimated that the southern region of the cluster has an average temperature of \mytilde 9~keV and the northern has \mytilde 4~keV. They show that the giant radio relic in the west has a filamentary emission.

\textbf{WL result:} The Subaru Suprime-Cam observations of A746 suffer from prominent ghosts from a nearby bright star. Subaru HSC observations were collected (PI: H. Cho) in 2022B with a careful planning to keep the bright star centered in the HSC field of view (for symmetry reasons) and with shorter exposure times per integration. The newly acquired HSC observations enabled a WL analysis that is presented in \cite{2023hyeonghan_a746}. The WL analysis (Figure \ref{fig:a746}) shows a similar complexity to that found in the X-ray and radio emission. A dominant mass peak is found that coincides with the BCG and two less significant mass peaks are found to the west and north. The total mass of the cluster is $M_{200}=6.3\pm1.5\times10^{14}$ $M_\odot$.

\textbf{Merger insight:} The complex features of A746 make it a difficult cluster to disentangle. The double relics suggest a merger in the west but X-ray emission and WL do not discern the two merger constituents. WL analysis with a telescope that can achieve a much higher number density of galaxies (i.e., HST, JWST, Euclid, or Roman) may provide details into this complex merging system.

\begin{figure*}[!ht]
    \centering
    \includegraphics[width=0.8\textwidth]{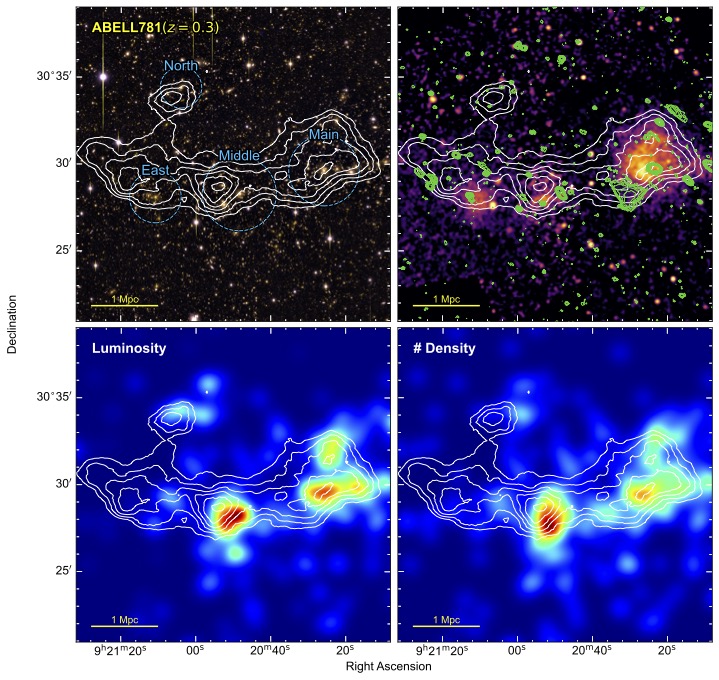}
    \caption{A781. A large-scale structure of clusters comprised of Main, Middle, and North. East is at a higher redshift. WL contours increase in steps of $0.5\sigma$.}
    \label{fig:a781}
\end{figure*}

\subsection{A781 \texorpdfstring{($z=0.297$)}{TEXT}}
A781 spans a large projected size on the sky (Figure \ref{fig:a781}). Here, we follow the naming scheme for clusters that was presented in \cite{2008sehgal_a781} and \cite{2014wittman}. The field of view contains 2 subclusters that are at $z\mytilde0.3$ (named Main and Middle), which are flanked by subclusters at $z\mytilde0.43$ called East and West (West is not shown in the figure). The candidate radio relic is situated between the Main and Middle subclusters \citep{2008venturi_a781, 2011venturi_a781, 2011govoni_a781}. \cite{2019botteon_a781} performed an in-depth analysis of X-ray and radio observations and suggested that the radio relic may be a combination of a radio galaxy and a shock. However, the polarimetric study of \cite{2023hugo_a781} concluded that it is likely not a radio relic.

The cluster is within the field of view of the Deep Lens Survey \citep[DLS;][]{2002wittman} and has a WL result \citep{2014wittman}. The DLS analysis detected the 4 subclusters and determined their masses. \cite{2012cook_a781} also performed a WL analysis and detected the East, Main, and Middle subclusters but were unable to detect the West subcluster, which hosts a strong-lensing arc. G19 separated the cluster galaxies into 4 subclusters but not the same subclusters as the previous WL results. Instead, they detected the Middle subcluster and then separated the Main subcluster into 3 additional subclusters.

\textbf{WL result:} Our WL result detects five subclusters: Main, Middle, East, West (not shown in the Figure), and one to the North. The Main subcluster coincides with the brightest X-ray emission and the BCG. The Main WL distribution is elongated in a east-west direction and has extensions to the west and north. These extensions are in agreement with the 3 substructures detected by G19. Both the Middle and East subclusters have X-ray emission counterparts. We remind the reader that the East subcluster is at a redshift of $z=0.43$. The North detection is coincident with a galaxy overdensity ($z\mytilde0.3$) and has X-ray emission as shown in \cite{2019botteon_a781}. Although it is not discussed, this North mass peak is also detected in the DLS analysis \citep{2014wittman}. We fit a four-halo model to the mass distribution and find masses of $3.5\pm0.8$, $4.2\pm0.9$, $3.3\pm0.8$, and $1.7\pm0.6$ $\times 10^{14}\ M_\odot$ for the Main, Middle, East, and North subclusters, respectively.

\textbf{Merger insight:} The location of the candidate radio relic and the X-ray emission hint that the merger-induced shock originated within the Main cluster. The elongation of the X-ray emission and the mass map agrees with this scenario. However, the resolution that is achievable with ground-based WL is insufficient to discern substructures in the Main subcluster and prevents us from constraining the mass of the collision that may have formed the relic.

\begin{figure*}[!ht]
    \centering
    \includegraphics[width=0.8\textwidth]{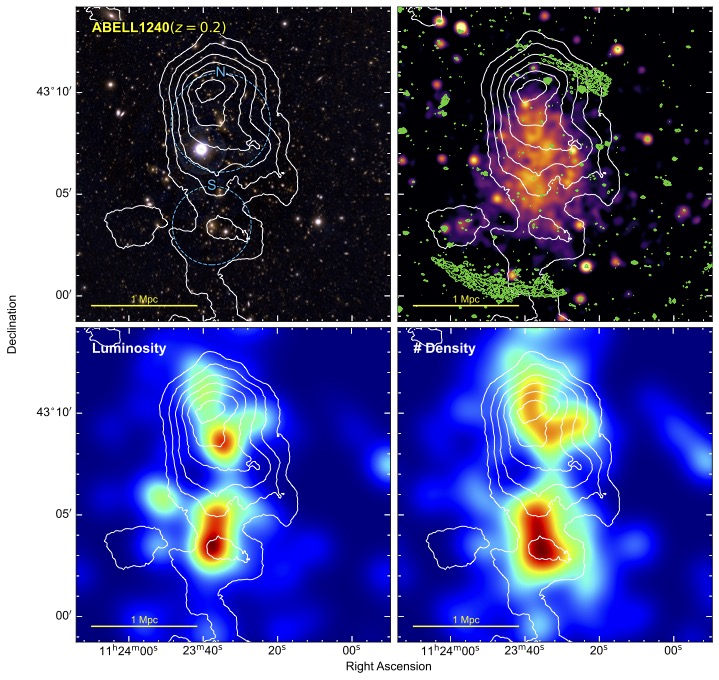}
    \caption{A1240. A head-on collision, dissociative merger with double radio relics.}
    \label{fig:a1240}
\end{figure*}

\subsection{A1240 \texorpdfstring{($z=0.195$)}{TEXT}}
A1240 is a double relic cluster with relics situated at opposing ends of the ICM distribution \citep{2009bonafede_a1240, 2018hoang_a1240}. From the radio spectral indices, \cite{2018hoang_a1240} derived Mach numbers of 2.4 and 2.3 for the north and south shocks, respectively. The X-ray emission spans the region between the radio relics and shows gas dissociation. \cite{2009barrena_a1240} detected two X-ray emission peaks in the Chandra observation and estimated the global temperature of the ICM to be 6~keV. \cite{2024sarkar_a1240} detected X-ray shocks at the locations of both radio relics and found them to have lower Mach numbers than radio ($M_{S}=1.5$ and $M_{N}=1.4$), which they suggested may be a sign of re-acceleration. G19 found that a two-halo model for the galaxy distribution was favored with a 1:1 mass ratio. A second cluster, A1237 ($z=0.194$), is located about 1.5 Mpc to the south of A1240.

\textbf{WL result:} This WL analysis was presented in \cite{2022cho}. The WL signal of A1240 (Figure \ref{fig:a1240}) shows the characteristic two peaks that are expected in bimodal mergers. The mass distribution is elongated along the merger axis that is represented by the X-ray and radio emission. The mass peak in the south is directly on the BCG, whereas the northern mass peak shows an offset but is statistically consistent with the BCG based on our bootstrapping. The signal from A1237 is also detected. We determine the masses to be approximately equal for the A1240 merger with $M_{200}=2.6\pm0.6\times10^{14}$ $M_\odot$ and $M_{200}=1.1\pm0.4\times10^{14}$ $M_\odot$ for the North and South subclusters, respectively. A bridge in the WL signal is found that runs between the A1240 and A1237 but at low significance.

\textbf{Merger insight:} A1240 is a bimodal merger between nearly equal-mass subclusters. \cite{2022cho} utilized the projected separation of the radio relics and the Monte Carlo Merger Analysis Code \citep[MCMAC;][]{2013dawson} to find that a merger phase that is returning from apocenter is favored with a time since collision of $1.7\pm0.2$ Gyr. \cite{2022cho} show that the A1240/1237 system is embedded in an \mytilde80 Mpc long filament as defined from SDSS galaxy positions. Further investigation of the connection of the merger to the filament would be interesting. The extreme dissociation of the gas for A1240 makes it a great candidate for further study of the nature of dark matter.

\begin{figure*}[!ht]
    \centering
    \includegraphics[width=0.8\textwidth]{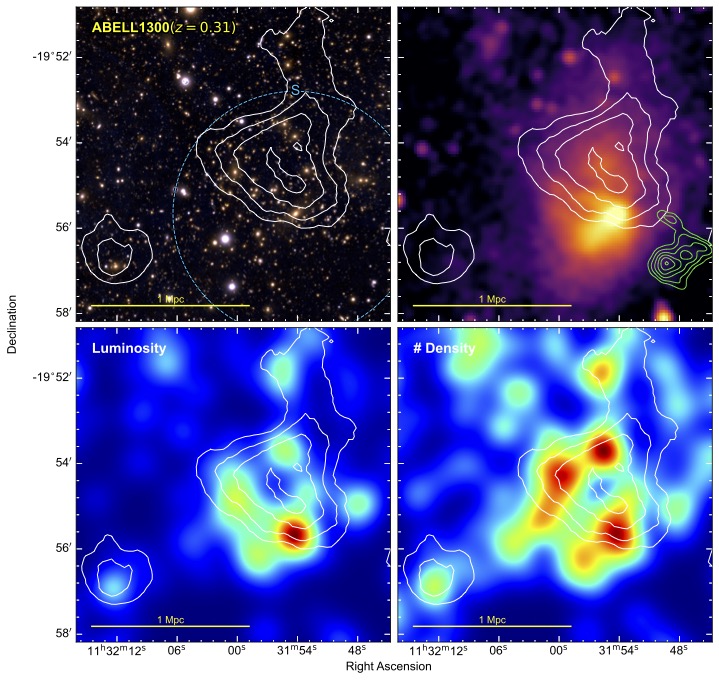}
    \caption{A1300. A complex cluster with an offset of the mass peak from the X-ray emission peak. An extension in the mass map traces an extension in the X-ray map to the north.}
    \label{fig:a1300}
\end{figure*}

\subsection{A1300 \texorpdfstring{($z=0.306$)}{TEXT}}
\cite{1999reid_a1300} detected two diffuse radio sources in A1300, a radio halo and relic \citep[both confirmed by][]{2011giacintucci_a1300}. \cite{1997pierre_a1300} performed a dynamical analysis and found that the velocity dispersion of the cluster galaxies had no significant departure from a Gaussian distribution. However, \cite{1997lemonon_a1300} highlighted the merging nature of A1300 from the structures seen in X-ray emission. \cite{2012ziparo} analyzed XMM-\textit{Newton} X-ray observations and showed that there are three primary ICM features: a bright peak in the south that elongates towards the southwest, a fainter peak about 250 kpc to the north, and an extension, possibly a filament, further north. The BCG lies directly on top of the bright X-ray peak in the south and has a semi-major axis along the same direction as the elongation of the X-ray emission. \cite{2021ternidegregory_a1300} presented the radio relic in the 1.3~GHz MeerKAT observation and noted that it has the morphology expected for radio emission from merger-induced shocks. G19 were unable to separate the cluster galaxies into multiple structures.

\textbf{WL result:} The WL map of A1300 (Figure \ref{fig:a1300}) does not follow our expectations based on the X-ray emission. Three features of the mass distribution stand out: a triangular-shaped clump in the center, a subcluster detached to the southeast, and a long extension to the north. The main clump has a primary peak that resides between the BCG and a bright galaxy immediately to its north. This peak is also offset from the X-ray brightness peak. The eastern vertex of the triangle has a nearby bright cluster galaxy and so does the northern vertex. It is likely that the complexity of the cluster is beyond the capabilities of the Subaru imaging. The extension to the north roughly follows cluster galaxies as can be seen in the luminosity and number density panels. It also follows the X-ray emission. The subcluster detected \mytilde1 Mpc to the southeast coincides with a bright galaxy and a faint X-ray detection. The agreement between the overall mass map and the X-ray emission is mixed. Since there is a lack of consistency between the WL substructures and the luminous tracers, we fit a single halo NFW model centered at the BCG and find the mass of the cluster to be $M_{200}=1.1\pm0.2\times10^{15}$ $M_\odot$.

\textbf{Merger insight:} The X-ray morphology is complicated and does not provide a clear feature that matches the position of the radio relic. The core of the X-ray emission has a bullet shape, which is sometimes a good indicator of the merger axis. However, like A115, it is hard to reconcile the bullet and the position of the radio relic. Perhaps, this is another case of a large impact parameter merger. The WL result seems to further complicate the interpretation because it is offset from the X-ray peak. However, the elongation of the core of the WL result does align with the radio relic. Similar to A746, the complexity of the merger seems to be beyond the ground-based Subaru imaging and may require higher resolution to discern the merging subclusters.

\begin{figure*}[!ht]
    \centering
    \includegraphics[width=0.8\textwidth]{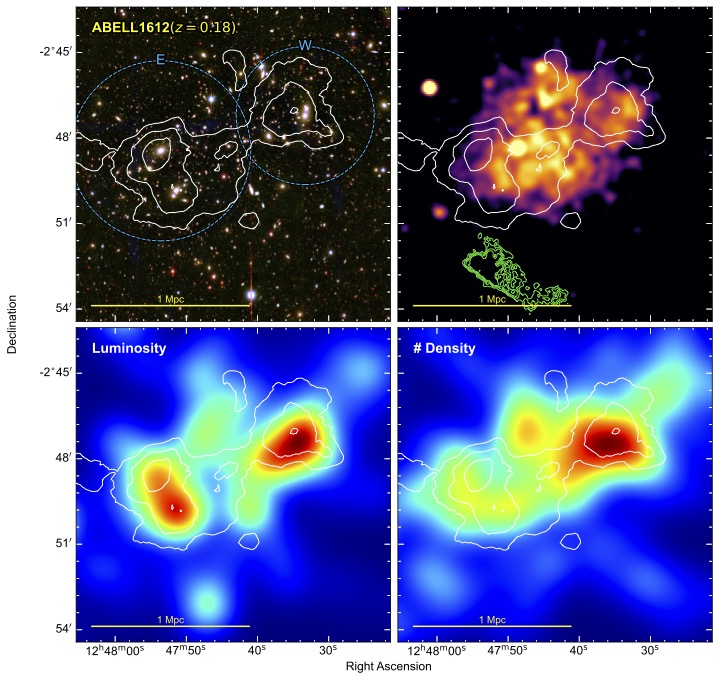}
    \caption{A1612. A merger between two equal-mass subclusters with a radio relic that is misaligned with the WL and X-ray elongated distributions. WL contours increase in steps of $0.5\sigma$.}
    \label{fig:a1612}
\end{figure*}

\subsection{A1612 \texorpdfstring{($z=0.182$)}{TEXT}}
A1612 has a single radio relic \citep{2011vanweeren} that is offset from the axis defined by the elongated X-ray emission. The X-ray emission does not show significant features, mostly because of the lack of X-ray photon counts.  The X-ray emission spans the region between the east and west BCGs. There is also X-ray emission detected to the north and a cavity (maybe due to low counts). G19 found two subclusters of galaxies that are each centered on the two BCGs.

\textbf{WL result:} The WL signal from A1612 shows 2 peaks separated by \mytilde 1~Mpc (Figure \ref{fig:a1612}). The east mass peak is coincident with the BCG and has a weaker S/N companion mass peak that is centered on the equally bright galaxy to its immediate south. The western peak is near the third BCG. The peaks are found at the ends of the elongated X-ray distribution. The $i$-band observations are shallow for this cluster and the WL signal is poor, even though the subclusters are resolved. A two-halo fit centered at the east and west BCGs finds masses of $4.0\pm1.0$ and $1.8\pm0.7 \times 10^{14}\ M_\odot$ for the E and W subclusters, respectively.

\textbf{Merger insight:} A1612 is a good candidate for a simple merger with a nearly equal mass ratio. However, the cluster has not garnered enough attention and lacks suffice multiwavelength observations to make strong conclusions about the collision. The radio relic is not in line with the elongation of the X-ray emission or the mass distribution, which may indicate a non-zero impact parameter of the collision. The eastern mass peak and its BCG do not have bright X-ray emission and thus A1612 may be a case of a dissociative merger.

\begin{figure*}[!ht]
    \centering
    \includegraphics[width=0.8\textwidth]{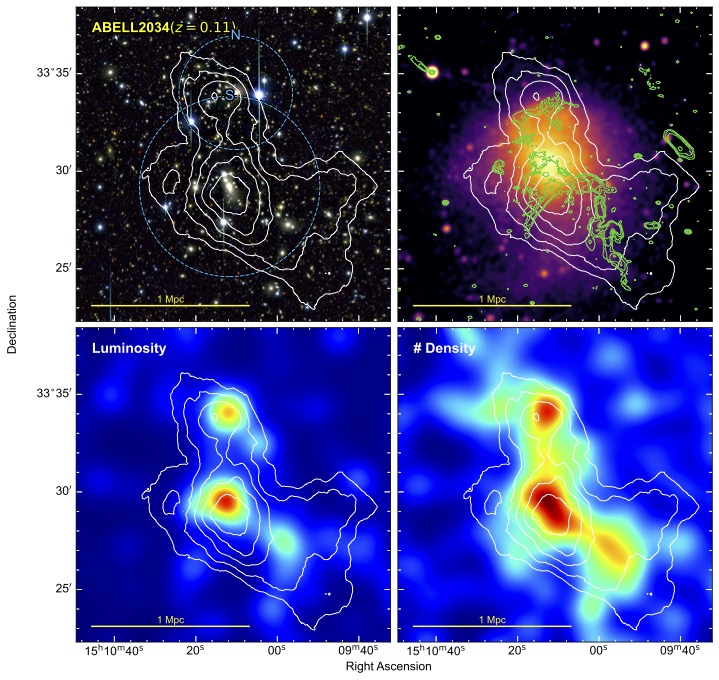}
    \caption{A2034. A dissociative merger with a bullet shape.}
    \label{fig:a2034}
\end{figure*}

\subsection{A2034 \texorpdfstring{($z=0.114$)}{TEXT} }
A2034 is a dissociative merger with a bullet-shaped morphology in the X-ray emission. A cold front was found in the Chandra observation and a candidate radio relic \citep{2001kempner_a2034, 2003kempner_a2034}. A shock ahead of the northern cold front was detected in \cite{2014owers} with a Mach number of $1.6\pm0.1$. For an in-depth discussion of the radio emission see the LOFAR work by \cite{2016shimwell}, which discussed two additional radio relic candidates. \cite{2008okabe_a2034} performed a WL analysis of the cluster and detected 6 significant peaks, of which 3 are slightly background to the cluster redshift. They suggested that these peaks may comprise a large-scale filament. \cite{2018monteiro} also presented a WL analysis of the cluster. Their mass map detected two significant peaks (with additional subpeaks) that are considered as the counterparts to the BCGs. In their work, the southern and strongest peak is situated to the south of the BCG and the northern peak is slightly offset to the east of the northern BCG. They estimate the masses of the clusters to be $2.4\pm1.0$ ($1.1\pm0.6$) $\times 10^{14}\ M_\odot$ for the south (north). \cite{2021moura_a2034} simulated the cluster with N-body hydrodynamical simulations using the measured properties from \cite{2018monteiro} as initial conditions. They concluded that the merger is near the plane of the sky with a low impact parameter and about 0.26 Gyr after collision. G19 found three galaxy overdensities that are centered in the north, south, and southwest regions. Their velocity dispersion measurements show that the north and south subclusters are comparable in mass and the southwest subcluster is minor.

\textbf{WL result:} Our mass map (Figure \ref{fig:a2034}) portrays a binary merger between the N and S subclusters. This is in agreement with the two previous WL studies of \cite{2010okabe} and \cite{2018monteiro} and the dynamical analysis of G19. The mass peak in the south is coincident with the BCG. The mass peak in the north is offset to the east of its BCG but within the WL statistical uncertainty. An extension is detected to the southwest that lies on top of the third BCG with a weakly detected peak nearby. Our mass estimates are $3.6\pm0.7$ ($0.9\pm0.4$) $\times 10^{14}\ M_\odot$ for the south (north) subclusters.

\textbf{Merger insight:} A2034 appears to be the most head-on merger in the sample. It has a large gas dissociation and thus is an ideal cluster for studying dark matter properties. There are some interesting features of this cluster that may warrant further investigation. The published WL analyses of this cluster show a consistent offset of the N mass peak to the east of its BCG. Is this a systematic of the data or an offset caused by the collision? The southern BCG is highly elongated along the merger axis. Is this a signature of the recent pericenter passage?  %Could it be that the northern BCG was offset from the mass peak during the merger and the interaction between the northern dark matter and the southern BCG caused the elongated southern BCG?

\begin{figure*}[!ht]
    \centering
    \includegraphics[width=0.8\textwidth]{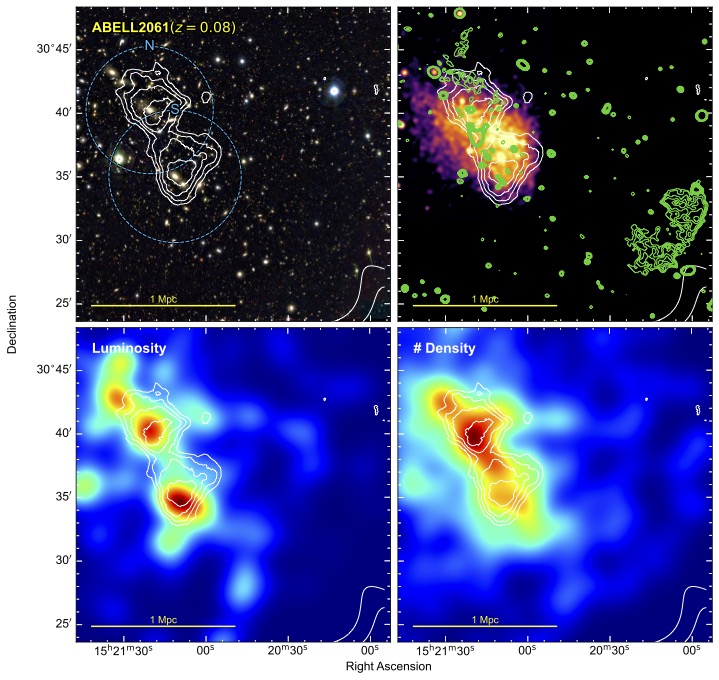}
    \caption{A2061. A dissociative merger with a radio relic that is very distant from the X-ray brightness peak. WL contours increase in steps of $0.5\sigma$.}
    \label{fig:a2061}
\end{figure*}

\subsection{A2061 \texorpdfstring{($z=0.078$)}{TEXT}}
A2061 is the lowest redshift cluster in the sample. It is part of the Corona Borealis supercluster \citep{1988postman, 2014pearson}. The X-ray emission peaks between the two BCGs suggesting that it is also a dissociative merger. Northeast of the cluster is a blob of X-ray emitting gas that may have been ejected from the main cluster \citep{2014sarazin_a2061}. The radio relic is approximately 2 Mpc southwest of the X-ray emission peak of the cluster \citep{2001kempner_a2034, 2011vanweeren}. G19 found that the cluster galaxies follow a single-Gaussian distribution with mass from velocity dispersion being $5.5\pm0.2\times 10^{14}\ M_\odot$.

\textbf{WL result:} The WL map of A2061 shows a bimodal distribution with mass peaks that lie directly on top of the BCGs (Figure \ref{fig:a2061}). The elongation of the mass map is slightly rotated from the elongation of the X-ray emission. Our two-halo fit estimates masses of $1.6\pm0.5$ ($2.0\pm0.6$) $\times 10^{14}\ M_\odot$ for the south (north), which suggests a 1:1 mass ratio merger of low total mass. The total mass found from lensing is in agreement with that of velocity dispersion.

\textbf{Merger insight:} The different orientations of elongation for the mass map and the X-ray emission is a possible signature of the impact parameter of the collision. The radio relic of A2061 is one of the farthest from the cluster when compared to the rest of this radio relic sample. The distance of the radio relic could be a hint that A2061 is an old merger. The galaxy distributions of A2061 show a group to the northeast and the WL signal has a slight elongation toward the direction of the group. It is unclear if this group was involved in the merger or is currently falling into the primary cluster. A2061 is a cluster that warrants further investigation to understand its dissociative nature and extreme distance to its radio relic.

\begin{figure*}[!ht]
    \centering
    \includegraphics[width=0.8\textwidth]{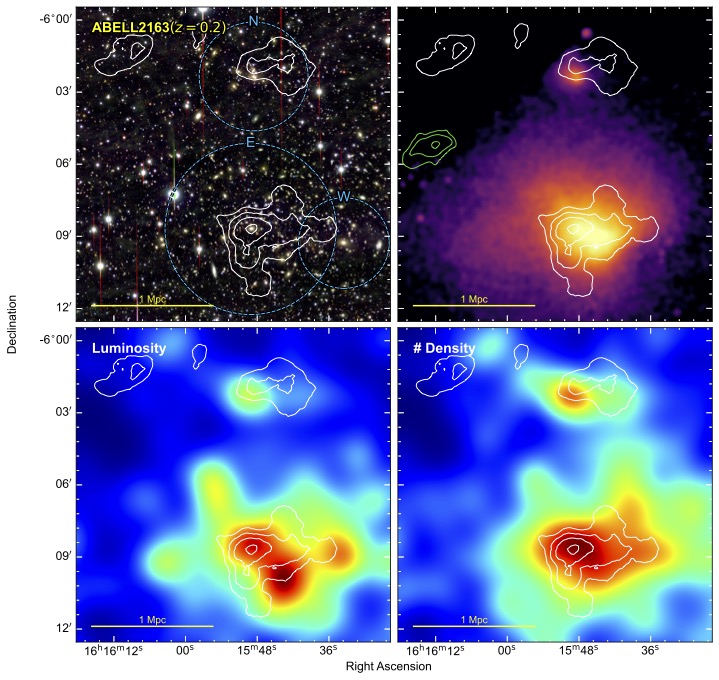}
    \caption{A2163. A low-mass subcluster (W) colliding with a massive primary cluster (E). }
    \label{fig:a2163}
\end{figure*}

\subsection{A2163 \texorpdfstring{($z=0.201$)}{TEXT}}
With a global temperature of $T=15$~keV \citep{1992arnaud}, A2163 is one of the hottest clusters in the sample. It is listed as the most massive cluster in the \cite{2016planck} catalog. The X-ray emission from A2163 is very extended, spanning approximately 2 Mpc in diameter in the south with an additional X-ray emitting subcluster in the north. A dynamical analysis by \cite{2008maurogordato_a2163} showed that the southern cluster is elongated in the east-west direction. G19 found that A2163 is composed of three structures with two located in the southern X-ray emission and one in the north. Based on their velocity dispersion estimates, the SE BCG is in the dominant subcluster and the SW is subordinate. Previous WL analyses have been consistent in detecting two mass peaks in the south and one in the north \citep{2008radovich_a2163, 2011okabe_a2163, 2012soucail}. The radio relic candidate in A2163 \citep{2004feretti_a2163} is found to the NE. \cite{2020shweta_a2163} provided evidence for a radio relic found in the center of the cluster. \cite{2018tholken_a2163} detected 3 shocks in Suzaku observations. One of the shocks is to the NE and coincides with the radio relic and the other two are found in the SW.

\textbf{WL result:} Our WL analysis (Figure \ref{fig:a2163}) detects the signal from both the north and the south clusters. The mass distribution of the southern cluster has a peak directly on the BCG (labeled as the E halo in Figure \ref{fig:a2163}). The mass distribution stretches to the west where it has a subordinate peak near the W BCG. The southern mass distribution also stretches to the south somewhat following the galaxy luminosity distribution, as shown in the bottom left panel. The northern cluster is detected with the mass peak slightly offset from the X-ray emission and the galaxy distributions. This offset is likely from noise. We fit a three-halo model and find that the masses are $9.2\pm1.8$, $0.8\pm0.6$, and $1.1\pm0.7 \times 10^{14}\ M_\odot$ for the E, W, and N subclusters, respectively. The mass estimate of the E subcluster agrees with that of \cite{2012soucail}.

\textbf{Merger insight:} A2163 is a complex cluster with many substructures. \cite{2008maurogordato_a2163} describe the many substructures of the cluster of A2613 found in galaxy overdensities. The detection of the mass distribution can provide support to some of the substructures. The cluster appears to be elongated east-west, which does not exactly agree with the position of the radio relic. However, the galaxy luminosity distribution shows an elongation extending from the BCG along the NE-SW axis that aligns with the position of the radio relic. The WL mass map also extends from the BCG along this axis but at low significance. A past collision between the main cluster and a subcluster directly to its southwest may be the cause of the radio relic.

\begin{figure*}[!ht]
    \centering
    \includegraphics[width=0.8\textwidth]{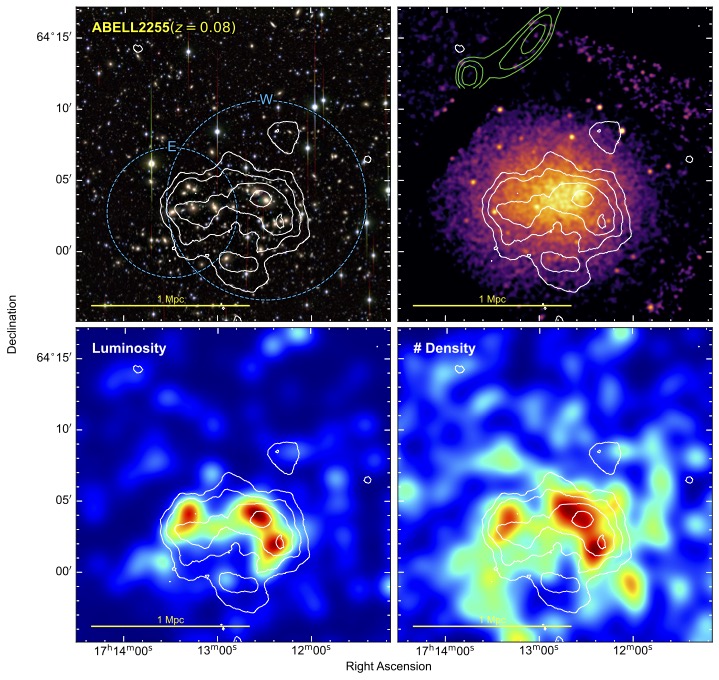}
    \caption{A2255. A relaxed-appearing X-ray emission with WL substructures.}
    \label{fig:a2255}
\end{figure*}

\subsection{A2255 \texorpdfstring{($z=0.08$)}{TEXT}}
A2255 has an X-ray morphology that is slightly elliptical with the major axis running east-west. The BCG is located near the western X-ray peak and there are bright galaxies related to the eastern X-ray emission. G19 was unable to resolve two subclusters in their dynamical analysis but did find a \mytilde 2000 km s$^{-1}$ difference in LOS velocity for the two BCGs in the west. \cite{2020botteon_a2255} described the ``beautiful mess'' of radio sources in A2255 and highlighted the NE radio relic. \cite{2017akamatsu} detected an X-ray shock at the location of the radio relic.

\textbf{WL result:} The WL map provides far more context to the merging system than the X-ray emission (Figure \ref{fig:a2255}). The mass distribution traces the bright cluster galaxies and is elongated along the east-west axis. A mass peak is coincident with the W BCG and an equally significant mass peak is detected about 100 kpc to its southwest. The WL signal extends to the east and is coincident with the eastern BCG. The proximity of the two western peaks prevent a three-halo model from being robustly fit. Our two-halo model gives masses of $4.7\pm1.1 \times 10^{14}\ M_\odot$ and $1.3\pm0.6 \times 10^{14}\ M_\odot$ for west and east, respectively.

\textbf{Merger insight:} There is great agreement between the WL mass map and the galaxy positions in A2255. The X-ray brightness peak is offset from the BCG and the mass peak, which makes this cluster a dissociative merger candidate. The mass map provides critical information that can explain the location of the radio relic. Rather than a merger between the E and W subclusters, it is more likely that a collision between the two substructures in the west led to the formation of the radio relic. Extending a line from the two substructures in the west bisects the radio relic. If these are the two subclusters that collided to form the radio relic, then their very small projected separation (\mytilde200 kpc) relative to the distance to the radio relic (\mytilde1 Mpc) is intriguing. It could indicate that the subclusters have had time to reach apocenter and begin their return to a second pericenter passage. Also, the large LOS velocity difference of the 2 BCGs in the west may indicate that projection may be important. This is an interesting case for the potential of a merger with a LOS component to the merger and a radio relic.

\begin{figure*}[!ht]
    \centering
    \includegraphics[width=0.8\textwidth]{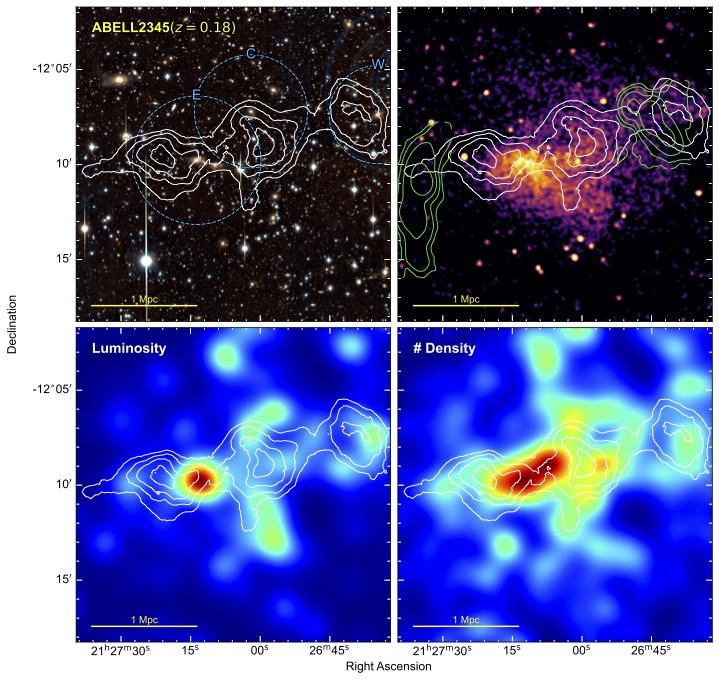}
    \caption{A2345. A chain of three subclusters in a merging system with two relics. WL contours increase in steps of $0.5\sigma$.}
    \label{fig:a2345}
\end{figure*}

\subsection{A2345 \texorpdfstring{($z=0.179$)}{TEXT}}
A2345 is a rare double relic cluster \citep{2009bonafede_a2345}. \cite{2021stuardi_a2345} described additional merger features in the X-ray and radio emission. The eastern relic is greater than one megaparsec in size and has the standard arc shape of a shock. The BCG is located in the east and is coincident with the X-ray brightness peak. The second brightest galaxy is 2 Mpc to the west of the BCG. The X-ray emission spans about 2 Mpc by 2 Mpc in projection. G19 were unable to separate subclusters in their analysis but they did present three peaks in the galaxy light distribution. The X-ray emission includes a subcluster that is located to the northwest of the second brightest galaxy. \cite{2002dahle} presented a WL result of A2345 that has three peaks with the eastern peak offset from the BCG by \mytilde1\farcs5 toward the east.

\textbf{WL result:} Our WL distribution of A2345 (Figure \ref{fig:a2345}) is tri-peaked in a chain that aligns with the axis connecting the radio relics. The distances between the central peak and the two other peaks are similar. As in \cite{2002dahle}, the eastern mass peak is offset from the BCG toward the eastern radio relic. The east mass peak is also detached from the bright X-ray emission indicating a potential dissociative merger. No bright galaxies are found in the region that the central peak is located but there is a galaxy overdensity. This detection is reminiscent of the dark core in A520, the Trainwreck cluster \citep{2007mahdavi_a520, 2012jee_a520}. The western peak is centered on the second BCG and the X-ray brightness peak in the region. Mass estimates for the three subclusters are $4.7\pm1.1$, $3.2\pm1.0$, and $2.1\pm0.8$ $\times 10^{14}\ M_\odot$ for east, central, and west, respectively.

\textbf{Merger insight:} Utilizing the information from the mass map, it is likely that the radio relics were formed by a merger between the east and central subclusters. These two mass peaks are approximately equal distance from the respective radio relics. The western mass peak is external to the radio relic. However, \cite{2021stuardi_a2345} describe a bullet feature that is coincident with this WL peak and has a tail to the east. This evidence would suggest that the W subcluster is moving westward, further complicating the merging scenario. For this subcluster to have formed the W relic and be positioned further westward it would have to overrun it. Assuming a collision between the E and C subclusters formed the radio relics, the remaining issue is understanding the central mass peak and its lack of BCG.

\begin{figure*}[!ht]
    \centering
    \includegraphics[width=0.8\textwidth]{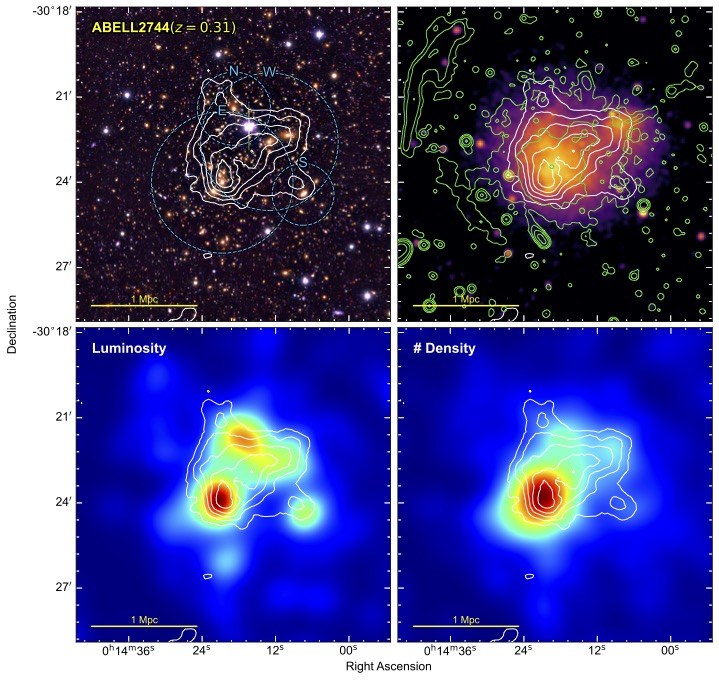}
    \caption{A2744. A well-studied, complex system of four subclusters.}
    \label{fig:a2744}
\end{figure*}

\subsection{A2744 \texorpdfstring{($z=0.306$)}{TEXT}}
A2744 (Pandora's cluster) is one of the most studied clusters in the sample. Many WL and SL analyses have been accomplished for this cluster. \cite{2016medezinski} analyzed the Subaru observations that are used in this paper with WL and were able to detect four substructures that roughly trace the galaxies in the cluster. The joint SL-WL analysis \citep{2016jauzac} traced the substructure of the cluster at high precision and discerned the substructures in the core. More recently, the joint SL-WL analysis of \cite{2024cha_a2744} used JWST to provide tight constraints on mass peaks that agree with their respective BCGs \cite[see][ for another JWST WL analysis]{2024harvey_a2744}. \cite{2024abriola_a2744} performed a WL analysis with Magellan observations and estimated the total mass of the cluster to be $M_{200}=2.56\pm0.26\times10^{14}\ M_\odot$.  The X-ray emission from A2744 has two peaks with the brighter near the BCG and the fainter to the northwest. \cite{1999giovannini_a2744} detected a radio halo and the NE radio relic in the NRAO VLA Sky Survey. \cite{2017pearce_a2744} identified three more radio relics in VLA observations. The radio features have been further analyzed by \cite{2021rajpurohit_a2744} and \cite{2022knowles}.

\textbf{WL result:} Our WL analysis detects four peaks (Figure \ref{fig:a2744}). The most significant peak is found on the E BCG. The second most significant peak is at the location of the W BCG. This peak and the BCG are offset eastward from the western X-ray emission peak. The two other mass-peak detections are in the north and south and are coincident with cluster galaxy populations. Our four-halo model results in mass estimates of $M_{200} = 7.2 \pm 1.8$, $6.6 \pm 2.3$, $1.0\pm0.8$, and $0.6\pm0.6 \times 10^{14}\ M_\odot$ for the E, W, N, S subclusters, respectively.

\textbf{Merger insight:} The position of the SE radio relic is easy to relate to the two bright X-ray peaks and the significant WL peaks of E and W. However, the NE and brighter relic is harder to reconcile with the detected substructures. It may be that one of the less significant subclusters passed through the system to form the NE radio relic. Both N and S mass peaks have better agreement with the morphology and position of the radio relic. Another scenario for the formation of the NE radio relic (but not the SE radio relic) could be that the more massive subclusters merged with a large impact parameter. The location of the BCG and the western mass peak inside the west X-ray brightness peak is perplexing and goes against the expectation of ram-pressure stripping for a dissociative merger. This configuration may occur at a later stage of merging while the subcluster is returning from apocenter. \cite{2011merten_a2744} discuss a scenario in which a ``ram-pressure slingshot'' effect has lead to the observed configuration of the X-ray emission and dark matter halo.

\begin{figure*}[!ht]
    \centering
    \includegraphics[width=0.8\textwidth]{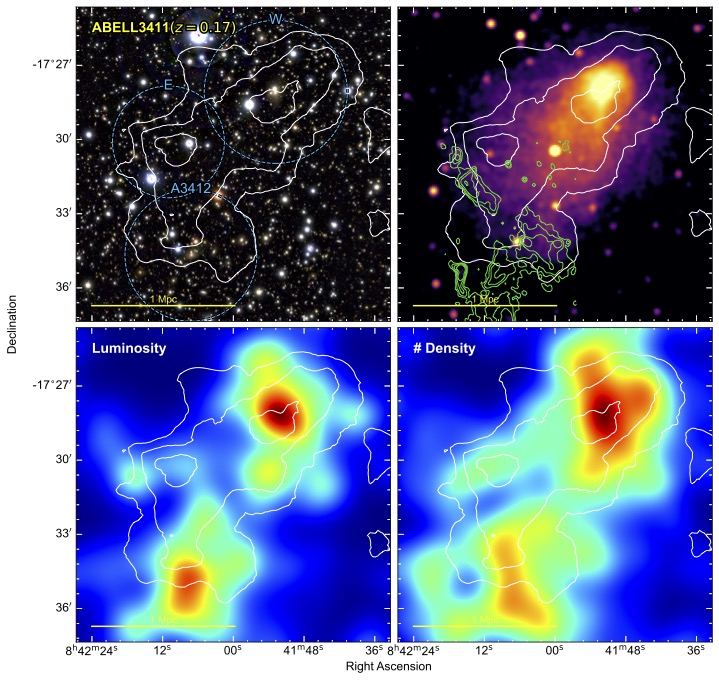}
    \caption{A3411. A merging system with a cool core and a connection between a radio galaxy and radio relic. WL contours increase in steps of $0.5\sigma$.}
    \label{fig:A3411}
\end{figure*}

\subsection{A3411 \texorpdfstring{($z=0.163$)}{TEXT}}
A3411 has a notable radio relic that was first detected in \cite{2013giovannini_a3411} and \cite{2013vanweeren_a3411}. The relic is notable because it is connected by a radio bridge to a nearby radio galaxy \citep{2017vanweeren}. The bridge has a spectral index that steepens away from the radio galaxy and then flattens along the radio relic, which they suggest is evidence for shock re-acceleration of the non-thermal population of charged particles that are seeded by the radio jet. \cite{2019andrade-santos} detected the shock at the relic location in deep Chandra observations and calculated a Mach number of $\lesssim1.15$. They estimated the mass of the cluster from X-ray observations to be $M_{500}=7.1\pm0.7\times10^{14} M_\odot$.  \cite{2020zhang} analyzed deep XMM-\textit{Newton} and Suzaku X-ray imaging and detected shocks in the southern edge and across the radio relic. Their X-ray temperature measurement of $T\mytilde5$~keV gives a mass of $M_{500}=5.1\times10^{14} M_\odot$. G19 found that the galaxy distribution shows two populations with one centered on the BCG in the NW and the other at A3412. The subclusters were found to have equally high velocity dispersion of$\mytilde 1200$ km s$^{-1}$.

\textbf{WL result:} The mass distribution of A3411 is marginally detected with $S/N\gtrsim3$ (Figure \ref{fig:A3411}). The mass distribution in the north peaks at the BCG and is elongated in the same direction as the X-ray distribution and toward the radio relic. Our WL analysis also detects A3412 but with low significance. The A3412 WL detection is offset from the position that G19 detected the galaxy overdensity but the offset is not statistically significant. Our mass estimates show that the W, E, and A3412 subclusters are $M_{200}= 2.3\pm1.0$, $1.8\pm1.0$, and $1.1\pm0.8 \times10^{14}\ M_\odot$, respectively. %These mass estimates project a total mass that is about $20\%$ lower than the total mass expected from the X-ray temperature measurement.

\textbf{Merger insight:} The WL result presented in this work provides weak evidence for the subclusters that collided to create the shock presented in \cite{2017vanweeren}. A merger between the E and W subclusters are the likely culprits. We do not expect A3412 to be the cause of the radio relic because of its position. New HSC observations (PI: H. Cho) have recently been completed and may provide an updated WL view of the massive structures in this galaxy cluster.

\begin{figure*}[!ht]
    \centering
    \includegraphics[width=0.8\textwidth]{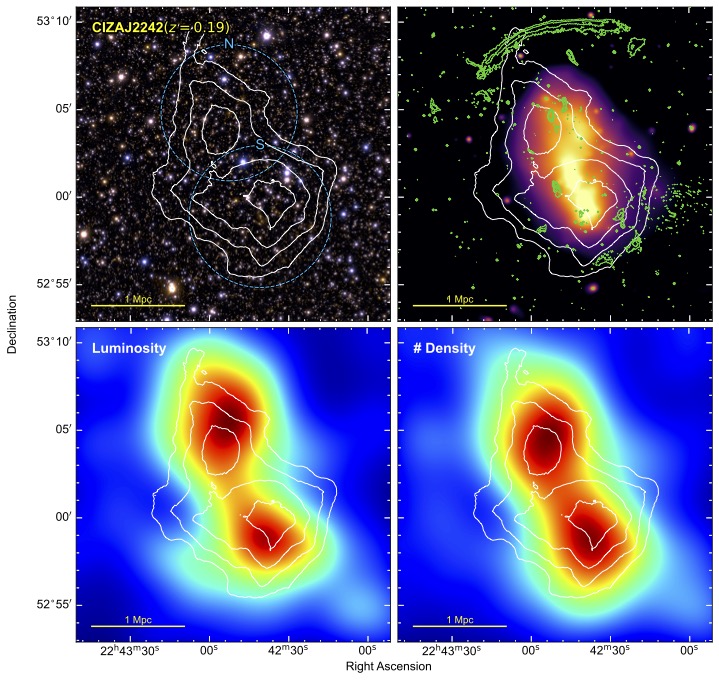}
    \caption{CIZAJ2242 (The Sausage Cluster). A system with a textbook, arc-shaped radio relic, complex X-ray emission, and a bimodal mass distribution.}
    \label{fig:cizaj2242}
\end{figure*}

\subsection{CIZA J2242.8+5301 \texorpdfstring{($z=0.189$)}{TEXT}}
CIZAJ2242 (The Sausage Cluster) contains the textbook example of a radio relic \citep{2010vanweeren_sausage} as well as radio relics to the south and east \citep[see][]{2011vanweeren_sausage, 2013stroe_sausage, 2014stroe_sausage, 2014stroe_sausage2, 2016stroe_sausage, 2017hoang, 2017loi_sausage, 2018digennaro_sausage}. The X-ray emission is elongated in the N-S direction with two peaks and an extension that stretches to the north. Shocks have been detected in X-ray at the location of the northern and southern radio relics as well as to the east \citep{2013ogrean_sausage, 2014ogrean_sausage, 2015akamatsu_sausage}. The archetypal relic in the north has spurred many simulations to attempt to recreate its properties as well as other ICM properties and features \citep{2011vanweeren_sausage, 2011matsukiyo_sausage, 2012kang_sausage, 2015kang_sausage, 2015fujita_sausage, 2016kang_sausage, 2016fujita_sausage, 2016donnert_sausage, 2017molnar_sausage, 2017donnert_sausage}. In addition, from SZ observations, \cite{2017rumsey_sausage} found a high pressure region that stretches perpendicular to the elongated X-ray emission. \cite{2015dawson_sausage} found the cluster galaxies to follow a bimodal distribution with close to equal velocity dispersion, which suggests equal mass subclusters.

Two previous WL analyses have been performed on CIZA2242. \cite{2015okabe} and \cite{2015jee} both mapped the WL signal of the Sausage cluster and found it to be bimodal. Both authors also found nearly 1:1 mass ratios (2:1 for \cite{2015okabe}) when simultaneously fitting the subclusters with \cite{2008duffy} $c-M$ models. However, the total mass in \cite{2015jee} is almost two times higher.

\textbf{WL result:} We utilize the \cite{2015jee} shape catalog for our analysis and unsurprisingly result in a similar mass map (Figure \ref{fig:cizaj2242}) to \cite{2015jee}. Fitting two NFW halos, we estimate a mass of $M_{200}= 8.2\pm1.8\ (9.3\pm1.8) \times10^{14}\ M_\odot$ for the north (south) subclusters, which are consistent with \cite{2015jee}.

\textbf{Merger insight:} The morphology and features of the X-ray emission from CIZA2242 are not as expected for a 1:1 mass ratio merger of two massive clusters, as pointed out by \cite{2013ogrean_sausage}. The brightest X-ray emission is located near the BCG in the south. A second brightness peak is located approximately 0.5 Mpc to the north and then an extension snakes a Mpc further to the north. The complexity of the features suggests that additional subclusters are likely to exist.  CIZAJ2242 is a dissociative merger with similarity to MACSJ1752 and ZwCl1856. An additional mystery is that the mass peaks in the north and south show offsets from their respective BCGs. The mystery may arise because ground-based WL cannot resolve such substructures and provide a poor constraint on the mass peaks because of the high extinction and stellar density of the zone of avoidance. High-resolution imaging is required to achieve the number density of source galaxies that is needed to constrain the mass peak positions and to resolve substructures.

\begin{figure*}[!ht]
    \centering
    \includegraphics[width=0.8\textwidth]{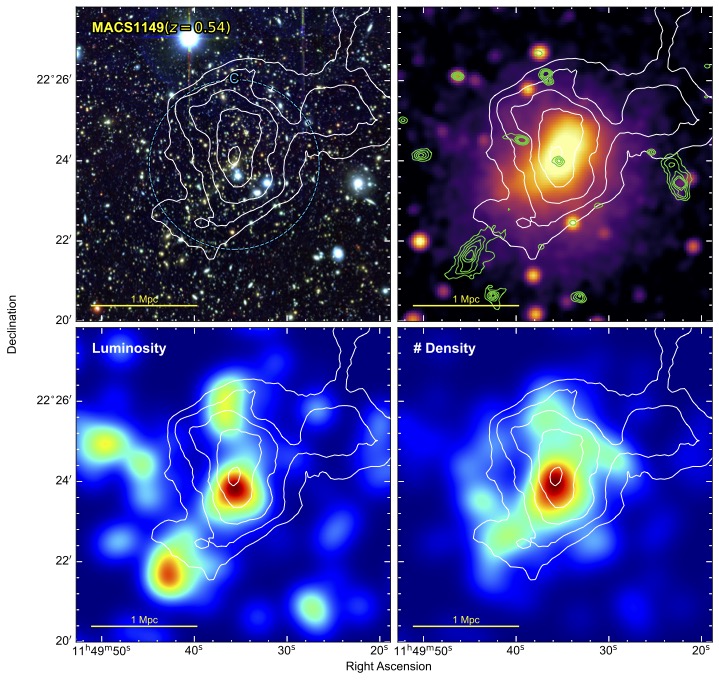}
    \caption{MACSJ1149. A high-z cluster with a N-S elongation in the mass and X-ray distributions. WL contours increase in steps of $0.5\sigma$.}
    \label{fig:macsj1149}
\end{figure*}

\subsection{MACS J1149.5+2223 \texorpdfstring{($z=0.544$)}{TEXT}}
MACSJ1149 is the highest redshift cluster in the sample. \cite{2012bonafede_macsj1149} pointed out two radio relic candidates in the system but one was shown to be a radio galaxy \citep{2020giovannini_macsj1149, 2021bruno_macsj1149}. It is one of the most X-ray luminous and hottest ($T=10.73^{+0.62}_{-0.43}$~keV) clusters known \citep{2016ogrean}. G19 found the cluster to have an extreme velocity dispersion of $1668$ km s$^{-1}$. The WL analysis by \cite{2014umetsu} showed an elongated signal that runs southeast from the BCG and the joint SL-WL analysis of \cite{2018finney} presented an elongated distribution that runs from the BCG to the north. The disparity is likely caused by the inclusion of strong lensing and the limited field of view of the HST observations used in \cite{2018finney}.

\textbf{WL result:} Our WL mass reconstruction (Figure \ref{fig:macsj1149}) has a peak directly on the BCG. The WL mass is elongated to the north from the BCG and encompasses the second BCG in similar fashion to the WL-SL analysis of \cite{2018finney}. There is a hint of an extension toward the southeast from the BCG (as is found in \cite{2014umetsu}) but it has no clear peak. We also get a detection in the northwest of the cluster that has a weak agreement with galaxy luminosity. Since no clear subclusters are detected, we fit a single NFW halo and find the mass to be $M_{200}= 16.4\pm3.1 \times10^{14}\ M_\odot$.

\textbf{Merger insight:} The low-significance substructures that are detected in our WL analysis do not provide additional evidence to the origin of the radio relic. As \cite{2021bruno_macsj1149} pointed out, the orientation of the relic and the mass distribution are difficult to reconcile.

\begin{figure*}[!ht]
    \centering
    \includegraphics[width=0.8\textwidth]{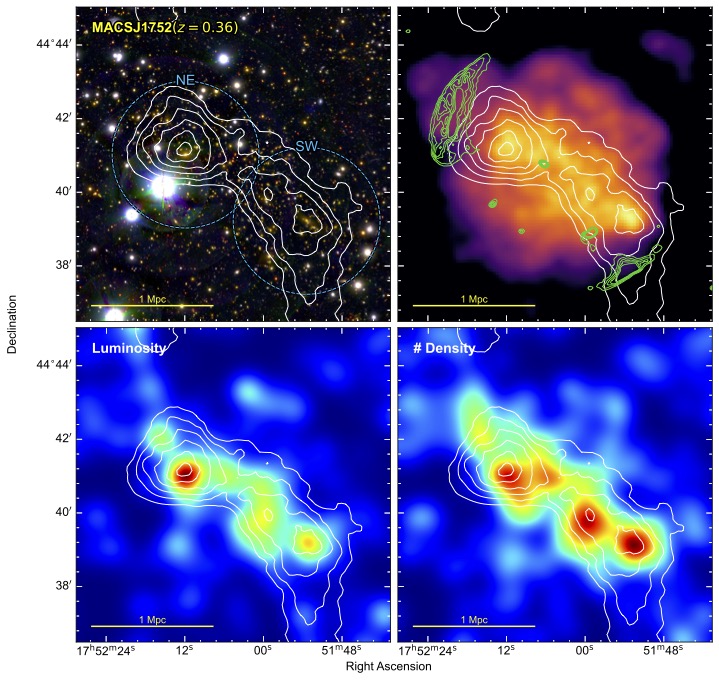}
    \caption{MACSJ1752. An equal-mass merger with ram-pressure stripped tails in the X-ray emission and double radio relics.}
    \label{fig:macsj1752}
\end{figure*}

\subsection{MACS J1752.0+4440 \texorpdfstring{($z=0.365$)}{TEXT}}
A multi-wavelength analysis of MACSJ1752 is presented in \cite{2021finner}. MACSJ1752 contains two bright and extended radio relics \citep{2012vanweeren_macsj1752, 2012bonafede_macsj1752}. The radio relics are arc-shaped and on opposing ends of the elongated X-ray distribution. \cite{2015degasperin} pointed out the extreme luminosity of the radio relics in MACSJ1752. The X-ray emission is double peaked with a bridge running between the peaks. \cite{2021finner} showed that these two X-ray brightness peaks are the sites of cold fronts. The elongated shape of the X-ray distribution resembles an ``S'' and is evidence of a small impact parameter collision. As a textbook example of a galaxy cluster merger, the cluster has been utilized in simulations to test the mass bias of WL \citep{2023lee} and cosmic ray acceleration \citep{2016vazza_macsj1752}.

\textbf{WL result:} In \cite{2021finner}, our WL detection (Figure \ref{fig:macsj1752}) revealed two mass peaks that coincide with the BCGs and X-ray peaks. In addition, we found a third substructure between the peaks that coincides with the third brightest galaxy and an X-ray peak. Our mass estimates show that the NE and SW subclusters have masses of $5.6\pm1.8 \times 10^{14}\ M_\odot$ and $5.6\pm1.7 \times 10^{14}\ M_\odot$ and the central substructure is $0.3^{+0.4}_{-0.1} \times 10^{14}\ M_\odot$. MACSJ1752 is among the most massive binary mergers in the sample.

\textbf{Merger Scenario:} The merger scenario of MACSJ1752 is straightforward. As shown in many idealized cluster simulations, a low impact parameter merger between equal-mass subclusters can recreate the observed X-ray morphology features such as the ``S'' shape and cold fronts. The radio relics are not perfectly symmetric but are among the cleanest example of double relics. An intriguing connection between a galaxy and the southern radio relic is noted in the WSRT observation in \cite{2021finner} but further spectral analysis is needed to see if it supports the re-acceleration mechanism.

\begin{figure*}[!ht]
    \centering
    \includegraphics[width=0.8\textwidth]{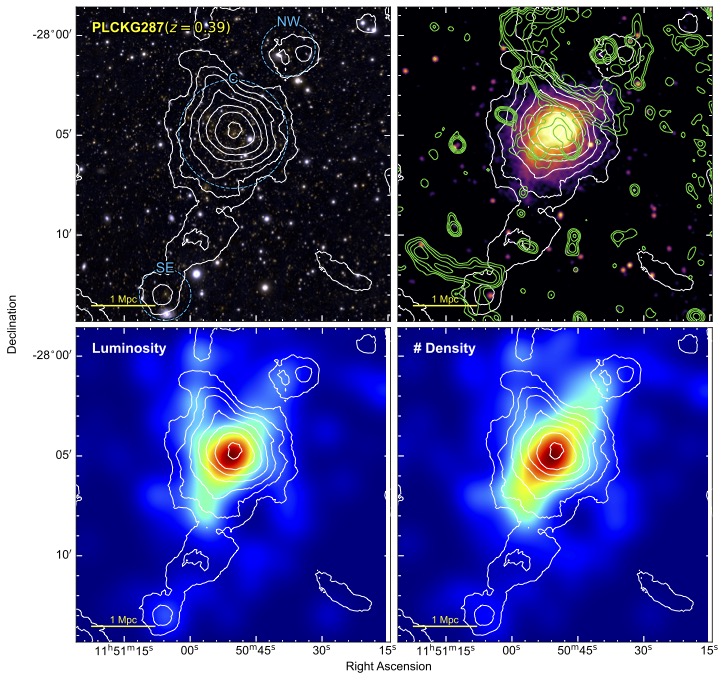}
    \caption{PLCKG287. A centrally dominant, massive cluster with low-mass subclusters and  double radio relics at asymmetric projected separation.}
    \label{fig:plckg287}
\end{figure*}

\subsection{PLCK G287.0+32.9 \texorpdfstring{($z=0.385$)}{TEXT}}
PLCKG287 is one of the more unusual clusters in the sample. The cluster has an X-ray distribution that contains only a single peak and is somewhat relaxed in appearance with a hint of an extension toward the southeast. The global temperature of the cluster is 13~keV and it is the second most massive cluster ($26\pm1.0 \times 10^{14}\ M_\odot$) in the Plank catalog \citep{2016planck}. There are a few tens of strong-lensing arcs found around the BCG and the cluster is one of the largest strong lenses in the universe that has been detected to date. G19 found the velocity dispersion of the cluster to be $1756$ km s$^{-1}$, the highest in this radio relic sample. Their analysis of the cluster galaxies separated the cluster into a massive primary with a secondary \mytilde2.5 Mpc to the south. It has two radio relics that are vastly different distances from the X-ray peak \citep{2011bagchi_plckg287, 2014bonafede_plckg287} and separated by about 3 Mpc.  \cite{2014bonafede} showed that the galaxy distribution follows a NW-SE layout and is likely a filament that is feeding the massive cluster.

\textbf{WL result:} Published in \cite{2017finner}, the WL signal of PLCKG287 is dominated by a central cluster with a mass peak that has excellent agreement with the BCG and the X-ray brightness peak (Figure \ref{fig:plckg287}).  Two additional subclusters are detected that lie along a NW - SE axis. A 3-halo fit to the mass distribution finds $20.4\pm1.9 \times 10^{14}\ M_\odot$, $1.7\pm0.7 \times 10^{14}\ M_\odot$, and $1.4\pm0.7 \times 10^{14}\ M_\odot$ for the C, SE, and NW subclusters, respectively.

\textbf{Merger insight:} It is expected that either one of the substructures merged to form both relics on two passages or that each of the two substructures were responsible for a relic. \cite{2014bonafede} estimates the time since collision (pericenter) for each of the relics to be approximately 0.7 and 0.1 Gyrs for the south and north radio relics, respectively. The BCG in the cluster has a nearly equally bright, nearby companion that may be valuable in interpreting the merger. It is also worth noting that the 3rd brightest cluster galaxy, that is \mytilde1 Mpc to the south east of the BCG, does not have a WL peak.

\begin{figure*}[!ht]
    \centering
    \includegraphics[width=0.8\textwidth]{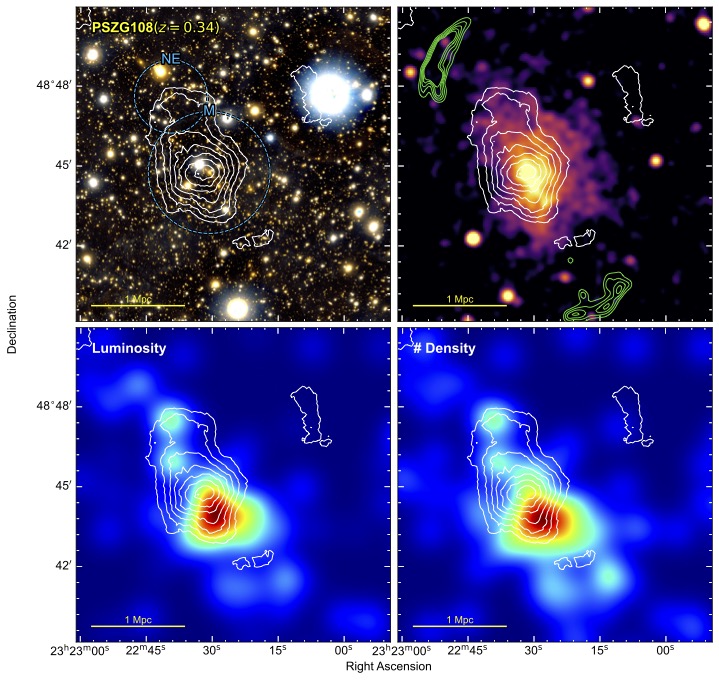}
    \caption{PSZ1G108. A centrally dominant cluster with large separation double radio relics. WL contours increase in steps of $0.5\sigma$.}
    \label{fig:psz1g108}
\end{figure*}

\subsection{PSZ1 G108.2-11.53 \texorpdfstring{($z=0.335$)}{TEXT}}
PSZG108 is a double radio relic cluster with a very large relic separation (\mytilde 3.5~Mpc). G19 identified two galaxy populations that are near the central point between the relics. However, only 40 galaxies had available redshifts and their GMM did not separate the galaxies into two subclusters. The X-ray emission in PSZG108 peaks near the BCG and is elongated along the axis that connects the radio relics. The second brightest galaxy is \mytilde 1.5~Mpc to the northeast of the BCG and is coincident with the northern radio relic. The radio relics of PSZG108 are two of the most powerful found to date \citep{2015degasperin}.

\textbf{WL result:} We analyzed the Subaru HSC observations of PSZG108. Our WL mass reconstruction (Figure \ref{fig:psz1g108}) detects the signal with a main peak that is coincident with the BCG. The WL signal elongates toward the NE following a similar distribution to the X-ray emission. The significant WL signal terminates at a galaxy overdensity that we consider the NE subcluster. We note that the second BCG is further to the NE. The WL signal also extends slightly to the SW where many of the bright clusters galaxies reside. We fit a two-halo NFW model and find the subcluster masses to be $6.9\pm2.1 \times 10^{14}\ M_\odot$ and $1.5\pm1.2 \times 10^{14}\ M_\odot$ for the main and NE subclusters.

\textbf{Merger insight:} The double-relic nature of PSZG108 can be used to constrain the merger axis to be approximately northeast to southwest. The X-ray emission and WL signal are elongated in the same direction. One scenario for the formation of the radio relics is a collision between the two subclusters that are detected in WL, which would have a mass ratio $\gtrsim$1:4. The large separation of the radio relics suggests a long time has passed since the collision if the shocks propagate at a constant velocity \citep{2018ha}. Close analysis of the HSC observations reveal that the BCG has many bright companion cluster galaxies (as is apparent from the luminosity and number density maps). An alternative scenario is that the collision that formed the radio relics occurred far enough in the past that the two subclusters are both situated in the BCG region. WL analysis would require a higher number density of galaxies to resolve subclusters in this region. Follow up observations with the HST or JWST would be valuable for SL-WL analysis of the cluster. There are large blue arcs at the core of the cluster that are likely strong-lensing images.

\begin{figure*}[!ht]
    \centering
    \includegraphics[width=0.8\textwidth]{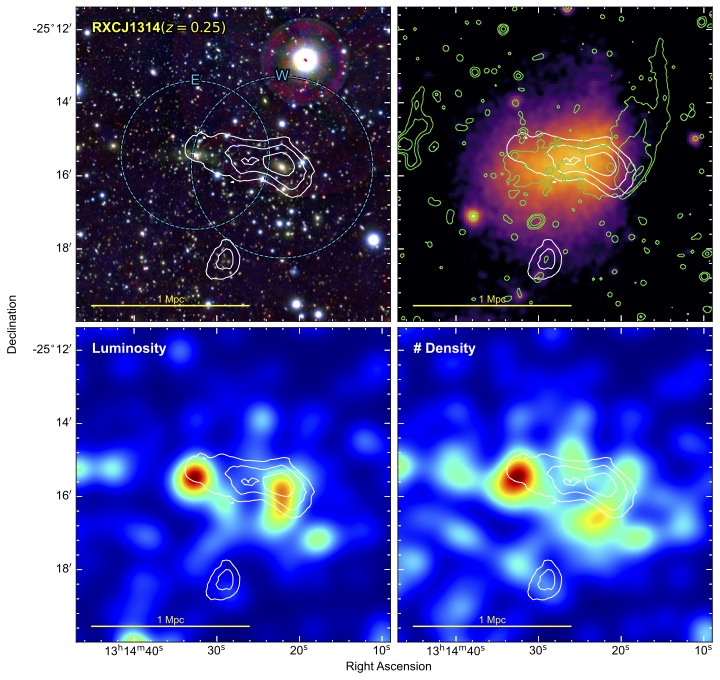}
    \caption{RXCJ1314. A merging system with a possibly large viewing angle and radio relics at asymmetric projected separation. WL contours increase in steps of $0.5\sigma$.}
    \label{fig:rxcj1314}
\end{figure*}

\subsection{RXC J1314.4-2515 \texorpdfstring{($z=0.247$)}{TEXT}}
RXCJ1314 is the second example of a two radio relic cluster that has a distinct difference in the distance of the radio relics from the barycenter of the subclusters (like PLCKG287). The imbalanced radio relic distribution could be caused by two merging events or it could also be a projection effect \citep[for recent radio work see][]{2019stuardi}. A shock was detected in XMM-\textit{Newton} observations at the location of the western relic \citep{2011mazzotta_rxcj1314}. G19 concluded that the member galaxies follow two Gaussian distributions with a \mytilde 1500 km s$^{-1}$ LOS velocity difference. Therefore, projection effects are important to consider for RXCJ1314. \cite{2019stuardi} came to a similar conclusion from vastly different rotation measures from the two radio relics and found that simulations suggest a 70 degree departure from the plane of sky. Nevertheless, the axis connecting the radio relics agrees with the elongated X-ray emission and defines a probable merging axis. The cluster was also recently observed with MeerKAT \citep{2022knowles}, which is presented in Figure \ref{fig:rxcj1314}.

\textbf{WL result:} The WL mass map (Figure \ref{fig:rxcj1314}) is elongated along the projected merger axis. The W mass peak is dominant and coincides with the BCG. The E mass peak is spatially consistent with the second BCG and a third mass clump is detected to the south. It is unclear whether this 3rd mass peak is at the redshift of the cluster or is background. Our two-halo fit finds a 2:1 mass ratio with masses of $4.2\pm1.3$ and $2.3\pm1.0 \times 10^{14}\ M_\odot$ for the E and W subclusters, respectively.

\textbf{Merger insight:} The X-ray emission, galaxies, and WL signal offer a simple merger geometry between the east and west subclusters. Similar to PLCKG287, it could be that two different mergers have occurred that created the two radio relics. However, our WL analysis does not detect an additional subcluster that may have been involved in the merger.

\begin{figure*}[!ht]
    \centering
    \includegraphics[width=0.8\textwidth]{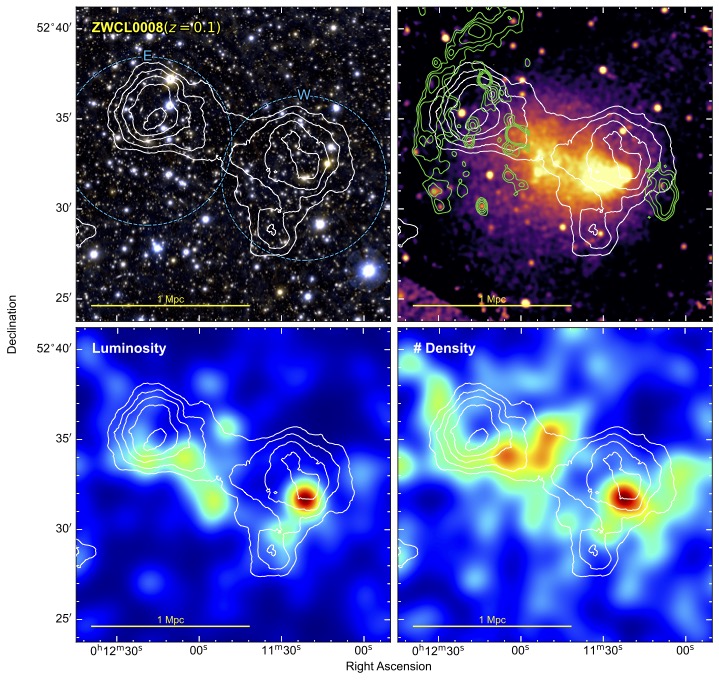}
    \caption{ZwCl0008. A dissociative merger of equal mass subclusters. WL contours increase in steps of $0.5\sigma$.}
    \label{fig:ZwCl0008}
\end{figure*}

\subsection{ZwCl 0008.9+5215 \texorpdfstring{($z=0.104$)}{TEXT}}
ZwCl0008 is another merging cluster with double radio relics \citep{2011vanweeren_zwcl0008}. The double relics are quite different from each other with the largest linear size of the east relic three times bigger than the west. A similar discrepancy is found in the Sausage relics. ZwCl0008 exhibits a bullet-shaped morphology in the ICM \citep{2017golovich}. It also has a very short radio relic standoff distance \citep[for a discussion of the standoff distance see][]{2019zhang}. These two features are signs of a recent merger. A WL analysis of the cluster was done with \textit{HST} and Subaru imaging in \cite{2017golovich} and found that the eastern subcluster is four times more massive than the western subcluster. The spectroscopic observations of G19 suggest a cluster merger closer to a 1:1 mass ratio. From Chandra and Suzaku observations, \cite{2019digennaro_zwcl0008} detected a shock at the western radio relic and measured a Mach number of $1.5\pm0.5$. They found no shock at the location of the eastern relic. \cite{2012kang_zwcl0008} simulated the collision and showed that the radio relic agrees with the combination of DSA and re-acceleration. \cite{2018molnar_zwcl0008} predicted that the cluster merger is an off-axis binary merger that is viewed shortly after first core passage.

\textbf{WL result:} Our WL analysis on solely the Subaru imaging detects the two subclusters (Figure \ref{fig:ZwCl0008}). Using our multiple NFW halo fitting technique, we find that the mass ratio is close to 1:1 with the eastern (western) subcluster mass being $M_{200} = 2.9\pm0.8\ (2.7\pm0.8) \times 10^{14}\ M_\odot$. This is in agreement with the velocity dispersion values in G19. In addition to the detection of the two subclusters, an elongation of the WL signal from the western subcluster towards the south is found. This detection is coincident with another bright cluster galaxy. Comparing the WL signal to the galaxy luminosity and number density distributions shows that there is likely another substructure.

\textbf{Merger Insight:} The merger scenario for ZwCl0008 is straightforward with a collision between the two components of a mass distribution that is dominated by two subclusters. The X-ray emission shows morphological features that suggest a small impact parameter. The short standoff distance of the radio relics is expected for a system that is observed shortly after collision.  ZwCl0008 is a great candidate for constraining the properties of dark matter.

\begin{figure*}[!ht]
    \centering
    \includegraphics[width=0.8\textwidth]{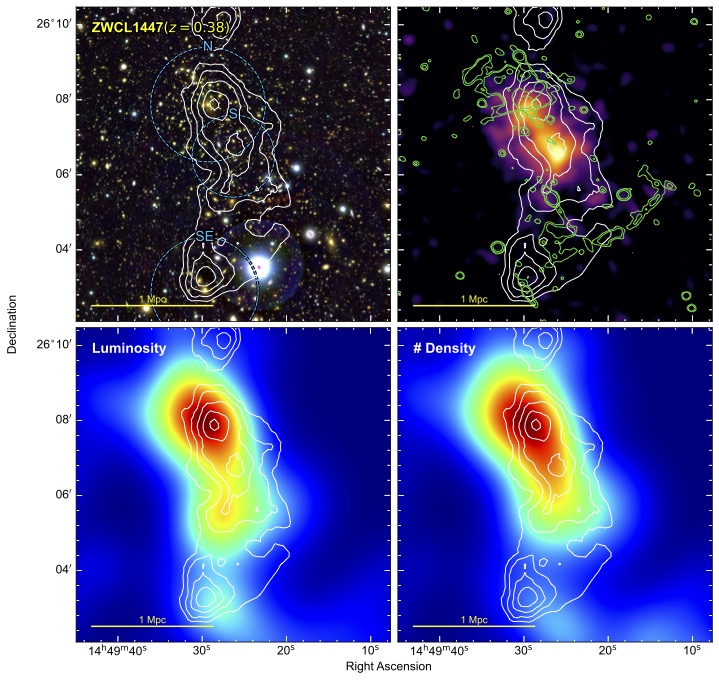}
    \caption{ZwCl1447. A merging system with double radio relics. WL contours increase in steps of $0.5\sigma$ and start at $3\sigma$.}
    \label{fig:ZwCl1447}
\end{figure*}

\subsection{ZwCl 1447+2619 \texorpdfstring{($z=0.376$)}{TEXT}}
ZwCl1447 is a \cite{1984butcher} galaxy cluster that has a poorly defined red sequence. Recent \textit{Chandra} observations revealed a double clump ICM distribution with a small (\mytilde 500 kpc) separation between the subclusters \citep{2022lee_zwcl1447}. \cite{2009giovannini} analyzed VLA 1.4 Ghz data and discovered a radio relic to the south of the cluster. They also found a radio relic candidate in the north but were undecided on whether it is a relic or halo. However, recent GMRT observations by \cite{2022lee_zwcl1447} have highlighted that the diffuse radio emission in the north is indeed a radio relic. They showed that the relic in the south is a textbook example of a bow shock (like the Sausage relic). In addition, they ruled out radio halo emission. G19 were unable to divide the spectroscopically confirmed cluster member galaxies into multiple distributions.

\textbf{WL result:} The WL analysis (Figure \ref{fig:ZwCl1447}) for this cluster was first presented in \cite{2022lee_zwcl1447}. We find that the WL signal consists of three distinct peaks. Two of the peaks are coincident with the X-ray brightness peaks. A third peak is located about 1 Mpc to the SE of the BCG and is associated with the second brightest galaxy in the region. As shown in Figure \ref{fig:ZwCl1447}, there is good agreement between galaxies and mass distribution in this cluster. The SE mass peak that is coincident with the second brightest galaxy in the field has no X-ray detection. One reason may be from the chip gap of the \textit{Chandra} detector. A second reason may be that the gas has been stripped from a past merger with the primary cluster. Better planned X-ray observations would be useful for understanding the nature of the SE substructure. Our three-halo fit to the distribution gives masses of $2.7\pm0.8$, $1.0\pm0.5$, and $2.2\pm0.7 \times 10^{14}\ M_\odot$ for the N, S, and SE subclusters, respectively.

\textbf{Merger Scenario:} The alignment of the two subclusters in the north is in great agreement with the radio relics. We suggest that these are the subclusters that merged to form the shocks. Furthermore, the radio relic in the south has a very uniform arc shape that may be indicating a head-on collision. The location of the SE mass peak suggests that it has not been involved in the merger, especially since it lies ahead of the radio relic. However, it is peculiar to not detect any X-ray emission from it. For detailed simulations of the cluster merger, see \cite{2022lee_zwcl1447}.

\begin{figure*}[!ht]
    \centering
    \includegraphics[width=0.8\textwidth]{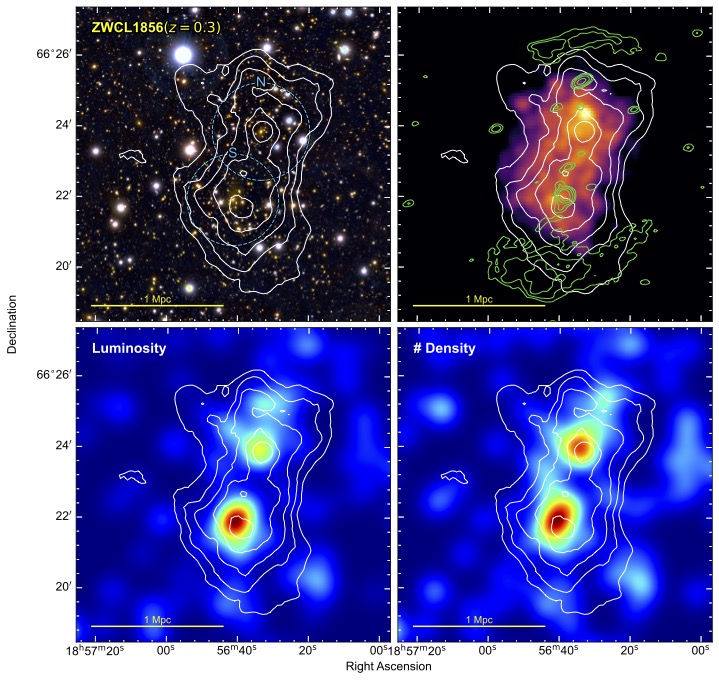}
    \caption{ZwCl1856. An equal-mass merger with low total mass and double radio relics. WL contours increase in steps of $0.5\sigma$.}
    \label{fig:ZwCl1856}
\end{figure*}

\subsection{ZwCl 1856.8+6616 \texorpdfstring{($z=0.334$)}{TEXT}}
ZwCl1856 is a pristine example of double radio relics that are nearly equal in shape, size, distance from the center, and orientation. The relics were first presented in \cite{2014degasperin_zwcl1856} and further analyzed in \cite{2021jones_zwcl1856}. Similar to MACSJ1752, the radio relics are found at opposing ends of an S-shaped X-ray morphology. However, in comparison to MACSJ1752, the location of the relics are in better agreement with the merger axis, being directly inline with the elongated X-ray. \cite{2021jones_zwcl1856} also detected a radio bridge (connection) between the southern radio relic and the southern BCG. G19 were able to separate the cluster galaxies into two populations of nearly equal velocity dispersion and suggested an equal mass merger.

\textbf{WL result:} The WL analysis (Figure \ref{fig:ZwCl1856}) of ZwCl1856 is presented in \cite{2021finner}. The mass distribution is dominated by two mass peaks that agree with their respective BCGs and align along the merger axis that is defined by the radio relics. A signature of a third, lower significance peak is found to the north that is located near the northern radio relic. We found the N (S) subclusters to have masses of $1.6\pm0.8$ ($1.5\pm0.7$) $\times 10^{14}\ M\odot$. Intriguingly, ZwCl1856 is a low-mass binary merger that hosts bright radio relics. The low mass is in agreement with the expected mass from our XMM-\textit{Newton} temperature estimate of $3.6^{+0.6}_{-0.5}$ keV \citep{2021finner}.

\textbf{Merger insight:} The S-shaped X-ray morphology suggests a merger with a small impact parameter. The mass peaks in both the north and south are slightly inside the X-ray brightness peaks, which could indicate that the cluster is near apocenter and the ICM has been slingshot through the potential. However, this interpretation, which is based on poor X-ray observations, should be remedied soon with upcoming deeper Chandra observations. MACSJ1752 and ZwCl1856 provide two examples of simple cluster mergers that have much different masses but host bright radio relics.

\begin{figure*}[!ht]
    \centering
    \includegraphics[width=0.8\textwidth]{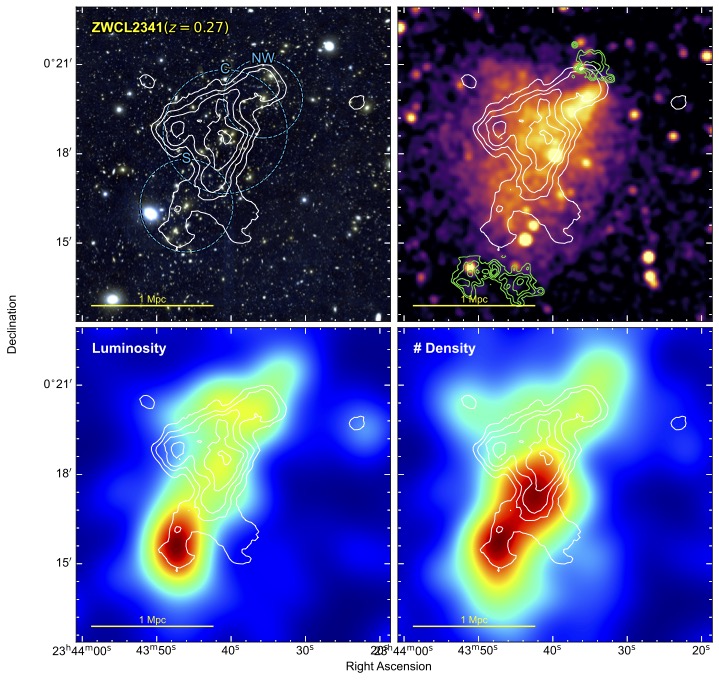}
    \caption{ZwCl2341.A complex cluster merger with a bullet-shaped X-ray emission. WL contours increase in steps of $0.5\sigma$.}
    \label{fig:ZwCl2341}
\end{figure*}

\subsection{ZwCl 2341+0000 \texorpdfstring{($z=0.27$)}{TEXT}}
The ICM of ZwCl2341 is rich in merging features. \cite{2009vanweeren_zwcl2341} found double radio relics, which were verified in \cite{2010giovannini_zwcl2341} and \cite{2022parekh_zwcl2341}. The polarization properties of the radio relics of ZwCl2341 were also studied in \cite{2022stuardi_zwcl2341}.  The radio relics are roughly aligned with each other and separated by about 2 Mpc. In the northwest, a bullet is apparent in X-ray emission. In front of the bullet shape is the northern radio relic with an X-ray shock detected in \cite{2014ogrean_zwcl2341}. The X-ray emission extends to the northeast and south. A similar X-ray distribution is seen in A3411. The southern ICM contains an X-ray shock and the southern radio relic. \cite{2021zhang_zwcl2341} go into detail about many of the merging features that are found in the Chandra observation. Cluster galaxies were identified in \cite{2013boschin_zwcl2341} and used to show the SE-NW elongation of the cluster. \cite{2017benson} performed a multiwavelength analysis of ZwCl2341. Their modeling of the cluster galaxies determined that the cluster is composed of three subclusters. G19 also detect three subclusters with two in the NW and one in the SE.

\cite{2017benson} also provide a WL analysis of the galaxy cluster. Their WL analysis on the available Subaru r-band imaging shows a single peak with a broad distribution that covers the brightest galaxies in the cluster.

\textbf{WL result:} Our WL analysis (Figure \ref{fig:ZwCl2341}) shows a peak in the center that traces an overdensity of cluster member galaxies. From the central peak there are extensions to the south, east, and north. The southern extension overlaps the brightest and densest galaxy region. The northern extension contains the bullet X-ray emission and more bright cluster galaxies. The eastern extension is the most peculiar. It seems to not trace many cluster galaxies. The peak of the eastern extension lies on top of a deep yellow elliptical galaxy. It may be possible that there is a background cluster in the east of the X-ray emission that is contaminating the WL signal. The histogram of spectroscopic redshifts in Figure 4 of \cite{2017benson} shows a possible background structure at $z=0.31$ but more spectroscopy would be needed to confirm it. We fit a three-halo model to the NW, C, and S subclusters and find masses of $0.8\pm0.6$, $3.1\pm1.2$, and $1.3\pm0.7 \times 10^{14}\ M\odot$.

\textbf{Merger insight:} ZwCl2341 is another example of a very complex cluster merger. The formation of the radio relics likely resulted from a collision between the central and southern subclusters, the central and northern clusters, or both collisions. The complexity of ZwCl2341 makes it a difficult cluster to simulate. A better approach may be to search for analogs from cosmological simulations.

\subsection{Non-detections}\label{sec:nondetections}
For three of the clusters that are analyzed in G19, we are unable to provide WL results. Our WL analyses of A3365 ($z=0.093)$ and RXC J1053.7+5452 ($z=0.072$) result in no detection. Further analysis of the imaging will be required to understand the failure. For RXCJ1053, bright stars in the Subaru imaging are the probable cause of no detection. Bright stars can be removed by careful subtraction of a 2D model from the stacked image. For example, observations with the HSC targeting A746 were planned to fix a bright star at the center of the focal plane, which allowed an easier subtraction of a symmetric model \citep{2023hyeonghan_a746}. Future imaging of RXCJ1053 could use this technique to alleviate the issue of the bright star. At the time of this analysis, A2443 ($z=0.110$) had no WL quality imaging. However, the cluster was recently observed with the HSC (PI: H. Cho) and the observations are starting to be analyzed. The WL result will be presented in Kim et al. in prep.

\section{DISCUSSION} \label{Section:discussion}
\subsection{WL Mass Estimate Comparison}\label{section:comparison}
The $c-M$ relation is typically derived from simulations that have a fixed volume and in some cases a fixed cosmology \citep[e.g.,][]{2008duffy, 2014dutton}, although, some are valid for any cosmology \cite[e.g.,][]{2019diemer, 2021ishiyama}. Under these constraints, the $c$ and $M$ of clusters in the simulation are limited and may not represent the full range of observed clusters. This may be particularly true for the merging cluster sample where recent, powerful gravitational interactions have taken place that could have a significant impact on the concentration of the dark matter distribution. To test whether the $c-M$ relation holds for merging clusters, it would be ideal to compare a measured concentration from the merging cluster sample to the expected concentration of the $c-M$ relation. However, concentrations are poorly constrained for most individual clusters \citep[see][for stacked WL constraints on concentration]{2020umetsu}. An alternative is to test the effect of a merger on the concentration by comparing the mass derived from a $c-M$ relation to that of an independent method such as our two-parameter fitting method (2PNFW).

\begin{figure}[!ht]
    \centering
    \includegraphics[width=0.45\textwidth]{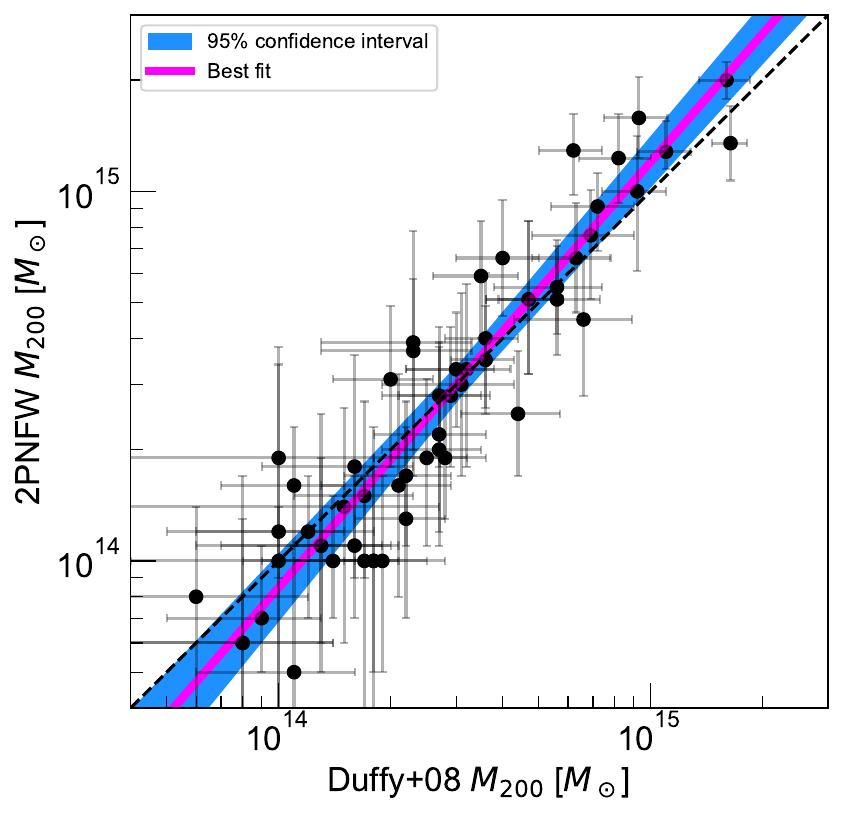}
    \caption{Comparison of the mass derived from the \cite{2008duffy} $c-M$ relation to the mass derived from the two-parameter fit (values from Table \ref{table:subaru_mass}). The black-dashed line represents the 1:1 relation. The violet line is the best-fit linear regression. Uncertainties are considered on both axes. The $c-M$ relation tends to overestimate/underestimate the mass of low-mass/high-mass clusters that exhibit radio relics.}
    \label{fig:duffy_vs_mcmc}
\end{figure}

Figure \ref{fig:duffy_vs_mcmc} compares the subcluster mass estimates from the \cite{2008duffy} $c-M$ relation to the 2PNFW method. The fuchsia line is a linear fit with the 95\% confidence interval filled blue. The best-fit line is steeper than the 1:1 ratio (black-dashed line), which suggests that the $c-M$ relation tends to provide higher (lower) mass than the 2PNFW method for low- (high-) mass merging clusters. The $2\sigma$ confidence interval of the linear fit overlaps with the 1:1 ratio at the low-mass end but is offset at the high-mass end.

The high-mass discrepancy is likely caused by the merging nature of the sample. If we assume that the mass discrepancy is directly related to the concentration of the subclusters, then the steeper relation hints that low-mass (high-mass) subclusters have lower (higher) concentration than expected from the power-law $c-M$ \cite{2008duffy} relation. This is not unexpected since \cite{2012ludlow, 2013meneghetti, 2019diemer} have all shown that unrelaxed clusters follow a different $c-M$ relation with an upturn in concentration at high mass. Therefore, it is not surprising that our sample of merging clusters departs from the \cite{2008duffy} $c-M$ relation at the high-mass end. However, this does not explain the low-mass end. One possibility is that the increased gravitational potential that the low-mass subcluster feels during pericenter passage increases the velocity dispersion of galaxies \citep[e.g.][]{1996pinkney, 2010takizawa} and dark matter, which in turn lowers the dark matter concentration and subsequently leads to an overestimation of mass from the \cite{2008duffy} $c-M$ relation. This would preferentially affect the low-mass subclusters since, in this sample, they are more likely to merge with a cluster of comparable or higher mass (see Section \ref{sec:mass_ratio} for mass ratios).

\cite{2023lee} investigated the bias of a WL mass estimate of merging clusters for both the $c-M$ relation and the 2PNFW method. Their simulations showed that the dark matter concentration increased after pericenter for both subclusters involved in the merger, with the lower mass subcluster having a larger increase than the higher mass subcluster. For the $c-M$ relation, this lead to an overestimation of the mass of both subclusters because of the low concentration. This effect persisted from pericenter to approximately 0.75 Gyrs after the collision. At 0.75 Gyrs, the mass bias switched to an underestimation for the low-mass subcluster while remaining an overestimation for the high-mass subcluster. On the other hand, the 2PNFW method resulted in a smaller overall bias shortly after collision (0.25 Gyr), but the mass bias for the lower-mass subcluster flipped from an overestimation to an underestimation by 0.5 Gyrs (earlier than the $c-M$ relation). We do not have a constraint on the time since the collision for the clusters in our sample and thus cannot make a definitive statement on whether the WL mass bias that \cite{2023lee} showed could be the cause of the steeper slope in Figure \ref{fig:duffy_vs_mcmc}. However, it is important to note that these biases could arise in these WL measurements.

\subsection{WL Mass - Velocity Dispersion Scaling Relation}
Merging galaxy clusters are the most energetic events since the big bang \citep{2002sarazin}. This energy is dissipated into the cluster and heats up the ICM. In addition, the increase in gravitational potential caused by the introduction of another cluster will in turn boost the velocity dispersion of galaxies and dark matter for a limited time \citep[e.g.][]{1996pinkney, 2010takizawa}. Therefore, in merging clusters that are observed shortly after collision it is expected that the mass derived from the velocity dispersion of galaxies would be biased high.

\begin{figure}[!ht]
    \centering
    \includegraphics[width=0.45\textwidth]{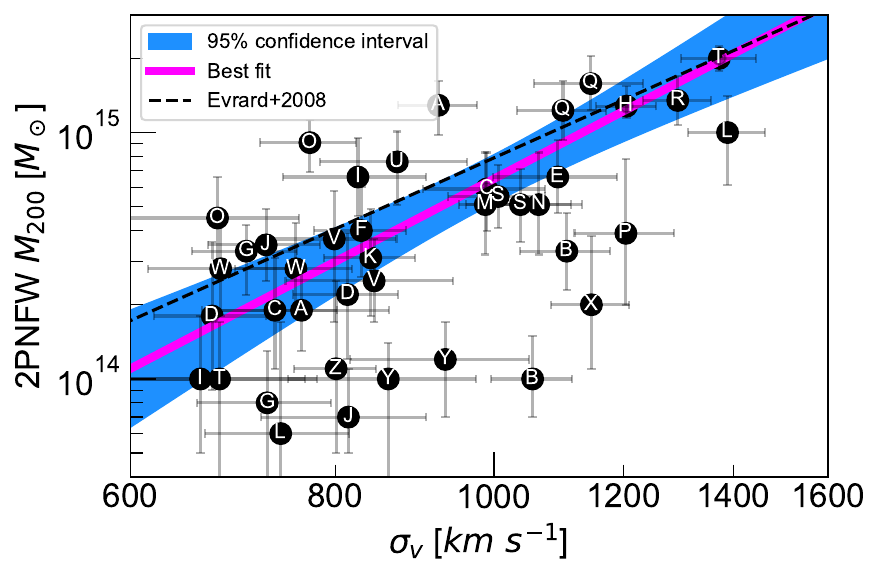}
    \caption{WL mass - $\sigma_v$ scaling relation for subclusters of merging systems that exhibit radio relics. Masses and system index letters are from Table \ref{table:subaru_mass}. The black-dashed line is the \cite{2008evrard} scaling relation. WL mass estimates are preferentially lower than the \cite{2008evrard} relation, especially at the low-mass end.}
    \label{fig:sigma_mass}
\end{figure}

In Figure \ref{fig:sigma_mass}, we plot our WL derived masses from the 2PNFW method against the velocity dispersion of galaxies that are reported in G19. The black-dashed line is the \cite{2008evrard} scaling relation. The best-fit line sits below the scaling relation, which is in agreement with the expectation that velocity dispersion is boosted by the recent merger. It is also interesting that the best-fit line is steeper than the \cite{2008evrard} scaling relation. This may be an indication that low-mass clusters in mergers are more affected than the high-mass counterparts. The scaling relation based on the best-fit line is:

\begin{equation}
    M_{200} = \left(\frac{\sigma_v}{1156\pm381\ \mathrm{km\ s}^{-1}}\right)^{3.8\pm0.4}\ 10^{15}\  \mathrm{M}_\odot
\end{equation}

An alternative explanation for the steeper slope could be systematics. Measuring the velocity dispersion relies on the assignment of galaxies into groups. In the crowded environment of a merging galaxy cluster it is difficult to perform this assignment. G19 used a GMM technique to perform the assignment. Naturally, it is more likely that a galaxy belongs to a high-mass halo than a low-mass halo. In addition, the global velocity dispersion will favor that of the high-mass halo. Thus, one would expect that the systematic of assigning galaxies to halos would preferentially affect low-mass halos, to the point that even some low-mass halos are completely missed.

\subsection{Mass ratio}\label{sec:mass_ratio}
The energy required to create megaparsec scale merger shocks is expected to be high and is dependent on cluster mass \citep[e.g.,][]{2014degasperin}. To better understand the types of mergers that are occurring that lead to the creation of radio relics, we plot the mass ratio of the merging subclusters.

\begin{figure}[!ht]
    \centering
    \includegraphics[width=0.45\textwidth]{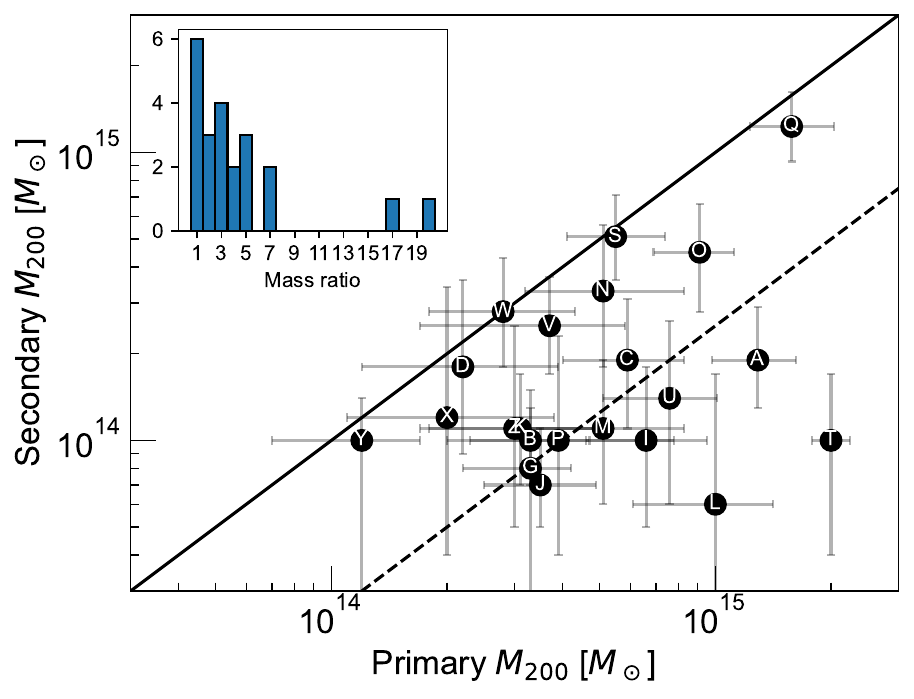}
    \caption{Mass ratio of merging clusters that exhibit radio relics. Masses are from the 2PNFW method. Masses and system index letters are from Table \ref{table:subaru_mass}. The solid (dashed) black line represents the 1:1 (1:4) mass ratio. Most of the radio relic systems are in major mergers (ratio $<4$). }
    \label{fig:mass_ratio}
\end{figure}

Figure \ref{fig:mass_ratio} compares the WL mass of the primary and secondary subclusters. These were selected from the substructures of each merging cluster by their alignment with the X-ray- and radio-defined merger axes, their agreement with BCG locations, and their WL $S/N$. The figure shows that many of the mergers are close to 1:1 mass ratio, independent of the mass scale. Furthermore, most of the clusters (\mytilde70\%) are in major mergers with mass ratio $<4$:1. The few outliers tend to be clusters with very massive primaries, such as PLCKG287 and A2163. Perhaps this is an indication that very massive major mergers are rare. However, there may be a selection effect because low-mass primaries ($\lesssim 10^{14} M_\odot$) would require very low-mass secondaries ($< 10^{14} M_\odot$), which are below the detection threshold of this Subaru WL dataset.

\subsection{Peak Separation}\label{sec:peak_separation}
Given the large number of subclusters analyzed in this work, we can compare the separation of various tracers of the dark matter potential, which may provide insight into the physics of the collision \citep[e.g.,][]{2017kim}. In Section \ref{section:convergence peaks}, we identified WL peaks by correlating them with nearby galaxies and/or X-ray brightness peaks. Therefore, this comparison is done knowing that the luminous tracer was initially used in identification of the WL peaks.

\begin{figure}
    \centering
    \includegraphics[width=0.5\textwidth]{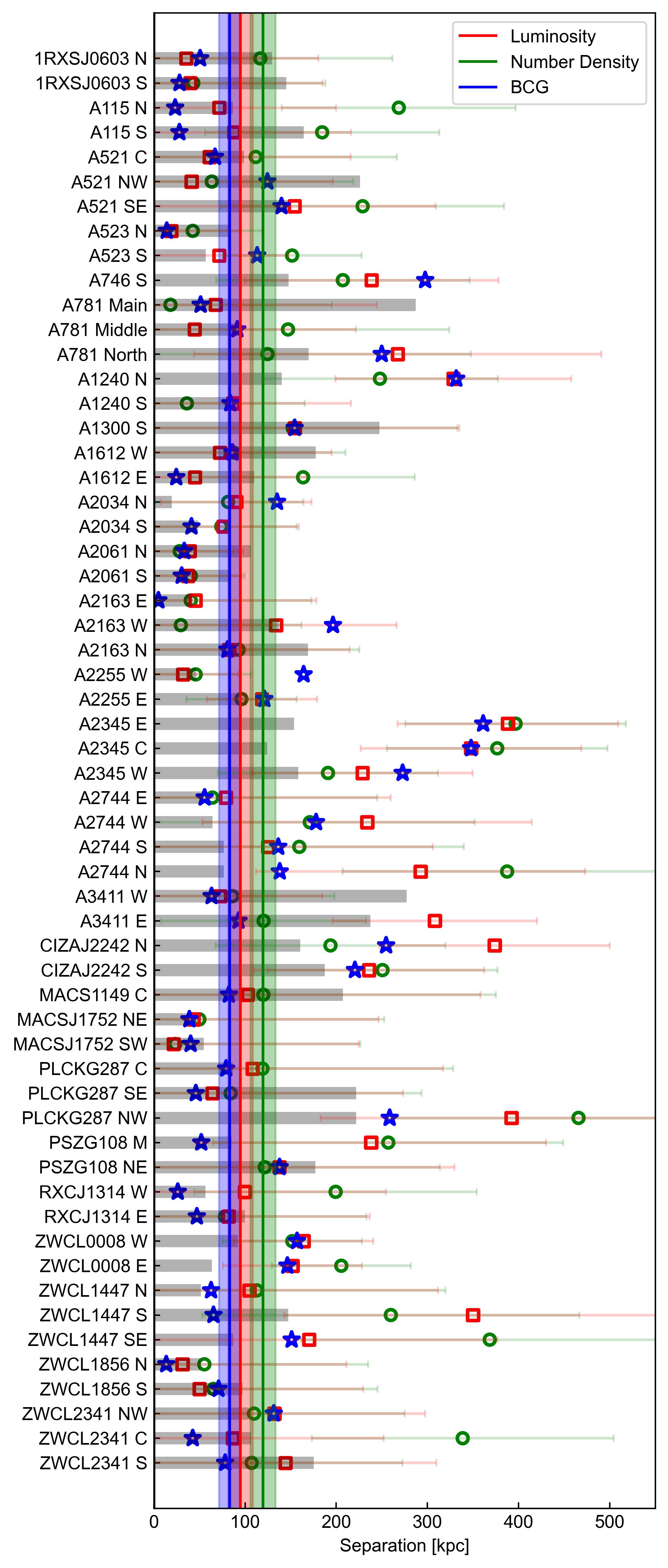}
    \caption{Projected separation of the BCGs (blue stars), luminosity peaks (red squares), and number density peaks (green ) from the WL peak. The horizontal grey bars represent the uncertainty on the mass peak position. The vertical colored lines show the median separations of each luminous tracer from the mass peak and their 68\% uncertainty.}
    \label{fig:peak_separations}
\end{figure}

Figure \ref{fig:peak_separations} shows the projected separation of the WL peak from the nearest BCG, luminosity peak, and number density peak for each subcluster. For each system, the uncertainty on the location of the mass peak for each subcluster was estimated by collecting the mass peak location in the 1000 bootstrapped mass maps. The peaks were then processed with a k-means clustering algorithm with the number of clusters fixed to the expected value (i.e., 2 for a merging cluster that has 2 subclusters). The distribution of each of the k-means-defined subclusters were then used to derive the error ellipse that contained 68\% of the peaks.

For the majority of subclusters, the uncertainty on the mass peak has a $1\sigma$ overlap with the positions of the luminous tracer. We find that the BCG is closer than the luminosity (number density) peak to the mass peak in 72\% (68\%) of the subclusters. The median values (vertical lines in Figure \ref{fig:peak_separations}) show that the BCG is the best tracer of the mass peak. The median projected separation of the BCG, the luminosity peak, and the number density peak from their respective WL peak is $79\pm14$, $90\pm15$, and $119\pm15$ kpc, respectively. Our median projected separation of the BCG from the mass peak agrees with \cite{2010oguri} who found that the separation follows a Gaussian distribution with $\sigma=90$ kpc. \cite{2012george} compared the WL mass peaks to BCGs in X-ray detected galaxy groups and found the average projected separation to be less than 75 kpc. Using the large sample of 10,000 SDSS clusters, \cite{2012zitrin} studied the offset of the BCG from the DM peak and found that the average offset is \mytilde13 kpc. For our sample, only 1 of the 58 subclusters has a BCG within 13 kpc of the mass peak. However, the uncertainties on the mass peak are on average approximately 100 kpc. Therefore, these results are not statistically conclusive.

An additional tracer of the DM potential that is beyond the scope of this work is the intracluster light (ICL). \cite{2022yoo, 2024yoo} show in simulations that the ICL is a good tracer of the DM potential. We will explore the possibility of using the ICL as a tracer of the DM potential in future work.

\subsection{The Golden Sample}\label{sec:gold_sample}
Dissociation of the ICM from the cluster potential by ram pressure during a cluster merger is a powerful feature that can be used to understand the merger scenario and constrain the physics of mergers \citep[for an example see][]{2019zhang}. However, the variety of gas dissociation in merging clusters is diverse. Clusters such as A115 have broad V-shaped X-ray emission tails that are as wide as they are long. In contrast, the tails behind the subclusters of MACSJ1752 extend approximately 0.5 Mpc but are only \mytilde100 kpc at their greatest width. The relation of X-ray brightness peaks and WL peaks also vary. In some dissociative mergers, like A115, the X-ray brightness peaks are retained in the WL peak. Conversely, in A2034, the X-ray brightness peaks trail behind the WL peaks. As demonstrated in simulations by \cite{2018zuhone}, the differences may be caused by impact parameter, pre-merger concentration of the ICM, total mass, concentration of the DM, or projection.

Achieving a sound understanding of cluster mergers relies on carefully recreating the collision on a computer. Cases like A2744 and A746 have multiple subclusters that are merging along many axes, which makes them difficult to reproduce. For constraining the properties of merging clusters, it is best to start with the simplest cases. G19 defined a ``gold sample'' of cluster mergers that have simple geometry. Having completed WL analyses of the systems, we provide an updated gold sample. These systems have distinct merger features and mostly bimodal mass distributions. Our updated gold sample is A1240, A2034, MACSJ1752, RXCJ1314, ZwCl0008, ZwCl1447 and ZwCl1856. CIZAJ2242 is also a strong candidate but, at this point, the existing optical/IR observations of the cluster cannot support a WL analysis of the precision that is needed to fully understand the complexity of the mass distribution. Aside from A2034, these systems all contain double radio relics, which provide critical constraints on the merger timescale and viewing angle. There are additional candidate systems that may be valuable, such as A1612, but the current multiwavelength data do not support a robust analysis.

\section{CONCLUSIONS}\label{Section:conclusions}
We present WL analyses of 29 merging galaxy clusters that exhibit megaparsec-sized radio relics. The WL analyses are performed on Subaru Suprime-Cam and Hyper-Suprime-Cam observations. Of the 29 systems, WL analyses are successfully achieved for 26 clusters and 3 clusters are unsuccessful.

Our WL analysis technique uses a PCA to create PSF models for each galaxy and a Gaussian model to fit the galaxy shape. Source galaxies are selected from CMDs utilizing the spectroscopically confirmed red-sequence galaxies as a guide. The source galaxies are used to generated high-resolution mass maps with the FIATMAP code. Comparing the mass maps with the vast multiwavelength supplementary data, significant (WL $S/N>3$) subclusters are identified.
Mass estimates for each subcluster of a system are determined by simultaneously fitting multiple NFW halos to the source galaxies. The mass estimates are performed by both fitting a $c-M$ relation \citep{2008duffy} and by MCMC sampling the $c$ and $M$ parameter space.  With subclusters identified and masses estimated, an in-depth discussion of each merging system is presented. From the analysis of each system and the analysis of the radio relic sample as a whole, the following conclusions are made.

\begin{itemize}
\item We combine our WL results with the multiwavelength observations from literature to develop a better understanding of each merging system. We show that WL analysis is critical to detecting and quantifying the substructures in merging clusters and in some cases discover subclusters that were previously undetected. Utilizing the new information provided through WL analysis, we update the merging scenarios for each system and give new insight into the past collisions. In addition, we highlight peculiar systems that warrant further investigation, such as the dissociative nature of Abell 2061 and its large radio relic distance.

\item We compare WL mass maps to the X-ray emission from the ICM. In cases such as MACSJ1752 and ZWCL1856, we find that the X-ray emission, WL mass maps, and radio relics paint a clear picture of the past merger. In others, the complexity of the system and the positions of the radio relics are less straightforward. Gas dissociation is clearly visible in many of the clusters. In general, the overall distribution of the mass maps is elongated in the same manner as the X-ray emission.

\item We compare the mass distributions to the galaxy light and galaxy number density distributions. A strong correlation between mass peaks and galaxies are obvious. In most cases, the most massive subcluster is co-spatial with the global luminosity and number density peaks of the system. This finding is not surprising if more massive halos are expected to have more galaxies. However, in clusters such as A1240 and A1612, the more massive subcluster is not associated with the global luminosity and number density peaks. The galaxy distributions in general show the same elongation as the X-ray and WL signals and in some cases stretch further along the axis of elongation.

\item We investigate the influence of method on the mass estimation by comparing the use of a $c-M$ relation to a mass estimate that samples both $c$ and $M$. We find that the two methods depart slightly from a 1:1 relation with the fixed $c-M$ mass estimate technique providing a smaller (larger) mass estimate for high- (low-) mass subclusters. We suggest the primary reason for this discrepancy is that the $c-M$ relation is derived from simulations that include both relaxed and unrelaxed clusters.

\item The WL-derived masses are compared to the velocity dispersion measurements, a proxy for mass, of each subcluster. We find that the scatter of the velocity dispersion with respect to the WL mass is large. Lower mass subclusters tend to have more inflated velocity dispersions than high mass systems when compared to the scaling relation from \cite{2008evrard}. We suggest that this could be caused by the increased gravitational effect that slow-mass subclusters feel during pericenter passage, which inflates the velocity dispersion and leads to the large scatter in comparison to WL mass.

\item For each system, primary and secondary subclusters that are responsible for the generation of the radio relics are distinguished by their mass and alignment to the radio relics. Comparing the mass of the primary to secondary subclusters, we find that most merging clusters that exhibit radio relics are major mergers with a mass ratio less than 4.

\item We analyze the separation of the peak of the dark matter distribution from other luminous peaks. On average, the BCG most closely traces the dark matter peak followed by the luminosity and then the number density peak. The average offset of the BCG from its mass peak is found to be $75\pm14$ kpc.

\item An updated sample of 8 ideal merging clusters were defined (a Golden sample). The ideal cases have clear X-ray features, such as bullet morphology that highlight the merger path, and have verified merger-induced radio relics. Additional important selection criteria come from the WL signal: robust detection of the WL signal, dominant bimodal distribution, and dissociation of the gas from the dark matter density peaks. This sample of merging clusters represent the ideal cases that can be reproduced through simulations. Furthermore, the large separation of the mass peaks from the gas peaks makes this sample idea for studying the nature of dark matter.

\end{itemize}

This work was supported by the National Research Foundation of Korea(NRF) grant funded by the Korea government(MSIT) (RS-2024-00340949). RJvW acknowledges support from the ERC Starting Grant ClusterWeb 804208. This research is based on data collected at the Subaru Telescope, which is operated by the National Astronomical Observatory of Japan. We are honored and grateful for the opportunity of observing the Universe from Maunakea, which has the cultural, historical, and natural significance in Hawaii. MGCLS data products were provided by the South African Radio Astronomy Observatory and the MGCLS team and were derived from observations with the MeerKAT radio telescope. The MeerKAT telescope is operated by the South African Radio Astronomy Observatory, which is a facility of the National Research Foundation, an agency of the Department of Science and Innovation. Based on observations obtained with XMM-\textit{Newton}, an ESA science mission with instruments and contributions directly funded by ESA Member States and NASA. This research has made use of data obtained from the Chandra Data Archive and the Chandra Source Catalog, and software provided by the Chandra X-ray Center (CXC) in the application packages CIAO and Sherpa. This paper makes use of LSST Science Pipelines software developed by the Vera C. Rubin Observatory. We thank the Rubin Observatory for making their code available as free software at https://pipelines.lsst.io.

\software{Astropy \citep{astropy},
          Matplotlib \citep{matplotlib},
          Scikit-learn \citep{scikit-learn},
          IRAF \citep{iraf},
          SDFRED1 \citep{sdfred1},
          SDFRED2 \citep{sdfred2},
          SExtractor \citep{1996bertin},
          SCAMP \citep{bertinscamp},
          SWarp \citep{bertinswarp},
          CIAO \citep{ciao}
          }
\bibliography{sample631}{}
\bibliographystyle{aasjournal}

\end{document}